\documentclass[apj]{emulateapj}
\epsscale{0.87}
\newcommand{\upscale}{\epsscale{.99}}
\newcommand{\downscale}{\epsscale{0.8}}
\newcommand{\returnscale}{\epsscale{0.87}}

\usepackage{epstopdf}

\newcommand{\kms}{\,km~s$^{-1}$}

\newcommand{\Msun}{\mbox{\,$M_{\odot}$}}

\def\spose#1{\hbox to 0pt{#1\hss}}
\def\simlt{\mathrel{\spose{\lower 3pt\hbox{$\mathchar"218$}}
     \raise 2.0pt\hbox{$\mathchar"13C$}}}
\def\simgt{\mathrel{\spose{\lower 3pt\hbox{$\mathchar"218$}}
     \raise 2.0pt\hbox{$\mathchar"13E$}}}

\shorttitle{NGC~205}
\shortauthors{Howley~et~al.}

\begin{document}

\title{Darwin Tames an Andromeda Dwarf: \\
Unraveling the Orbit of NGC~205 Using a Genetic Algorithm}

\author{K.~M.\ Howley\altaffilmark{1}, M.\ Geha\altaffilmark{2}, P.\ Guhathakurta\altaffilmark{1}, R.~M.\ Montgomery\altaffilmark{1}, G.\ Laughlin\altaffilmark{1}, K.~V.\ Johnston\altaffilmark{3}}

\altaffiltext{1}{UCO/Lick Observatory, University of California
    Santa Cruz, 1156 High Street, Santa Cruz, CA~95064; {\tt [kirsten, raja, rmontgom, laugh]@ucolick.org}}
\altaffiltext{2}{Astronomy Department, Yale University, New Haven, CT 06510; {\tt marla.geha@yale.edu}}
\altaffiltext{3}{Department of Astronomy, Columbia University, 
550 West 120th Street, New York, NY~10027; {\tt kvj@astro.columbia.edu}}
 
   \begin{abstract}
\renewcommand{\thefootnote}{\fnsymbol{footnote}}

NGC~205, a close satellite of the M31 galaxy, is our nearest example of a dwarf elliptical galaxy.  
Photometric and kinematic observations strongly suggest that NGC~205 is currently undergoing tidal distortion as a result of its interaction with M31.  
Despite earlier attempts, the orbit and progenitor properties of NGC~205 are not well known.
In this paper, we present an optimized search for these unknowns by combining a genetic algorithm with restricted $N$-body simulations of the interaction. 
This approach, coupled with photometric and kinematic observations as constraints, allows for an effective exploration of a 10-dimensional parameter space.
We represent the gravitational potential of M31 as a static analytic bulge-disk-halo model.  
NGC~205 is modeled as a static Hernquist potential with embedded  mass-less test particles that serve as tracers of surface brightness. 
We explore three distinct, initially stable configurations of test particles: a cold rotating disk (no velocity dispersion), a warm rotating disk (small amount of velocity dispersion), and a hot, pressure-supported spheroid (isotropic velocity distribution with no rotation).  
Each of these models is able to reproduce some, but not all, of the observed
features of NGC~205.  This leads us to speculate that a rotating
progenitor with substantial pressure support could match all of the
observables.
Furthermore, plausible combinations of mass and scale length for the
pressure-supported spheroid model of the progenitor of NGC 205 reproduce the
observed velocity dispersion profile.
For all three models, we find that NGC~205's line-of-sight distance and proper motion are well constrained by NGC~205's kinematic profile and surface brightness distribution.   
Orbits that best match the observables place the satellite $11\pm9$ kpc behind M31 moving at very large velocities: $300-500$ \kms\  on primarily radial orbits.  Given that the observed radial component is only 54 \kms, this implies a large tangential motion for NGC~205, moving from the north-west towards the south-east, that translates into a predicted proper motion of $\sim 0.1$ mas yr$^{-1}$.  These results suggest that NGC~205 is not associated with the stellar arc observed to the north of M31 and to the north-east of NGC~205.  Furthermore, NGC~205's velocity appears to be near or greater than its escape velocity, signifying that the satellite is likely on its first M31 passage.

\end{abstract}
 
 
   \section{INTRODUCTION}\label{intro_sec}
\renewcommand{\thefootnote}{\fnsymbol{footnote}}

In the hierarchical model of galaxy formation, low mass dwarf-sized
galaxies collapse and merge to form the massive
galactic structures that we observe today.  Minor mergers in the
present-day Universe continue to influence the properties of stellar
halos \citep{bul05}, and perhaps the appearance of
galactic disks \citep{gil02, aba03, yoa05}.  How
this process of minor merging works in detail is not well understood.  
The Local Group is a prime locale for studying minor merging events.
In the Milky Way, three satellites are undergoing
tidal disruption and accretion: the Magellanic Stream \citep*{mat74,con06},
Sagittarius dwarf \citep*{iba94,maj03,new03}, and Canis Major dwarf \citep{mar05}.
The Andromeda Spiral Galaxy (M31), our nearest large neighbor, also hosts a number of
tidally distorted and disrupted satellites. These include the dwarf elliptical galaxy NGC~205 \citep*{ken87, cho02, geh06}, the compact elliptical galaxy M32 \citep{ken87, cho02}, the Giant Southern Stream \citep{iba01, fon06, far07}, and possibly the satellite galaxy Andromeda VIII \citep{mor03}.  
In this paper, we present a detailed study of the tidal interaction between M31 and the dwarf elliptical galaxy NGC~205.

In addition to being building blocks of larger galaxies, dwarf galaxies 
are interesting in their own right.
Of the various dwarf galaxy types, dwarf elliptical (dE) galaxies are 
the least understood, and their origin is heavily debated.  
What is known about dEs is that they are a galaxy population found 
exclusively in denser regions, accounting for more than 75\% of the 
objects in cluster environments \citep{bin87,tre02}.  
They are typically dark matter poor in their inner regions and predominantly gas poor.
The two general dE formation theories postulate that these galaxies are primordial
building blocks or an evolved/transformed population, but neither theory is 
capable of explaning all the observables.  
The first theory, based on Cold Dark Matter models, asserts that these are low
mass, primordial building blocks that formed early ($z>6$) in their
present locale.  If dEs formed by this scenario, we could gain insight
into the very first galaxies to form by studying these relics.     
The second theory postulates that dEs are born as more massive spiral
galaxies which are subsequently altered as a result of interactions
with other galaxies through tidal effects \citep*{moo98, ric05}.  
If dEs are transformed spirals then this transformation is expected to 
cause an increase in the velocity dispersion and induce some rotation 
\citep{moo98}.  However, this scenario is unable to explain the observed 
distribution of rotation speeds and velocity dispersions among Virgo 
Cluster dEs.  Specifically, it cannot explain the presence of a 
substantial population of dEs with no detectable rotation and relatively 
low velocity dispersion \citep*{geh02, geh03}.  To complicate matters 
further, the faint underlying disks are found in both rotating and 
non-rotating dEs \citep{geh03, lis06}.
It is also possible that the dEs result from some combination of the
above two mechanisms.
A complete model explaining dE formation is
still needed.  Local group dE galaxies offer an excellent resource for the
exploration of these and other formation scenarios.

Unlike the Milky Way which does not host any dE satellites, M31 hosts a total of three: NGC~185, NGC~147, and NGC~205.  
NGC~185 and NGC~147 lie far enough away from M31 to escape present tidal distortion.  NGC~205 lies a projected $37\arcmin$ from M31 and is our nearest example of a tidally distorted dE galaxy.  
It is a gas poor, low luminosity, early-type galaxy with an exponential surface brightness profile \citep{cho02}, defining features of dE galaxies.  
The little gas and dust in NGC~205 is concentrated within a $1\arcmin$ radius, beyond which the satellite is essentially gas and dust-free \citep*{you97, wel98}.  
The current line-of-sight distance estimate between NGC~205 and M31 is about 39 kpc, with M31 residing $785 \pm 25$ kpc and NGC~205 at  $824 \pm 27$ \citep{mcc05}.  Throughout this paper, we adopt this distance of 824 kpc to NGC~205.
The absolute $V$-band magnitude of NGC~205 is $M_V = -16.5$, corresponding to a few percent of M31's luminosity \citep{hod92, cho02}.
\citet*{ben91} reports a mass-to-light ratio for NGC~205 of $(M/L) \approx 7$.   
This value is consistent with dynamical models of dEs that suggest a global mass-to-light ratio of $3 \le \Gamma_{V} \le 6$ and is indicative of an intermediate to old stellar population containing little to no dark matter within an effective radius \citep{geh02}. 
\citet{der06} finds a $B$-band mass-to-light ratio within $2R_{e}=260\arcsec$ (1.04 kpc) of $(M/L)_B = 4.5^{+1.5}_{-1.0} (M/L)_{\Sun}$.  This supports the conclusion that there is little dark matter contained within the inner regions of NGC~205.  
It is not known whether NGC~205 resides in an {\it extended\/} dark matter halo.

There is strong evidence that NGC~205 is tidally distorted by M31.  
\citet{cho02} measure twisting of elliptical isophotes, along with a subtle break in the surface brightness profile, providing confirmation of tidal distortion found in earlier studies.  A radial velocity survey of red giant branch stars (RGB) in NGC~205 by \citet{geh06} notes an abrupt turnover in the semi-major axis velocity profile coincident with the tidal distortion reported by \citet{cho02}.   
\citet{mcc04} postulates that the $\sim1^\circ$ (15 kpc) long, arc-like feature seen in the northwest quadrant of M31 could be a stellar stream associated with NGC~205.
\citet*{dem03} report possible evidence of tidal debris to the west of NGC~205 beyond its tidal radius in their study of C stars in the system.
Within an effective radius, NGC~205 appears to be undistorted.  However, beyond this radius, there is strong evidence for distortion.

In this paper we investigate the large parameter space defining the orbit and history of NGC~205 using a restricted $N$-body code and a genetic algorithm for optimization.  
Modeling satellite/parent galaxy interactions is extremely difficult when there are no gaseous or stellar streams to directly constrain the path of the satellite's orbit.  In the absence of tidal trails, as is the case for NGC 205, there is a limited set of observable quantities that can be used as constraints: stellar radial velocities, projected sky position, and constraints on the line-of-sight distance.  To complicate matters further, large parameter spaces with inherent degeneracies typically define galaxy encounters.  Hence, in weakly distorted systems, it is difficult to find unique, qualified fits to the observables.  For this reason, few optimized $N$-body studies of mildly distorted galaxies have been carried out.  In our simulations we employ a genetic algorithm to achieve optimization \citep{hol75, gol89}.  
This particular optimization technique has been successfully used in similar studies of  more heavily distorted dwarf galaxies \citep*{the01, the2001}.   Using a restricted $N$-body code, we explore three distinct, initially stable configurations of 1000 test particles representing NGC~205: a (cold) rotating disk without velocity dispersion, a (warm) rotating disk with velocity dispersion, and a (hot) spheroid supported completely by randomly distributed, isotropic velocities.  Detailed photometric and kinematic observations enable this unique study by providing strong constraints on the simulated system \citep{cho02, geh06}.

The paper is laid out as follows: 
in \S\,\ref{sec_observ} we discuss the photometric and kinematic observational data used to constrain our system.  
In \S\,\ref{sec_numerical} we give the details of the parameters, galaxy models, and methods used in our numerical simulations.
In \S\,\ref{sec_results} we present the results of the numerical simulations.
In \S\,\ref{sec_discussion} we discuss the implications of our results. 
Finally, in \S\,\ref{sec_future} we discuss the observational and computational improvements that could be applied to this study.  

   \section{OBSERVATIONAL DATA}\label{sec_observ}
\begin{figure*} \includegraphics[trim=0in 4.4in 0in 0in,clip]{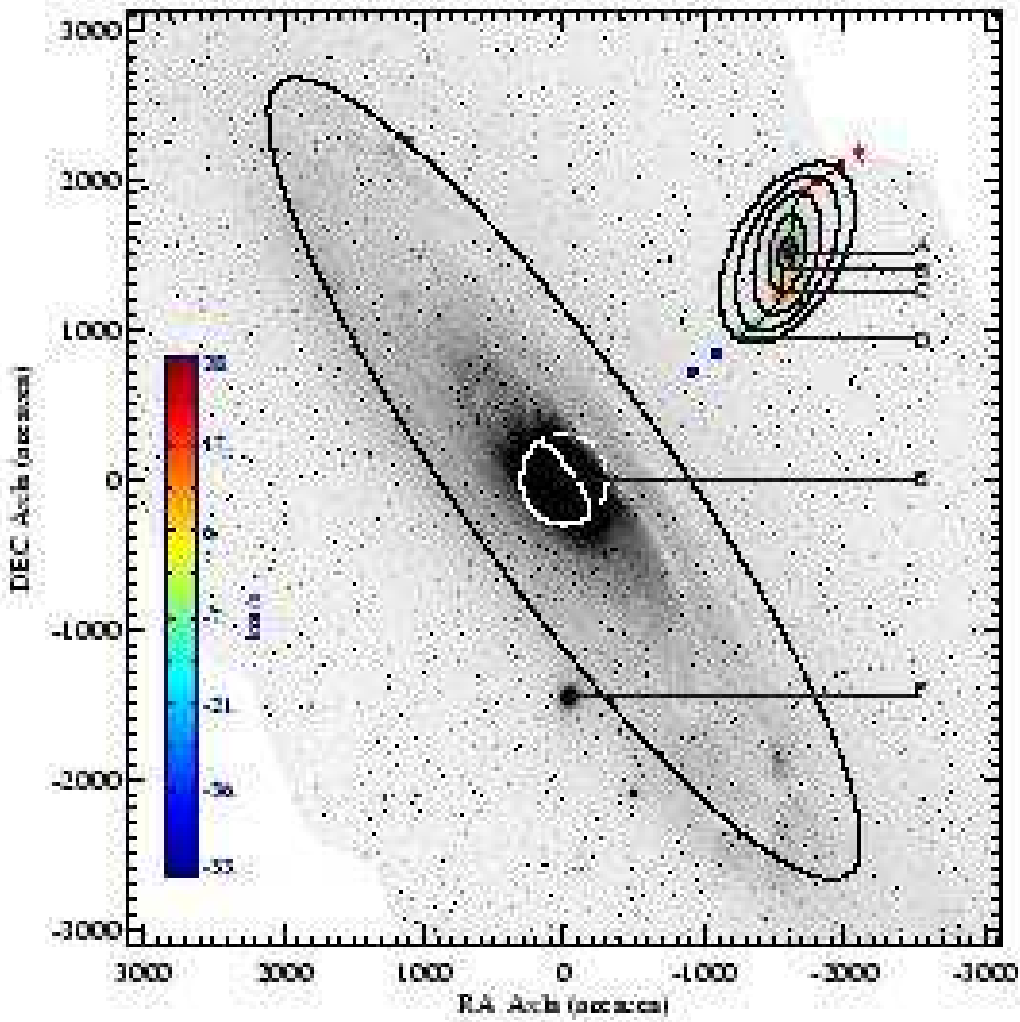}
\caption{Observed properties of the M31-NGC~205 system.  The colored dots represent NGC~205's semi-major axis velocity relative to its systemic velocity of $-$246 \kms\  as measured by \citet{geh06}.  Note the turnover in velocity at the semi-major axis distance of $\sim 270\arcsec$ (1.08 kpc) is coincident with the isophotal twisting radius measured by \citet{cho02}.  The labels A--F are defined as follows: A=center of NGC~205, B=radius at which the use \citet{cho02} data begins (internal to this we use an exponential surface brightness profile), C=radius at which isophotal twisting and radial velocity turnover occurs, D=radius at which the use of \citet{cho02} data ends (external to this we use an exponential law), E=center of M31, F=center of M32. The \textit{projected} distance between NGC~205 and M31 (points A and E) is $36.6 \arcmin$, or 8.8 kpc if we take NGC~205 to be at a distance of 824 kpc \citep{mcc05}.  The photometric contours B, C, \& D lie at semi-major axis distances of $100 \arcsec$ (0.40 kpc), $260 \arcsec$ (1.04 kpc), and $672 \arcsec$ (2.68 kpc), respectively.  (Image courtesy of Phil Choi)}  
\label{fig:m31_n205}
\end{figure*}

Our study of NGC~205 is unique because we enforce both photometric and kinematic constraints on our simulated system.  Although this cannot ensure that we converge on the correct orbital history of NGC~205, it does remove a significant amount of the degeneracy inherent in studies that only use photometric constraints.  Strong evidence of tidal distortion is seen both in the photometry and the kinematics of NGC~205 at a major-axis distance of $r \sim 4.5\arcmin - 5\arcmin$ ($1.1 - 1.2$ kpc) \citep{cho02, geh06}.  A discussion of the surface photometry by \citet{cho02} is included in \S\,\ref{ssec_photo}, and a discussion of the internal kinematics by \citet{geh06} is included in \S\,\ref{ssec_kinematic}.  These observations are summarized in Figure \ref{fig:m31_n205}.
\downscale
\subsection{Photometric Data}\label{ssec_photo}
The surface photometry of NGC~205 in the \textit{B}- and \textit{I}-bands was obtained by \citet{cho02} using the Kitt Peak National Observatory 0.9/0.6 meter Burrell Schmidt telescope.  Due to the incompleteness of the \textit{I}-band coverage, \citet{cho02} employed the \textit{B}-band data and IRAF's ELLIPSE task to determine the surface brightness contours of NGC~205 out to a limiting magnitude of $\mu_{B} = 27$ mag arcsec$^{-2}$, coinciding with a semi-major axis distance of 720\arcsec (2.88 kpc).

\citet{cho02} reports that NGC~205 is well fit by two different exponential profiles at various semi-major axis radii.  One profile, covering a semi-major axis range of $75 \arcsec < r < 250\arcsec$ ($0.30 < r < 1.00$ kpc), has a disk scale length of $r_{0}^{\rm exp}=150\arcsec$ (0.60 kpc), where $r_{0}$ is defined as the radius at which the intensity has decreased by a factor of $e$; the other profile, covering a semi-major axis range of $150 \arcsec < r < 250\arcsec$ ($0.60 < r < 1.00$ kpc) and $r>500\arcsec$ (2.00 kpc), has a disk scale length of $r_{0}^{\rm exp}=170\arcsec$ (0.68 kpc).  Also reported was a ``subtle downward break" at $r=300\arcsec$ (1.20 kpc), which correlates nicely with the semi-major axis distance of $r=260\arcsec$ (1.04 kpc) that marks the location where the isophotal position angle and ellipticity stop increasing and begin decreasing, leading to a \textit{S}-shaped semi-major axis profile for NGC~205.  The location of this isophotal twisting matches the radial velocity turnover radius of $4.5\arcmin$ ($270\arcsec$ or 1.08 kpc) observed by \citet{geh06}, thus supporting the conclusion that NGC~205 is interacting tidally with M31.  

\begin{figure*}\includegraphics[trim=0in 0in 0in 5.5in,clip]{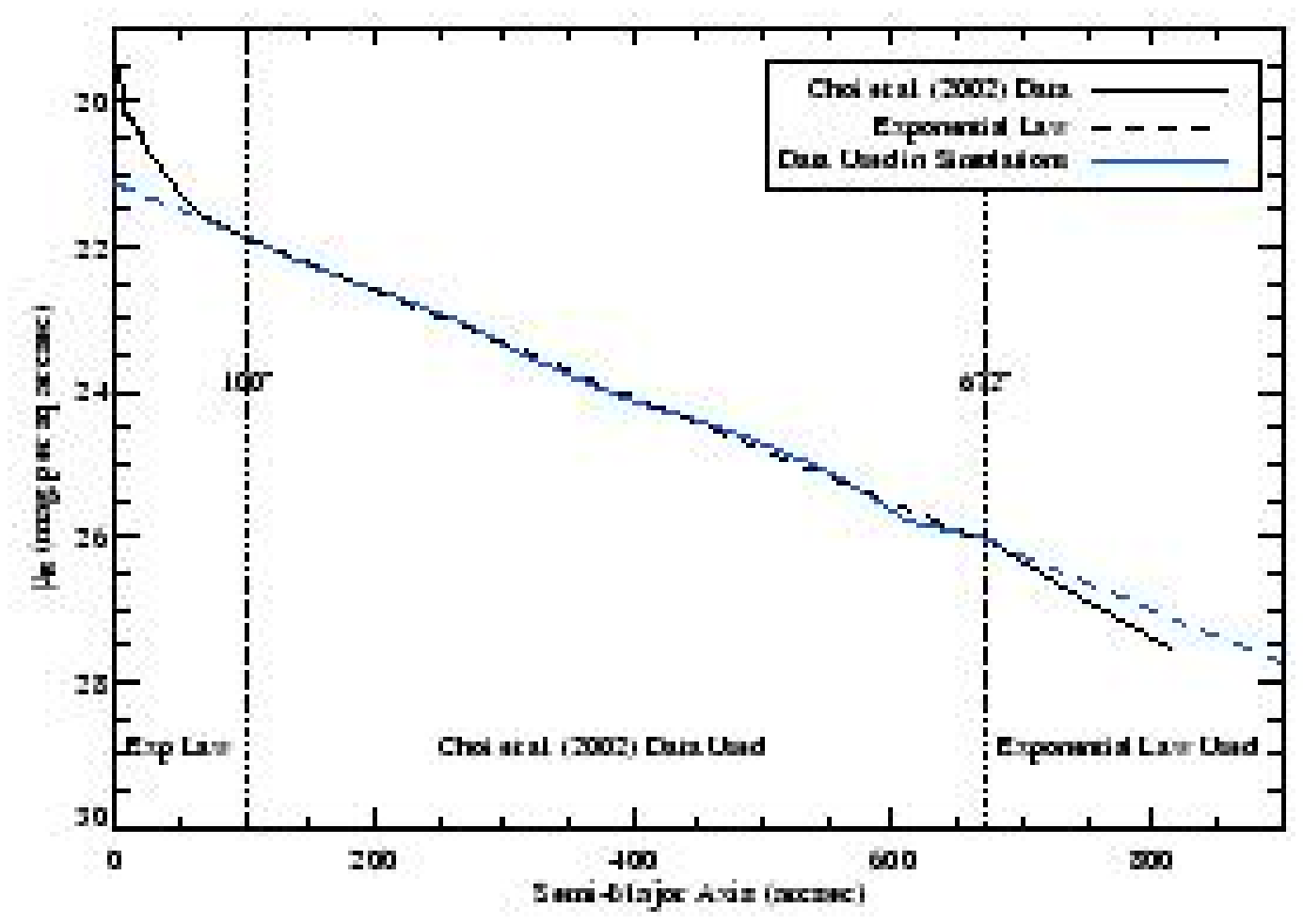}
\caption{Surface brightness profile of NGC~205 in magnitudes per square arcsecond versus semi-major axis distance.  The solid line illustrates the brightness profile in B magnitudes measured by \citet{cho02}, the dashed line represents the exponential fit performed using the intermediate regions of \citet{cho02} data ($100 \arcsec$ to $672 \arcsec$), and the blue line defines the data used in our simulations.  Note that in the intermediate regions ($100 \arcsec$ to $672 \arcsec$) we use \citet{cho02} measurements, and in the inner ($>100 \arcsec$) and outer ($<672 \arcsec$) regions we use our exponential fit.  This effectively removes the nucleus from the inner region and models the data in the outer region where the surface brightness of the galaxy and the sky brightness become comparable.}  
\label{fig:intensity}
\end{figure*}

We perform our own exponential fit of the form:
\begin{equation}
I(r) = I_{0} e^{-r/r_{0}}
\label{eq:eprofile}
\end{equation} 
to the intermediate parts of \citet{cho02} intensity profile in order to estimate the disk scale length of the galaxy if it were tidally undistorted.  We exclude the nucleus (interior to 100\arcsec, or 0.40 kpc) where the data follows neither an exponential law (exp[${-r/r_{0}}$]) nor a de Vaucouleurs law (exp[${-kr^{1/4}}$]), and the outer regions (exterior to 672\arcsec, or 2.68 kpc) where the surface brightness and sky become comparable.  We measure a disk scale length of $r_{0}=148\arcsec$ (0.59 kpc) and a central (nucleus-free) surface brightness of $I_{0}=21.1$ magnitudes per square arcsecond.  These results are used as a basis for generating an undistorted NGC~205 and modeling the inner and outer regions of NGC~205 in its final, integrated state.  Hence, we are able to remove the nucleus from NGC~205 by extrapolating the exponential surface brightness profile from the intermediate region ($100\arcsec - 672\arcsec$) to the central region ($0\arcsec - 100\arcsec$).  The surface brightness profile of \citet{cho02}, our exponential surface brightness fit, and the data used in our simulations are summarized in Figure \ref{fig:intensity}.
\subsection{Kinematic Data}\label{ssec_kinematic}
A recent kinematic study of NGC~205 by \citet{geh06} using the Keck/DEIMOS multislit spectrograph resulted in the collection of radial velocities for 725 RGB stars.  
The stellar spectra were obtained using 4 masks, each covering an area of $\approx 16\arcmin \times 4\arcmin$ ($3.83 \times 0.96$ kpc).  Two of these masks were centered on NGC~205 along the major axis, while the other two were placed off center along the tidally distorted major axis, one in south-east (SE) and one in north-west (NW).  These observations extend out to $20\arcmin$ (4.79 kpc), well beyond the tidal radius of NGC~205, and are spatially well distributed.  The contamination by M31 field stars is estimated to be only a few percent.  

The systemic velocity of NGC~205, as measured by \citet{geh06}, is $-246 \pm 1$\kms, while the systemic velocity of M31 is $-300 \pm 4$\kms\  \citep{dev91}.  This results in a velocity for NGC~205 of $54 \pm 5$\kms\  relative to M31.  The velocities of the NW tail of the satellite are more positive, while the SE tail, which is closer to M31, are more negative as compared to NGC~205's systemic velocity.  The opposite is true in the central regions ($< 270\arcsec$).

\begin{figure*}\includegraphics[trim=.5in 2.5in 0in 2.5in,clip]{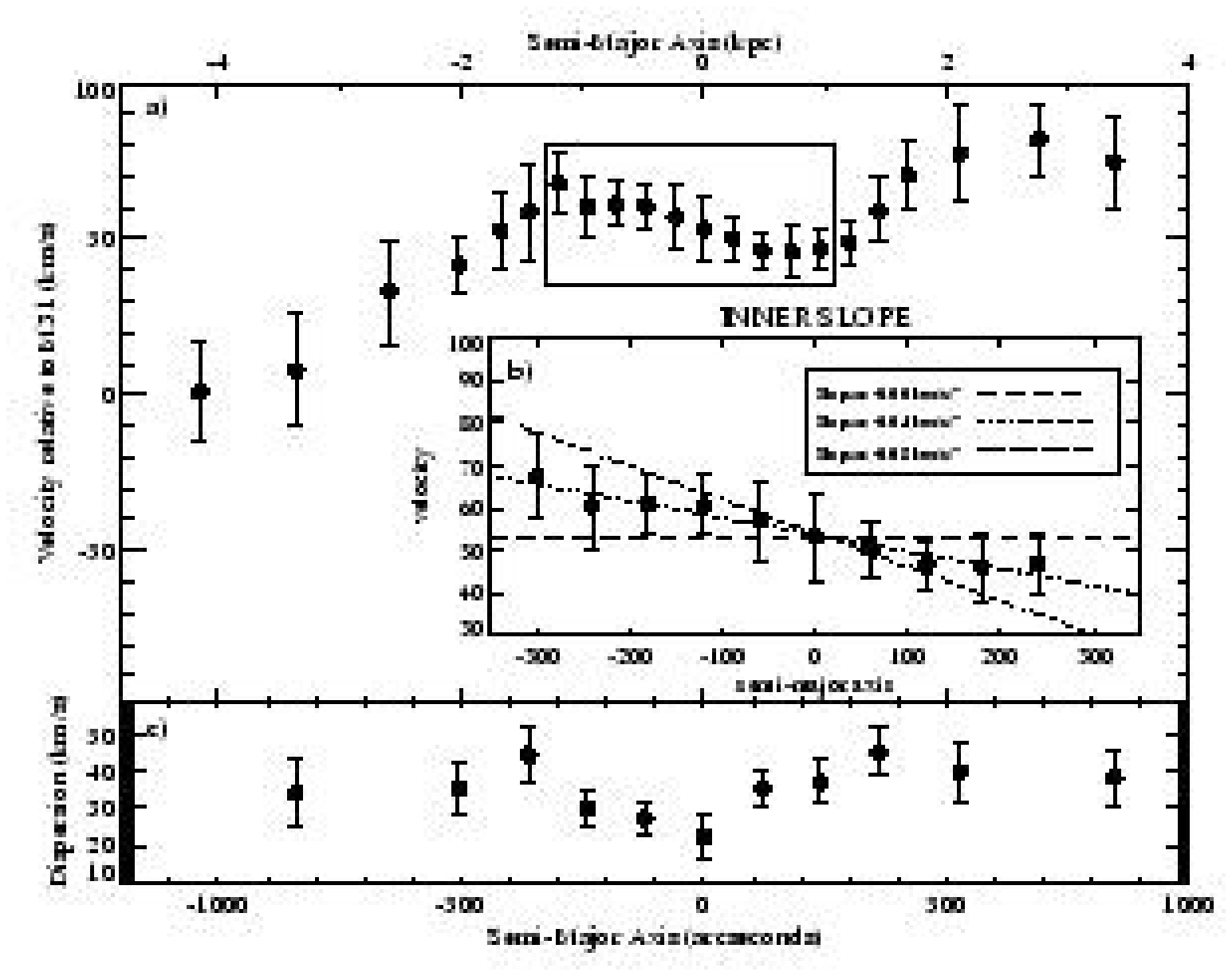}
\caption{NGC~205's semi-major axis radial velocity (relative to M31) and dispersion profile as reported by \citet{geh06}.  (a) The black dots represent the combined velocity measurements of RGB stars.   At distances of $\sim 270\arcsec$ (1.08 kpc) from the center of NGC~205 tidal distortion becomes evident.  (b) The slope interior to the distortion of NGC~205.  The ranges used to constrain our simulations are illustrated in the insert and correspond to a lower limit of $-$0.08 \kms\ arcsec$^{-1}$ and an upper limit slope of 0.0 \kms\ arcsec$^{-1}$ (or $-20$ \kms\ kpc$^{-1}$ and $0$ \kms\ kpc$^{-1}$, respectively, when NGC~205 lies at a distance of 824 kpc).  (c) The velocity dispersion of NGC~205 along the semi-major axis.}  
\label{fig:velocity}
\end{figure*}

\citet{geh06} maps the semi-major axis velocity profile of NGC~205 by binning the RGB stars perpendicular to the \textit{S}-shaped major axis into $\ge 1\arcmin$ (0.24 kpc) radial bins.  The velocity of each bin, containing a minimum of 25 stars, is determined by fitting a Gaussian profile to its stars.  The resulting fit is plotted in Figure \ref{fig:velocity}.  The major-axis rotation curve of NGC~205 shows a distinct velocity turnover (with $v_{\rm max} = 9 \pm 4$ \kms) at $4.5\arcmin$, which is coincident with the estimated tidal radius ($\sim 4\arcmin$) and onset of isophotal twisting (see \S\,\ref{ssec_photo}).  \citet{geh06} suggests that this feature is due to gravitational interaction between NGC~205 and M31, and infers that NGC~205 is on a prograde encounter.  
If correct, over half of \citet{geh06} evenly sampled RGB stars lie outside NGC~205's tidal radius, meaning they have been tidally stripped from the satellite and are no longer bound.  The ratio of the maximum rotational velocity to the average velocity dispersion of NGC~205 (assuming an average ellipticity of $\epsilon=0.43$) is $v_{\rm max}/\sigma = 0.21$ \citep{geh06}.  Given that the expected ratio for an isotropic, oblate, rotational-flattened galaxy with $\epsilon=0.43$ is $v_{0}/\sigma_{0} \approx 1.1$ and that similar anisotropic Virgo dE's on average have $v_{0}/\sigma_{0} < 0.1$, NGC~205 falls somewhere between a rotational supported body and an anisotropic object \citep{bin87, geh06}. 
 
We perform a linear fit to the the interior regions ($<270\arcsec$) of NGC~205 where significant tidal distortion has not yet occurred.  Using \citet{geh06} error bars for each velocity measurement, we determine a maximum and minimum slope for the satellite's central velocity profile of 0.0 \kms\ arcsec$^{-1}$ and $-$0.08 \kms\ arcsec$^{-1}$, respectively.  This significance level corresponds to $1.83\sigma$ and is plotted in Figure \ref{fig:velocity}. 
We use both this subset and the full dataset to constrain the motion of test particles in our numerical simulations (see \S\,\ref{sssec_slope} and \S\,\ref{sssec_wtvelo}).
 
   \section{NUMERICAL SIMULATIONS}\label{sec_numerical}
The previous section discusses the photometric and kinematic observations of NGC~205.  In order to reproduce the observed features of NGC~205, we trace NGC~205's center of mass back half an orbit, set up an initial configuration of test particles (to represent the undistorted NGC~205), and allow it to interact under the influence of both NGC~205's and M31's potential. The final, integrated galaxy is then compared to photometric and kinematic observations.  Strict upper and lower limits are placed on the variable parameters and a genetic algorithm is employed to effectively search our parameter space.  In the following sections we discuss the explorable parameters of the system (\S\,\ref{ssec_initial}), the potentials of both galaxies (\S\,\ref{ssec_potential}), the initial configurations of NGC~205's test particles (\S\,\ref{ssec_particles}), the integration scheme (\S\,\ref{ssec_nbody}), the genetic algorithm used to improve on the initial guesses (\S\,\ref{ssec_GA}), and, finally, our method of comparing the integrated system to observations (\S\,\ref{ssec_fitness}).  The goal of these simulations is to determine the orbital parameters of NGC~205.

For our simulations, the units are chosen such that the distance unit is $2503\arcsec$ (or 10 kpc if we assume NGC~205 lies at a distance of 824 kpc), the velocity unit is 280 \kms, and $G=1$.  This roughly results in a mass unit of 1.8 $\times 10^{11}\Msun$ and a time unit of 35 Myr. 

\subsection{System Parameters and Constraints}\label{ssec_initial}

There are a total of 10 free parameters in our numerical simulation.  Six of these parameters define the initial conditions of the undistorted NGC~205 system, while the other four define the final conditions of the present day observed system.  The six initial condition parameters are  
the mass of NGC~205 ($M_{205}$), the disk scale length ($r_{0,i}$), the relative orientation (in $\phi$, $\theta$) of NGC~205's initial, undistorted disk, whether NGC~205 is on a prograde or retrograde orbit about M31 ($\Delta\theta$), and the Hernquist (1990) scale length (the radius at which a quarter mass is enclosed) of 
NGC~205's potential ($a_{205}$).  The four final constraints are the three-component final velocity vector of NGC~205's center of mass relative to M31's center of mass ($v_x, v_y, v_z$), and the line-of-sight distance between NGC~205 and M31 ($z$).  A detailed discussion of the upper and lower limits placed on each of these parameters is motivated and discussed in detail below.  The parameters are discretized into 256 steps within their given constraints, with the sole exception of the prograde/retrograde parameter, $\Delta\theta$, which can take on a value of either 0 or 1.  This number of steps is chosen such that the genetic algorithm can thoroughly explore our defined parameter space, while limiting the time needed to perform the simulations.
These given parameter values define approximately $10^{22}$ possible orbits.  In the continuous real world, there are an infinite number of orbits.  A summary of the constraints is given in Table~\ref{table:N205param}.  

\begin{deluxetable*}{l c c c}
\tablecaption{NGC~205 Genes}
\tablewidth{0pt}
\tablecolumns{4}
\tablehead{
\colhead{Parameter} &
\colhead{Symbol} &
\colhead{Lower Limit} &
\colhead{Upper Limit}}
\startdata
    Mass & $M_{205}$ & $8.2\times 10^{8}\Msun$ & $5.0 \times 10^{9}\Msun$\\
    Disk Scale Length (disk only): &&&\\
            \hspace{0.25in}...angular size  & $r_{0,i}$ & $15\arcsec$ & $313\arcsec$\\
            \hspace{0.25in}...at distance of 824 kpc & $r_{0,i}$ & 0.06 kpc & 1.25 kpc\\    
    Clockwise Rotation (disk only) & $\phi$ & 0$^{\circ}$ & 180$^{\circ}$\\
    Inclination (disk only) & $\theta$ & $-$180$^{\circ}$ & 180$^{\circ}$\\
     Prograde vs. Retrograde (disk only) & $\theta - \Delta\theta$ & $\Delta\theta \equiv 0^{\circ}$ & $\Delta\theta \equiv 180^{\circ}$ \\
    Hernquist Scale Length: &&&\\
            \hspace{0.25in}...angular size  & $a_{205}$ & $0\arcsec$ & $2503\arcsec$\\ 
            \hspace{0.25in}...at distance of 824 kpc & $a_{205}$ & 0 kpc & 10 kpc\\       
    Distance from M31 & $z$ & 2 kpc & 76 kpc\\
    Center of Mass x-Velocity & $v_x$ & $-$500  \kms\  & 500 \kms\  \\
    Center of Mass y-Velocity & $v_y$ & $-$500  \kms\  & 500 \kms\  \\
    Center of Mass z-Velocity & $v_z$ & 49 \kms\  & 59 \kms\  \\
\enddata
\tablecomments{Constrained parameters (genes) searched by the genetic algorithm in the numerical simulations.  Each parameter contains 256 steps between the upper and lower limits, with the exception of $\theta-\Delta\theta$ which can take on a value of 0 or 1.}
\label{table:N205param}
\end{deluxetable*} 

     	\textbf{Mass.}  The mass of NGC~205 is not currently well known.  A recent constraint was placed on the mass contained within a 2$R_e$ radius sphere by \citet{der06} of $M_{205}(2R_e) = 10.2_{-2.0}^{+3.3} \times 10^8$\Msun.  We use De \citet{der06} minimum of $M_{205} \ge 8.2 \times 10^8$\Msun as a lower limit on the total mass of NGC~205 (i.e. if all the mass were contained within 2$R_e$) and set a very liberal upper limit of $M_{205} \le 5.0 \times 10^{9}$\Msun.   
	
	\textbf{Disk scale length.}  This parameter is used when NGC~205 is modeled as a disk.  In \S\,\ref{ssec_photo} we determined that the observable disk scale length of NGC~205 is  $r_{0}=148\arcsec$.   We assume that at earlier times NGC~205's surface brightness profile followed an exponential law governed by a different $r_0$, which has since been affected by its interaction with M31.  We therefore allow the genetic algorithm to explore the range $15\arcsec \le$ $r_{0,i} \le 313\arcsec$ for the initial disk scale length in order to model the undistorted NGC~205.  If NGC~205 indeed lies at a distance of 824 kpc, then this $r_0$ range corresponds to $0.0625$ kpc $\le$ $r_{0,i} \le1.25$ kpc.  The disk scale length is also used to determine the maximum distance an initial test particle can be placed in the disk of NGC~205.  We chose a maximum distance of 8$r_{0,i}$; however, at this distance we expect to find very few particles since the initial distribution is subject to follow an exponential law.

	\textbf{Initial disk orientation.}  These orientation parameters are used when NGC~205 is modeled as a disk.  We assume axisymmetry for NGC~205's initial, undistorted disk and describe its orientation in 3-D space by two of the three Euler angles.  We define the plane of the sky as the $x$-$y$ plane centered on M31 with the positive $x$-axis in the direction of increasing RA, the positive $z$-axis pointing towards the observed NGC~205 and the negative $z$-axis pointing towards us.  
In our defined coordinate system, the Euler angle $\phi$ describes turning the galaxy's $x$-axis clockwise 
in the $x$-$y$ plane and is also allowed to vary from $0^{\circ}$ to $180^{\circ}$.  
The inclination angle $\theta$ describes tilting the galaxy about the rotated $x$-axis and is allowed to vary from $0^{\circ}$ to $180^{\circ}$.  An inclination angle of $90^{\circ}$ means the initial galaxy appears edge on, while an inclination angle of either $0^{\circ}$ or $180^{\circ}$ means the initial galaxy appears face on.  Note the visual symmetry about $180^{\circ}$, which does \underline{not} translate into dynamical symmetry.   
The Euler angle $\psi$ describes revolution about the rotation axis and is meaningless due to the axisymmetry of NGC~205's disk.  
Similarly, the axisymmetry of NGC~205 makes visual searches in the range $180^{\circ}$ to $360^{\circ}$ for $\phi$ and $\theta$ redundant.
Thus, the position (but \underline{not} the dynamics) of NGC~205's initial disk is completely described by the parameters $\phi$ and $\theta$ in the range $0^{\circ}$ or $180^{\circ}$. 

	\textbf{Prograde vs. Retrograde.}  This parameter is used when NGC~205 is modeled as a disk.  Since the initial sense of rotation of NGC~205 is unknown, it is important to explore both prograde and retrograde orbits.  A prograde orbit means that the galaxy is rotating in the direction of the encounter, while a retrograde orbit means it is rotating in the direction against the encounter.  As mentioned in the previous paragraph, exploring $\theta$ in the range of $0^{\circ}$ to $180^{\circ}$ provides a complete visual search, but not a complete dynamical search.  For completeness, we introduce the rotation parameter $\Delta\theta$ into our simulation whose purpose is to determine whether the Euler angle $\theta$ will be flipped about $180^{\circ}$.  The rotation parameter can be set to either $0^{\circ}$ or $180^{\circ}$, where $\Delta\theta=0^{\circ}$ means $\theta$ is unchanged and $\Delta\theta=180^{\circ}$ changes $\theta$ to  $\theta -180^{\circ}$.  Hence, this parameter essentially allows $\theta$ to vary from $-180^{\circ}$ to $180^{\circ}$ so that both prograde and retrograde orbits can be explored.
The motivation for using this separate parameter, $\Delta\theta$, rather than just allowing $\theta$ to vary initially from $-180^{\circ}$ to $180^{\circ}$ is that $\Delta\theta$'s binary nature provides an easy means for the genetic algorithm to instantly change the direction of the encounter via mutation or  reproduction (see \S\,\ref{ssec_GA} for more information on the genetic algorithm).  That is, the genetic algorithm is capable of changing the rotation of the disk by merely turning this parameter off or on.  Note, this effect could have also been achieved using the $\phi$ parameter instead.

	\textbf{Hernquist scale length.}  We model NGC~205's potential as a spherical symmetric distribution of matter governed by a Hernquist profile (\S\,\ref{ssec_potential}, Eqn~\ref{eq:hernquist}).  
The concentration of the combined dark and baryonic matter in NGC~205 is governed by Hernquist scale length parameter, $a_{205}$ (or $a_{0}$ in Eqn~\ref{eq:hernquist}).  We set a lower limit on $a_{205}$ of $0\arcsec$, corresponding to a Keplerian potential, and an upper limit of $2503\arcsec$, which at a distance of 824 kpc would correspond to a scale length of 10 kpc. 
  
     	\textbf{Velocity.}  The mean radial velocities of M31 and NGC~205 are $-300 \pm 4$ \kms\  \citep{dev91}  and $-246 \pm 1$ \kms\  \citep{geh06}, respectively, resulting in a relative radial velocity for NGC~205 of 54 $\pm$ 5  \kms\  with respect to M31 (see \S\,\ref{ssec_kinematic}).  Thus the radial velocity component, $v_z$, is given a lower limit of 49  \kms\  and an upper limit of 59 \kms .  The transverse velocity components, $v_x$ and $v_y$, are given upper and lower limits dictated by the escape velocity of NGC~205's center of mass relative to M31, and are therefore not only functions of $z$ and $v_z$, but also of M31's mass.  In order to explore both bound and unbound orbits, we allow $v_x$ and $v_y$ to explore the full range of NGC~205's escape velocity, with a maximum escape velocity of 500 \kms\  if NGC~205's line-of-sight distance and velocity ($z$ and $v_z$) are at their minimums.

	\textbf{Line-of-sight distance.}  The reported distances to M31 and NGC~205 are $785 \pm 25$ kpc and $824 \pm 27$ kpc, respectively \citep{mcc04}. By adding the errors in quadrature, we approximate the line-of-sight distance to be $39 \pm 37$ kpc and allow the genetic algorithm to search within the range of 2 kpc to 76 kpc for the final position of NGC~205.  We recognize that this is a generous range to search, but decide to err on the side conservatism. 

\subsection{Galaxy Potential Models}\label{ssec_potential}
The close proximity of M31 to NGC~205 stipulates that fairly accurate potentials are necessary in modeling the interaction between the galaxies.  However, there is an implied limit to the level of complexity of the models resulting from the large number of orbits searched by the genetic algorithm.  That is, using too simple of a model can result in incorrect final integrated positions and velocities, and, if the model is too complex, the genetic algorithm will spend large amounts of time attempting to complete a single orbit.  Hence it is essential to find balance between the two requirements in order to come up with an optimal model.  In this subsection we discuss the potential models used in our simulations.  We assume that the galaxy potentials are not changing significantly during the course of the interaction and thus remain static.

Previous studies suggest dwarf elliptical galaxies do not contain a significant amount of dark matter in their inner regions.  The amount of dark matter in the outer regions of dEs is completely unknown.  Thus, one of the goals of this project is to probe the dark matter content of NGC~205.  
We make the simple, but appropriate, approximation that the potential of NGC~205 is a spherically symmetric spheroid with a Hernquist profile (Hernquist 1990).
However, it should be noted that the ideal model for NGC~205 is a stellar population orbiting in a fully consistent potential with a brightness profile lying somewhere in between an exponential disk and exponential bulge (i.e. a hot, exponential disk).  The difficulty of constructing such a model is demonstrated by the complex photometry discussed in \S\,\ref{ssec_photo}.  
For the purpose of this paper we model NGC~205's potential as,
\begin{equation}
\Phi_{b}(r) = -\frac{GM}{a_{0} + r}
\label{eq:hernquist}
\end{equation}
where $M$ is the total mass of NGC~205 (stars, interstellar gas, dark matter), $a_0$ is the Hernquist scale length that defines the concentration of matter in NGC~205, and $r$ is the distance from the center of NGC~205.  Both the mass and the Hernquist scale length are variables probed by the genetic algorithm.

The potential of M31 is best represented by a three-component model containing a bulge, a disk and a halo.  \citet{gee05} derives a ``Simple Analytic Bulge-Disk-Halo Model''  using a spherical symmetric Hernquist profile for the bulge (Eqn~\ref{eq:hernquist}), an infinitesimally thin, exponential disk 
and a spherically symmetric NFW profile for the extended dark halo given by
\begin{equation}
\Phi_{h}(r) = -4\pi G\delta_{c}\rho_{c}r_{h}^{2} \left( \frac{r_{h}}{r} \right) \ln \left[ \frac{r+r_{h}}{r_h}\right] 
\label{eq:NFW}
\end{equation}
where $\delta_c$ is a dimensionless density parameter, $\rho_{c}=277.72h^{2}$\Msun kpc$^{-3}$ is today's critical density with Hubble constant $h=0.71$ in units of 100 km/s/Mpc, and $r_{h}$ is the halo scale radius \citep*{nav96}. 

The exponential disk model used by \citet{gee05} 
is presently without a known analytic solution.  Although the equation and its spatial derivatives can easily be solved numerically, the amount of time required to integrate a set of particles through a complete orbit becomes unreasonable for our purposes.  An alternative approach is to construct a large table of values for the spatial derivatives and perform a bi-cubic interpolation in both $R$ and $z$ in order to find the force on a single particle at each step in the integration. Unfortunately, this latter method is unable to significantly decrease the amount of computation time required for a single orbit.  Thus, we use a Miyamoto \& Nagai (1975) disk which provides results that are comparable to those 
from the exponential disk used by \citet{gee05}:
\begin{equation}
\Phi_{d}(R,z) = \frac{-GM_{d}}{\sqrt{R^{2}+(R_{d}+\sqrt{z^{2}+b^{2}})^{2}}}
\label{eq:miyamoto}
\end{equation}
where $R_{d}$ is the disk scale length and $b$ is the vertical scale factor.

We use the ``Best-fit Model" values derived by \citet{gee05} to describe the various parameters of M31, with the sole exception of $b$, the vertical scale factor, which was not a reported parameter.  For $b$ we use the vertical scale height of the dust at a value of 0.1 kpc \citep{hat97}. The values reported by \citet{gee05} include the 
bulge mass, with $M_{b}=3.3 \times 10^{10}$\Msun, the 
bulge scale factor, with $a_{0}=r_{b}=0.61$ kpc, the 
disk central surface density, with $\Sigma_0=4.6 \times 10^{8}\Msun$ kpc$^{-2}$, the 
disk scale radius, with $R_d=5.4$ kpc, the 
halo scale radius, with $r_h=8.18$ kpc, and the 
total mass enclosed inside 125 kpc, with $M(<125$ kpc)$= 5.6 \times 10^{11}$\Msun.  
To compute the mass of the disk we use,
\begin{equation}
M_{d} = 2\pi R_{d}^{2}\Sigma_{0} = 8.4 \times 10^{10} \Msun
\label{eq:diskmass}
\end{equation}
for the halo mass we use,
\begin{equation}
M_{h} = M_{(<125 \rm{ kpc})} - M_{d} - M_{b} = 4.4 \times 10^{11} \Msun
\label{eq:halomass}
\end{equation}
and to find the dimensionless density parameter $\delta_c$ we use the values for $r$ and $M_{h}$ at $r=125$ kpc, and solve the mass profile for a NFW halo,
\begin{equation}
M_{h} = 4\pi G\delta_{c}\rho_{c}r_{h}^{3}\left[\ln \left(\frac{r+r_{h}}{r_h}\right) - \frac{r}{r+r_{h}}\right]
\label{eq:NFW_mass}
\end{equation}
This results in a dimensionless density parameter value of,
\begin{displaymath}
\delta_{c} = \frac{M_{h}}{4\pi\rho_{c} r_{h}^{3} (\ln \left[(r+ r_{h})/r_{h}\right] - r/(r_{h}+r))} 
\end{displaymath}
\begin{equation}
 = 24.8 \times 10^{4}.
\label{eq:deltac}
\end{equation}
A summary of these parameters is given in Table~\ref{table:M31param}.
\begin{deluxetable*}{l c c}
\tablecolumns{3}
\tablecaption{M31}
\tablewidth{0pt}
\tablehead{
\colhead{Parameter} &
\colhead{Symbol} &
\colhead{Value} }
\startdata
    Bulge: &&\\
            \hspace{0.25in} Mass & $M_b$ & $3.3\times 10^{10}$\Msun\\
            \hspace{0.25in} Scale Radius & $r_b$ & 0.61 kpc\\  
   Disk: &&\\
            \hspace{0.25in} Central Surface Density & $\Sigma_0$ & $4.6\times 10^{8}$\Msun\ kpc$^{-2}$\\   
            \hspace{0.25in} Mass & $M_d$ & $8.4\times 10^{10}$\Msun\\
            \hspace{0.25in} Scale Radius & $R_d$ & 5.4 kpc\\
            \hspace{0.25in} Vertical Scale Height & $b$ & 0.1 kpc\\            
   Halo: &&\\
            \hspace{0.25in} Mass & $M_h$ & $4.4\times 10^{11}$\Msun\\
            \hspace{0.25in} Scale Radius & $r_h$ & 8.18 kpc\\       
            \hspace{0.25in} Density Parameter & $\delta_c$ & $24.8 \times 10^4$\\ 
   Total: &&\\
            \hspace{0.25in} Mass ($<125\ {\rm kpc}$) & $M_{(<125\ {\rm kpc})}$ & $5.6\times 10^{11}$\Msun\\                                             
\enddata
\tablecomments{M31 parameter values used in the numerical simulations adopted from the values in \citet{gee05}.}
\label{table:M31param}
\end{deluxetable*} 
\subsection{Arrangement of NGC~205's Test Particles}\label{ssec_particles}  

In addition to the parameters discussed in \S\,\ref{ssec_initial} and the potentials outlined in \S\,\ref{ssec_potential}, the satellite's initial, undistorted particle configuration (both in position and velocity space) directly affects the appearance of the final, integrated distribution of particles.   
The particles are used as tracers of surface brightness and to map the radial velocity profile of the galaxy.  The present true brightness profile of NGC~205 lies in between an exponential and $r^{1/4}$ law \citep{ken87, cho02}, while the galaxy's support is thought to come from a mix of both rotational and anisotropic velocities \citep{geh06}.  As mentioned in \S\,\ref{ssec_kinematic}, the current observed ratio of the maximum rotational velocity to the average velocity dispersion is $v_{\rm max}/\sigma = 0.21$.  Hence the current structure of particles in the satellite lies somewhere between an exponential bulge and exponential disk.

The morphology of NGC~205 before tidal distortion is unknown, while a wide range of internal dynamics has been observed for cluster dEs, ranging from non-rotating to rotational flattened dEs \citep{geh03}.  Furthermore, it is unclear which features of NGC~205 are due to intrinsic properties, projection, or tidally interaction.  For this reason, we test three configurations for NGC~205: a rotating cold exponential disk (\S\,\ref{sssec_colddisk}), a non-rotating hot spheroid (\S\,\ref{sssec_hotspheroid}) , and a rotating warm exponential disk (\S\,\ref{sssec_warmdisk}).   These configurations explore the two extremes, a rotationally supported satellite with $v_{\rm max}/\sigma=\infty$ and an isotropically supported satellite with $v_{\rm max}/\sigma=0$, as well as an intermediate construction, a warm disk with $1.05 < v_{\rm max}/\sigma < 4.25$.  These three models are outlined in detail below.

\subsubsection{Rotating Cold Disk}\label{sssec_colddisk}
Our most basic model for NGC~205 places mass-less test particles in an infinitely thin, flat disk.  
The particles are randomly distributed in accordance with an exponential surface brightness profile, as given in Eqn \ref{eq:eprofile}.
We assume that the central (nucleus free) surface brightness, $I_{0}$, has not evolved and that the undistorted NGC~205 follows an exponential characterized by a variable scale length, $r_{0,i}$. 
We set the maximum radius of NGC~205 to $8r_{0,i}$, a location at which very few particles are expected.  
In addition to $r_{0,i}$, the particle positions in NGC~205's initial disk are also defined by the disk's clockwise rotation in the $x$-$y$ plane, $\phi$, and inclination angle, $\theta$.  The values of these three parameters ($r_{0,i}$, $\phi$, $\theta$) are obtained from the genetic algorithm.  

Initially, the cold disk is completely supported by rotation, such that $v_{\rm max}/\sigma=\infty$.  Each particle moves in a circular orbit about the satellite's center under the influence of a Hernquist profile.  This results in a circular velocity, in the plane of NGC~205's disk, of
\begin{equation}
v_{c} = \sqrt{\displaystyle\frac{G M r}{(r + a_{0})^2}}
\label{eq:vc}
\end{equation}
where $M$ is the mass of NGC~205, $a_{0}$ is the Hernquist scale length of NGC~205, and $r$ is the particle's distance from the center of NGC~205.  The direction of rotation in the disk is determined by the parameter $\theta - \Delta\theta$.  Hence, the particle's circular velocities are very simply defined by their distance from the satellite's center and by the three variable parameters: $M_{205}$, $a_{205}$, and $\theta - \Delta\theta$.  Any dispersion found in the final, integrated system is a direct result of tidal interaction with M31.

\subsubsection{Rotating Warm Disk}\label{sssec_warmdisk}
A more realistic model for NGC~205 arranges mass-less test particles in a disk with some vertical thickness.  We employ the method prescribed by \citet{her93} using cylindrical coordinates and distribute the particles in an exponentially decaying density profile given by,
\begin{equation}
\rho = \displaystyle\frac{M}{4\pi r_{0}^2 z_{0}} \exp (-R/r_{0}) {\rm  sech}^{2} \left(\frac{z}{z_0}\right)
\end{equation}
where $R$ is the cylindrical radius, $M$ is the mass of the satellite, $r_0$ is the disk scale length, and $z_0$ is the vertical scale thickness.   For our simulations, we set vertical scale thickness, $z_0$, equal to the scale radius, $r_{0,i}$, in order to create a ``puffier" disk.  As with the cold disk model outlined in \S\,\ref{sssec_colddisk}, the maximum radius of NGC~205 is constrained to $8r_{0,i}$ and the disk is oriented using the $\phi$ and $\theta$ parameters obtained from the genetic algorithm.

The warm disk is supported by a combination of rotational and anisotropic velocities.    
Using the method outlined by \citet{her93}, self consistent disks are approximated using the moments of the collisionless Boltzmann equation.   In order to avoid imaginary streaming velocities ($ \sigma_{\phi}^{2} < 0$) near the center, the softened version of the radial dispersion equation is used,
\begin{equation}
{\sigma_{R}^{2}} = A \exp (-\sqrt{R^2 + 2a_{s}^{2}}/r_{0})
\end{equation}
where $a_s$ is the disk softening radius and $A$ is the normalization constant.  As prescribed by \citet{her93}, the softening radius is set to $r_{0}/4$.  The normalization factor, $A$, can then be easily computed by specifying ${\sigma_{R}^2}$ at $R=r_0$,
\begin{equation}
 A = {\sigma_{R}^{2}}_{=r_{0}} \exp(\sqrt{r_{0}^{2} + 2a_{s}^{2}}/r_{0})
\end{equation}
However, we cannot simply set the radial dispersion at one scale length equal to the observed dispersion in NGC~205 of $35\pm 5$\kms\  \citep{geh06}.  This would cause the streaming velocities to become imaginary for much of the parameter space. 
To avoid this potential problem, the maximum allowed radial dispersion that imposes $\sigma_{\phi}^{2} \ge 0$ is used, with the caveat that $\sigma_R$ cannot exceed 29 \kms.
The vertical dispersion perpendicular to the plane of the disk is given by,
\begin{equation}
{\sigma_{z}^{2}} = \pi G \Sigma(R)z_{0}
\end{equation}
where $\Sigma(R)$ is the disk surface density given by the local Hernquist density \citep{her90},
\begin{equation}
\Sigma(R) = 2\rho z_{0} = \displaystyle\frac{M a_{0} z_{0}}{\pi R (R+a_{0})^{3}}
\end{equation}
and $a_{0}$ is the Hernquist scale length.
Using the epicyclic approximation, the azimuthal dispersion can be related to the radial dispersion,
\begin{equation}
\sigma_{\phi}^{2} = {\sigma_{R}^{2}} \displaystyle\frac{\kappa^{2}}{4\Omega^{2}}
\end{equation}
where the epicyclic and angular frequencies are respectively given by,
\begin{equation}
\kappa^{2} = \displaystyle{\frac{3}{R}\frac{\partial\Phi}{\partial R} + \frac{\partial^{2}\Phi}{\partial R^{2}}}
\end{equation}
\begin{equation}
\Omega^{2} = \displaystyle\frac{v_{c}^{2}}{R^2}
\end{equation}
with the circular velocity $v_{c}$ given by Eqn \ref{eq:vc} and the potential energy $\Phi$ of a Hernquist profile given by Eqn \ref{eq:hernquist}.  If we assume that the velocity ellipsoid stays aligned with the coordinate axes, the streaming velocity becomes,
\begin{equation}
{v_{\phi}}^{2}=v_{c}^{2}+{\sigma_{R}^{2}}(1 - \displaystyle\frac{\kappa^2}{4\Omega^2} - 2\displaystyle\frac{R}{r_{0}}) 
\end{equation}

Once these low order moments are determined, the radial and vertical velocities for individual particles can be drawn randomly from Gaussians with widths $\sigma_{R}$ and $\sigma_{z}$, while the azimuthal velocity is determined by adding the streaming velocity, $v_{\phi}$, to the velocity value drawn from a Gaussian with width $\sigma_\phi$.   This configuration results in observable ratios of the maximum rotational velocity to the average velocity dispersion dispersions of $1.05 < v_{\rm max}/\sigma < 4.25$.
Since the satellite still contains rotational support, the parameter $\theta - \Delta\theta$ is used to specify the direction of this rotation.

\subsubsection{Non-Rotating Hot Spheroid}\label{sssec_hotspheroid}
The third model for NGC~205 configures the satellite as a dynamically hot, pressure supported spheroid.  Hence, the velocities of this model are completely isotropic.  This construction is essentially the counter to the rotationally supported, cold disk model outlined in \S\,\ref{sssec_colddisk}.  The density profile follows the analytical model for spherical galaxies outlined by \citet{her90},
\begin{equation}
\rho(r) = \displaystyle\frac{M}{2\pi}\frac{a_{0}}{r}\frac{1}{(r+a_{0})^{3}}
\label{eq:hern_rho}
\end{equation}
where $M$ is the mass of the satellite, $a_{0}$ is the Hernquist scale length of NGC~205, and $r$ is the particle's distance from the center of NGC~205.  As with the previous two models, the maximum radius of NGC~205 is constrained to $8r_{0,i}$.

Using the energy dependent distribution function, $f(E)$, for a Hernquist profile, a spheroid supported by isotropic velocities can be constructed using random realizations \citep{her90, asc05}.  
The phase-space density as a function of specific energy is, 
\begin{displaymath}
f(E) = \displaystyle\frac{M/a_{0}^{3}}{4 \pi^{3} (2GM/a_{0})^{3/2}}
\end{displaymath}
\begin{equation}
 \times \frac{3 \sin^{-1}(q) + q\sqrt{(1-q^{2})}(1-2q^2)(8q^{4} - 8q^{2} -3)}{(1-q^{2})^{5/2}}
\label{eq:df}
\end{equation}
with the dimensionless variable $q$ defined as,
\begin{equation}
q \equiv \displaystyle\sqrt{-\frac{a_{0}}{GM}E}
\end{equation}
The specific energy $E$ of a particle at position $r$ and velocity $v$ is,
\begin{equation}
E = \frac{1}{2} v^{2} + \Phi(r)
\end{equation}
with $\Phi(r)$, the potential energy for a Hernquist profile, given by Eqn \ref{eq:hernquist}.
We normalize the distribution function in a volume element $d\vec{r} d\vec{v}$, centered on ($\vec{r}$, $\vec{v}$), by dividing $f d\vec{r} d\vec{v}$ by its maximum.  This maximum corresponds to the point in parameter space containing the largest amount of mass (and in our case, mass-less test particles).  Using spherical symmetry, $d\vec{r} d\vec{v}$ becomes $4\pi r^{2} dr 4\pi v^{2} dv$.  Hence, the maximum of $f d\vec{r} d\vec{v}$ is easily found numerically and resides at $r_{\rm peak}=0.638a_{0}$ and $v=v_{c}(r_{\rm peak})$, where $v_{c}$ is the circular velocity given by Eqn \ref{eq:vc}.  

Using the normalized $f d\vec{r} d\vec{v}$, particles are placed in phase space using a von Neumann rejection technique \citep{pre86}.  
Tentative values for $r$ and $v$ are randomly chosen within the range $0 \le r \le 8r_{0,i}$ and $0 \le v \le v_{\rm esc}=\sqrt{-2\Phi(r)}$, and a tentative $f d\vec{r} d\vec{v}$ is computed.  
Next, a random number $k$ between 0 and 1 is selected (where 1 corresponds to the normalized peak of $f d\vec{r} d\vec{v}$).  The provisional values for $r$ and $v$ are kept only if $k \le f d\vec{r} d\vec{v}$, else the process is reinitiated until the condition is satisfied.  Once $r$ and $v$ are generated, position and velocity angular coordinates are randomly chosen from $0 \le \phi \le 2\pi$ and $ -1 \le \cos\theta \le 1$, and the particle is placed in the satellite. The procedure is then repeated until the satellite is completely populated with the desired number of particles.  
This configuration results in a galaxy with a density profile given by Eqn \ref{eq:hern_rho} and that is completely supported by isotropic velocities ($v_{\rm max}/\sigma = 0$).  
Note that, unlike the rotating cold disk and warm disk, the symmetry of the non-rotating spheroid negates the need for parameters $\phi$, $\theta$, and $\theta - \Delta\theta$, while the density profile's dependence on $a_{0}$ reduces $r_{0,i}$'s use to setting the maximum radius for placing particles.  

\subsection{Restricted $N$-Body Simulations}\label{ssec_nbody}  

The simulations are carried out using a restricted $N$-body approach \citep{pfl61,too72} in a Cartesian coordinate system.
The use of a Cartesian coordinate system is motivated by the format of the available observations: \citet{cho02} photometric observations are defined in the $x$-$y$ plane and \citet{geh06} velocity observations are defined in the $z$ direction. We center our coordinate system on M31 and define the $x$-$y$ plane as the plane of the sky, with $y$ pointing north, along a line of increasing declination (DEC), and $x$ pointing east, along a line of increasing right ascension (RA).  The positive $z$-axis, which is also centered on M31, is therefore along our line-of-sight pointing away from us. By this definition, an object moving toward us is defined to have a negative $z$-velocity, while an object moving away from us is defined to have a positive $z$-velocity.

The center of M31 lies at the center of this coordinate system $(0,0,0)$.  The spherical symmetry of the bulge and the halo require no rotation, while the axisymmetry of the disk requires a rotation of $i=77.5^\circ$ around the $x$-axis followed by a $\theta=37.7^\circ$ rotation about the $z$-axis.  Thus, if we define a disk as edge on along the $x$-axis at angles of $i=90^{\circ}$ and $\theta=0^{\circ}$ then the orientation of the M31 disk in the projected plane of the sky is given by:
\begin{equation}
\left( 
\begin{array}{c}
x' \\ y' \\ z' 
\end{array}
\right) = \left( 
\begin{array}{ccc}
\cos i \cos\theta & \sin\theta & 0 \\ -\cos i \sin\theta & \cos\theta & 0\\ \sin i & 0 & 0
\end{array}
\right) \left( 
\begin{array}{c}
x \\ y \\ z
\end{array}
\right)
\label{eq:rotation}
\end{equation}
For computational purposes, the inverse of the matrix above is used to rotate each particle into the frame of M31 (with the disk lying in the $x$-$y$ plane) in order to compute the particle acceleration due to the disk, and then rotated back into our coordinate system as defined above.
\upscale
\begin{figure*}\includegraphics[trim=.8in 4.2in 0in 3in,clip]{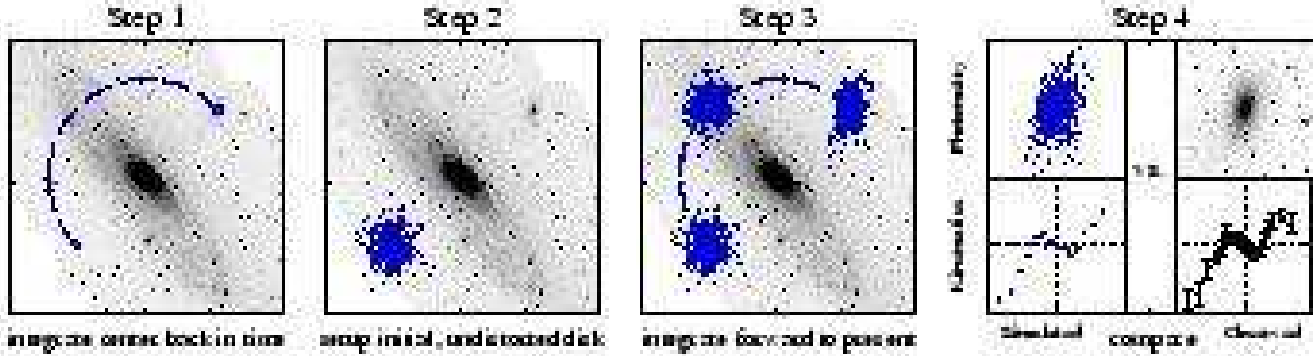}
\caption{Schematic overview of the simulated interaction between a cold NGC~205 and M31.  In Step 1, the center of mass of NGC~205 is integrated back in time approximately half an orbit.  In Step 2, a cold disk representing the undistorted NGC~205 is created.  In Step 3, the undistorted disk is integrated forward in time in response to NGC~205's and M31's potentials.  In Step 4, a photometric and kinematic comparison is done between the integrated system and the observed NGC~205 system.}
\label{fig:steps}
\end{figure*}
\returnscale
There are five major steps involved in the simulation.  For each individual orbit, the values of the unknown parameters described in \S\,\ref{ssec_initial} are obtained from the genetic algorithm as discussed in \S\,\ref{ssec_GA}.  Each step in a simulated orbit is described below:

1.  First, the center of mass (COM) of NGC~205 is integrated back in time approximately half an orbit using a Bulirsch Stoer integration scheme \citep{pre86}.  
The presence of M31's disk presents difficulties for performing straightforward calculations of the orbital period.  Hence, we approximate the orbital period as twice the circular velocity period of NGC~205.  Although this approximation is far from exact, we intentionally overcompensate for the period to ensure that the satellite travels at least half an orbit or descends from a reasonably large radii.
If the integration time is longer than 6.8 Gyr the orbit is fixed to 6.7 Gyr, since such orbits are much longer than the dynamical time scale for NGC~205.  Since only the COM's $x$ and $y$ positions are known precisely, the final $z$ position and the velocity vector ($v_{x}, v_{y}, v_{z}$) must be taken from the genetic algorithm's initial parameter space. The acceleration is computed using a  three-component potential model for M31 consisting of a bulge, a disk and a halo as described in \S\,\ref{ssec_potential}.  
We set the required accuracy per time step to one part in $10^{7}$. 

2. At this earlier time, we assume tidal distortion of NGC~205 has not yet taken place.
After successful integration back in time, the initial, undistorted NGC~205 galaxy is created and populated with 1,000 mass-less test particles in a preselected configuration, as discussed in \S\,\ref{ssec_particles}.  The distribution of test particles is not intended to be a self consistent tracer of the mass, rather, it is constructed to be a tracer of the expected surface brightness profile of NGC~205 prior to its distortion.  

3. Next, the undistorted system of particles is integrated forward in time to the present day under the influence of both galaxies' potentials.  As discussed in \S\,\ref{ssec_potential}, we model NGC~205 as a spherically symmetric static potential and M31 as a three-component static potential.  The mass, $M_{205}$, and Hernquist scale length, $a_{205}$, used to construct NGC~205's potential are unknown parameters and are obtained from the genetic algorithm.  
We set the required accuracy per particle per time step in our integration to one part in $10^{7}$.

4.  After the integration is complete, the final, simulated system is compared to the known, observed system.  The comparison involves a total of 6--7 tests, depending on the model used for NGC~205.  Although, individually, these tests report a probability measurement, their true purpose is to provide information to the GA instructing it on how to improve subsequent guesses.  
An outline of each test is provided in \S\,\ref{ssec_fitness}.

5. Steps 1--4 are repeated until the desired number of orbits is reached. We then use a genetic algorithm to improve our initial conditions and repeat the process until we can match our observations. 
For each orbit, parameters resulting in a goodness-of-fit value higher than the previous value are recorded.  Steps 1--5 are then repeated using a new guesses for the free parameters.

\subsection{The Genetic Algorithm (GA)}\label{ssec_GA}

Our 10-dimensional parameter space discussed in \S\,\ref{ssec_initial} contains approximately $10^{22}$ possible orbits.  At an average of 6.25 CPU seconds an orbit for our simplest model, the cold disk, an iterative search of the space would take on order $10^{6}$ CPU Gyr.  For this reason, we must efficiently search the parameter space with an optimization algorithm.  We have chosen to use the optimization technique invented by \citet{hol75} called a genetic algorithm (GA) for two pragmatic reasons.  The first is that GAs have the distinct advantage of not getting stuck in local optima, a problem that plagues hill climbing techniques which exploit gradient information.  The second benefit is that GAs are reasonably effective at reducing the amount of searching needed to converge on a solution in a given parameter space.  Hence, GAs are especially useful in problems where iterative searches are computational unreasonable.  These two features make genetic algorithms ideal for searching parameter spaces containing an unknown distribution of local optima.

The inspiration behind the GA is biological evolution.   The idea follows that evolution is a process determined by chromosomes and natural selection.  In the GA application, an individual point in parameter space is equivalent to a \textit{chromosome}, and the parameters making up that chromosome are analogous to \textit{genes}.
As dictated by biology, evolution occurs at the point of reproduction and individuals with genes advantageous to their species are more likely to reproduce.  
Mutations of genes and recombination of chromosomes (crossover) can result in offspring that are very different than their biological parents, and, if these mutations and recombinations result in a higher level of success for the offspring, they become more likely to reproduce.  With each new generation, a new population is created and the old population is replaced.

There are three general modules that define and drive all genetic algorithms.  
These are the \textit{Evaluation, Population and Reproduction Modules}.  
Within any particular application, the details of each module can vary.  
The purpose of the \textit{Evaluation Module} is to determine the \textit{fitness} of a particular individual.  Fitness is a measure of an individual's feasibility as a solution to the problem at hand and is commonly determined by a mathematical function.  The fitness is the only information fed into the genetic algorithm and is used in the \textit{Population} and \textit{Reproduction Modules}.  
At any given time, the \textit{Population Module} hosts the current population of individuals, called a generation.  Encoded in this module are the existing population's chromosomes and the methods used for creating new generations (e.g. how many children are produced by a set of parents).  The method of parent selection (e.g. roulette-wheel, stochastic tournament, deterministic sampling) is often contained in this module as well, but can alternatively be a component in the \textit{Reproduction Module}.
The \textit{Population Module} begins by creating an initial population (the first generation), which generally consists of random, single-point guesses in parameter space.  The success of a given GA run only weakly depends on this initial first guess.  
The successive replacements of this population by the \textit{Reproduction Module} results in new generations.  
The \textit{Reproduction Module} determines the method by which new chromosomes will be created.  
The amount of mutation and the method of crossover (the way the parent's chromosomes are combined to produce offspring) are defined in this module.  The resulting offspring becomes an individual of the next generation \citep{gol89, dav91}.

We use a micro genetic algorithm (micro-GA) driver designed by \citet{car01} which employs a tournament selection scheme with a shuffling technique to select individuals for reproduction.  
Carroll's micro-GA is characterized by a small population (5 individuals), with uniform crossover, elitism and no mutation or niching.
The algorithm works by first arranging individuals in the current population in some arbitrary order (shuffling).
Two individuals are then randomly chosen from the population, and the fittest amongst the pair is selected as a parent (tournament selection).  Tournament selection is then repeated to select the second parent.  Next, genes are picked at random from the first or second parent to create an offspring (uniform crossover), which subsequently becomes a new member of the next generation.  The process of selection and reproduction continues until the new generation is completely populated.  In Carroll's micro-GA configuration, elitism ensures that the best individual from the previous generation is replicated into the new generation and the absence of niching means that only one local optima can exist in a population at any given time.  

\subsection{Measuring Fitness (The \textit{Evaluation Module})}\label{ssec_fitness}

As previously discussed, the only interface between the GA and our $N$-body simulation is the \textit{Evaluation Module}, which determines the fitness of an individual.  For our problem, we use the photometric and kinematic observations discussed in \S\,\ref{sec_observ} to measure the fitness of each simulated orbit.  Since the fitness is a solitary number fed into the genetic algorithm, each simulated galaxy has its photometric and kinematic goodness-of-fit measurements weighted and combined into a single value.  The goal of the fitness value is to tell the GA when and how to improve.  Hence, it is not advantageous to use a straightforward $\chi^{2}$ method (especially since the simulated photometric and kinematic results are dependent on the same parameters), it is, in fact, more effective to weight and combine individual tests that measure the qualities of a given orbit.  In our simulation, a total of 6--7 tests are used.  After testing many orders and weighting schemes we find the following configuration to be most effective:  
two-dimensional two-sided Kolmogorov-Smirnov (2DKS) test with a weight of $1\%$(\S\,\ref{sssec_kstest}), slope measurement with a weight of $9\%$ (\S\,\ref{sssec_slope}), weighted velocity profile $\chi^2$ probability with a weight of $18\%$ (\S\,\ref{sssec_wtvelo}), and surface brightness profile $\chi^2$ probabilities  with weights of $18\%$ each (\S\,\ref{sssec_photo}).  The weights and criteria for each of these evaluations is discussed below and summarized in Table \ref{table:fitness}.
\begin{deluxetable*}{l l c c | l l}
\tabletypesize{\scriptsize}
\tablewidth{0pt}
\tablecolumns{6}
\tablecaption{Determining Fitness, $F$}
\tablewidth{0pt}
\tablehead{
\colhead{} &
\colhead{Test} &
\colhead{Passing} &
\colhead{Test's} &
\colhead{Fitness if} &
\colhead{Fitness if} \\
\colhead{} &
\colhead{} &
\colhead{Criteria} &
\colhead{Weight} &
\colhead{Orbit \textbf{Fails} Test} &
\colhead{Orbit \textbf{Passes} Test} }
\startdata
1. 	& 2DKS Probability & $\ge 0.01$ & $1\%$ 
		& $F = \rm{P(D > observed)}$\tablenotemark{a} & $f_{1}$ = 0.01 \hspace{.3in} \\[.1in]\hline
& & & & & \\
2.\tablenotemark{b}  	& Slope, m  
 		& $0.0 \ge \rm{m} \ge $ $-$0.80 & $9\%$ 
		& $F= f_{1} + 0.09\delta |\displaystyle\frac{\rm{m_{mid}}}
			{\rm{m - m_{mid}}}|\tablenotemark{c}$ & $f_{2} = f_{1} + 0.09$ \\
	 & (in \kms\ arcsec$^{-1}$)& & & & \\[.1in]\hline
& & & & & \\
3.	&Weighted Velocity & $\ge 10^{-10}$ & $18\%$ 
		& $F = f_{2} - \displaystyle\frac{1.0}{{\rm log}[Q(\chi^{2}|\nu)]}$\tablenotemark{d} 
		& $f_{3} = f_{2} + 0.10 + 0.08\times Q(\chi^{2}|\nu)$ \\
 	&  $\chi^2$ Probability & & & & \\[.2in]\hline
& & & & & \\
4.	&Surface Brightness & & 
		&\multicolumn{2}{c}{$F = f_3 + f_4 + f_5 + f_6 + f_7$}\\[.05in]
	& $\chi^2$ Probabilities: & & & & \\[.05in]
     	& \hspace{.05in} Radial & $\ge 10^{-10}$ & $18\%$ 
		&  \multicolumn{1}{c}{$f_{4} =\displaystyle\frac{-1.0}{{\rm log}[Q_{4}(\chi^{2}|\nu)]}$}
		&  \multicolumn{1}{c}{$f_{4} = 0.10 + 0.08 \times Q_{4}(\chi^{2}|\nu)$}  \\[.1in]
     	& \hspace{.05in} Angular & $\ge 10^{-10}$& $18\%$ 
		&  \multicolumn{1}{c}{$f_{5}=\displaystyle\frac{-1.0}{{\rm log}[Q_{5}(\chi^{2}|\nu)]}$} 
		&  \multicolumn{1}{c}{$f_{5} = 0.10 + 0.08 \times Q_{5}(\chi^{2}|\nu)$}  \\[.1in]
     	& \hspace{.05in} Weighted Radial & $\ge 10^{-10}$& $18\%$ 
		&  \multicolumn{1}{c}{$f_{6}=\displaystyle\frac{-1.0}{{\rm log}[Q_{6}(\chi^{2}|\nu)]}$}
		&  \multicolumn{1}{c}{$f_{6} = 0.10 + 0.08 \times Q_{6}(\chi^{2}|\nu)$}  \\[.1in]
     	& \hspace{.05in} Weighted Angular & $\ge 10^{-10}$& $18\%$ 
		&  \multicolumn{1}{c}{$f_{7}=\displaystyle\frac{-1.0}{{\rm log}[Q_{7}(\chi^{2}|\nu)]}$}
		&  \multicolumn{1}{c}{$f_{7} = 0.10 + 0.08 \times Q_{7}(\chi^{2}|\nu)$} \\[.1in]              
\enddata
\label{table:fitness}
\tablenotetext{a}{ Where the 2DKS probability P(D $>$ observed) is given in \citet{pre86}}
\tablenotetext{b} { Used only in the exponential disk models.  For the non-rotating spheroid model, $f_{2}$ is set to 0.10}
\tablenotetext{c} {  Where $\rm{m_{mid}} = $ $-$0.40 \kms\ arcsec$^{-1}$ and 
 	$\delta = \left\{ \begin{array}{ll}
	0,  &\mbox{if $|\displaystyle\frac{ \rm{m_{prev} - m_{mid}}}{\rm{m - m_{mid}}}| > 1$} \\
	1, &\mbox{if $|\displaystyle\frac{ \rm{m_{prev} - m_{mid}}}{\rm{m - m_{mid}}}| \le 1$} \\
	\end{array}\right.$}
\tablenotetext{d} {  Where $Q(\chi^{2}|\nu)$ is the $\chi^2$ probability function from \citet{pre86}}
\end{deluxetable*} 
\subsubsection{Two-Dimensional Two-Sided Kolmogorov-Smirnov (2DKS) Test}\label{sssec_kstest}
Our first evaluation uses the two-dimensional two-sided Kolmogorov-Smirnov (2DKS) test \citep{pre86}. The observed surface brightness profile of NGC~205 system is modeled as a distribution of particles placed as tracers of the projected surface brightness profile.  The number of particles in the `observed' distribution is chosen such that it equals the number of particles in the simulated distribution.  
The 2DKS test probability, ${\rm P}_{(\rm{D > obs})}$, is measured using the statistic from Numerical Recipes \citep{pre86}.

Simulations resulting a probability $\ge 0.01$ are given a fitness value of 0.01 and allowed to proceed to the next test.  Those with values $< 0.01$ return a fitness value equal to the 2DKS probability value.  The 2DKS test fitness value is thus,
\begin{equation}
\rm{Fitness} = {\rm min[P_{(D>obs)}}, 0.1]
\end{equation}
A weight of only $1\%$ (or a maximum fitness value of 0.01) is given to the 2DKS test because it is most effective at recognizing bad fits, rather than quantifying good fits.  That is, the two disadvantages of this test are that it has only a 0.20 significance level and a bias towards the center (since this is where most of the particles are located).

\subsubsection{Interior Velocity Slope (Disk Models Only)}\label{sssec_slope}

We ensure that NGC~205 has the correct tidally \textit{un}distorted central semi-major axis velocity profile by using the velocity slope interior to $\sim270\arcsec$ for the second fitness evaluation.  This test is performed only on the disk models, since the non-rotating spheroid contains no preservable internal rotation.   
To determine the simulated semi-major axis velocity profile (and measure its slope), the minimum distance between each test particle and the $S$-shaped major axis of NGC~205 is computed.  
The slit-masks used by \citet{geh06} to observe NGC~205's velocity profile are adjacent to the semi-major axis, with dimensions $16\arcmin$ (3.8 kpc) long by $4\arcmin$ (1.0 kpc) wide.  Based on the \citet{geh06} collection method, simulated particles whose perpendicular distance to the major axis is less than $250\arcsec$ (1 kpc) are selected, and their velocities are binned and average with neighboring particles to create a simulated velocity profile.  The radial size of each velocity bin along the semi-major axis is roughly $1\arcmin$ (0.24 kpc).

The allowed slope (m) range is determined from \citet{geh06} velocity observations of NGC~205 (see \S\,\ref{ssec_kinematic}). We set a generous slope range for our simulations of $-$0.80 \kms\ arcsec$^{-1}$ to 0.0 \kms\ arcsec$^{-1}$ (see Figure \ref{fig:velocity}), corresponding to a $-$20 \kms\ kpc$^{-1}$ to 0.0 \kms\ kpc$^{-1}$ gradient, if NGC~205 lies at a distance of 824 kpc.  The significance level is chosen such that the slope has the maximum range to vary, without ever going positive, and corresponds to 1.83$\sigma$.

If the slope falls within the range specified above, it is given a probability of 1 and a weight of $9\%$ (or a value of 0.09).  This is then added to the 2DKS result of 0.01 (for a total fitness value of 0.10) and the code proceeds to the next test.  If it does not fall within the range, it is compared with the slope of a previous run in order to determine if the GA is improving and assigns a fitness value.  
Improvement is defined by how the inner slope is changing with respect to previous orbits and is symbolized by the parameter $\delta$.
The parameter $\delta$ is set equal to 1 if the absolute value of the slope minus the observationally best fit slope of $-$0.40 \kms\ arcsec$^{-1}$ is \textit{less} than the previous orbit's value 
(i.e. $|$m$_{\rm new} - (-$0.40 \kms\ arcsec$^{-1})| < |$m$_{\rm prev} - (-$0.40 \kms\ arcsec$^{-1})|$), and is otherwise set equal to 0.  The fitness value at this step is thus given by,
\begin{equation}
\rm{Fitness} =
\left\{ \begin{array}{lc}
0.10, & \hspace{-.7in} \mbox{if $-0.8 \le$ m $\le 0.0$ \kms\ arcsec$^{-1}$}\\[.13in]
0.01 + 0.09\delta\displaystyle\left|\frac{-0.40}{\rm{m} - (-0.40)}\right|, & \mbox{otherwise} \\
\end{array}\right.
\end{equation}
This results in a minimum fitness value of 0.01 for a slope of $\infty$, and a maximum fitness value of 0.10 for a slope very close to 0.0 \kms\ arcsec$^{-1}$ or 0.80 \kms\ arcsec$^{-1}$.  
If, instead, the absolute value of the slope minus the best fit slope is \textit{greater} than the previous orbit's value (i.e. 
$|$m$_{\rm new} - (-$0.40 \kms\ arcsec$^{-1})| > |$m$_{\rm prev} - (-$0.40 \kms\ arcsec$^{-1})|$
, then the parameter $\delta=0$ and the GA returns a fitness value of 0.01.  The ultimate goal of this evaluation is to inform the GA if it is improving on the internal velocity profile.  In the case of the non-rotating spheroid, the Fitness is automatically set to 0.10 before proceeding to the next test.

\subsubsection{Weighted Semi-Major Axis Radial Velocity Profile}\label{sssec_wtvelo}

If the 2DKS test and central slope requirements are satisfied, a weighted $\chi^2$ statistic is determined for the semi-major axis velocity profile and a probability $Q$ inferred.  
Similar to the method outlined in \S\,\ref{sssec_slope}, simulated particles with a perpendicular distance to NGC~205's semi-major axis less than $250\arcsec$ (1 kpc) are collected and binned.  
The only difference is that the radial length of the velocity bins are slightly larger than those used to measure the slope.  Each bin is chosen to be in accordance with those used by \citet{geh06}, resulting in an average bin width of approximately $2\arcmin$ (0.48 kpc). 

The $\chi^2$ weights are constructed so that more weight is given to the tidally distorted regions beyond $\sim270\arcsec$.  Ergo, this externally weighted velocity profile compliments the interior velocity slope discussed in \S\,\ref{sssec_slope} to provide a complete description of the semi-major axis velocity profile.  We use the weighting function,   
\begin{equation}
w(r) = \left\{ \begin{array}{ll}
 \frac{1}{2} \exp\left[\frac{\begin{array}{c}r - r_{1/2}\end{array}}{\begin{array}{c}\Delta\end{array}}\right]               
 &\mbox{  if  $r<r_{1/2}$} \\
1 - \frac{1}{2} \exp\left[\frac{\begin{array}{c}r_{1/2} - r\end{array}}{\begin{array}{c}\Delta\end{array}}\right] 
&\mbox{  if  $r\ge r_{1/2}$} \\
\end{array}\right.
\label{eq:weight}
\end{equation}
where $\Delta$ determines steepness of the function, and $r_{1/2}$ corresponds to the semi-major axis distance at which the weight is equal to $\frac{1}{2}$.  In our model, we set $\Delta=50$ and $r_{1/2}=500\arcsec$ in order to produce $\chi^2$ weights that turn on at $\sim270\arcsec$ and rise rapidly.  
Since the weights are discretely sampled in accordance with \citet{geh06} bins, we normalize the weights to the number of available bins,
\begin{equation}
w_{i} = \frac{N \times w(r_{i})}{\displaystyle\sum_{i=1}^{N} w(r_{i})}
\label{eq:nweight}
\end{equation}
where $r_{i}$ is the semi-major axis value corresponding to the center of each velocity bin, and $N$ is the total number of bins.
These weights are then used to determine the $\chi^2$ statistic,
\begin{equation}
\chi^2 = \displaystyle\sum_{i=1}^{N} \frac{w_{i}(S_{i} - O_{i})^{2}}{\sigma_{i}^{2}}
\label{eq:chi}
\end{equation}
where $S_{i}$ is the average radial velocity found by the \textit{simulation} in the $i$th bin, and both $O_{i}$ and $\sigma_{i}$ are the \textit{observed} average radial velocity and error, respectively, in the $i$th bin.

Once the $\chi^2$ value is obtained, its statistical significance can be determined using the $\chi^2$ probability function, $Q(\chi^{2}|\nu)$ \citep{pre86}, where the degrees of freedom, $\nu$, is given by $N-1$.    
If the weighted radial velocity profile probability is $\ge 10^{-10}$ it is considered an adequate fit and a value of $0.1 + 0.08\times$probability is added to the slope and 2DKS fitness'.  The code then proceeds to the next set of tests.  
However, if the weighted radial velocity profile probability is $< 10^{-10}$, then a new fitness is  computed by subtracting the inverse logarithm of the probability from the slope and 2DKS fitness'.  So, based on the probability, the orbit has an updated fitness value of,

\begin{equation}
\rm{Fitness} =  
\left\{ \begin{array}{cl}
f_{3}\equiv 0.2 + 0.08 \times Q(\chi^{2}|\nu), & \mbox{if $Q \ge 10^{-10}$} \\[.13in]
0.10 - \displaystyle\frac{1.0}{\rm{log}[Q(\chi^{2}|\nu) ]}, & \mbox{if $Q< 10^{-10}$} \\[.08in]
\end{array}\right.
\label{eq:fit_velo}
\end{equation}
This results in a minimum fitness value of 0.10 for a $Q(\chi^{2}|\nu)$ probability of 0, and a maximum fitness value of 0.28 for a $Q(\chi^{2}|\nu)$ probability of 1.  
The contribution of the weighted radial velocity profile test to the the total fitness is 18\%.

\subsubsection{Weighted \& Unweighted Radial \& Angular Brightness Profile}\label{sssec_photo}

If the previous three tests (2DKS, slope, weighted radial velocity profile) are satisfied, the weighted and unweighted $\chi^2$ statistics for the projected surface brightness profile are obtained and probabilities, $Q$, inferred.  
Since NGC~205 is undergoing tidal distortion, information from both the radial and angular profiles can be used as constraints.
It is advantageous to look at the weighted and unweighted, radial and angular profiles independently since they each provide unique information about the satellite's orbit.  
That is, the weighted $\chi^2$ statistics give information about the exterior, tidally distorted regions of the satellite, whereas the unweighted $\chi^2$ statistics inform about the interior, undistorted regions of NGC~205.  Likewise, the radial profile reveals the extent that the satellite is being compressed or expanded, while the angular profile describes the twisting and elongation of NGC~205.

In our simulation, mass-less test particles are used as tracers of surface brightness.  The weighted and unweighted angular and radial distribution of test particles is compared with the surface brightness profile given in Figure \ref{fig:intensity}, which uses the data from \citet{cho02} in the range $100\arcsec$ to $672\arcsec$ and our fitted exponential surface brightness profile interior to $100\arcsec$ and exterior to $672\arcsec$.  In addition to this surface brightness profile, we use the position angles and ellipticities measured by \citet{cho02} from $100\arcsec$ to $672\arcsec$.  The position angle and ellipticities of the isophotes internal to $100\arcsec$ and external to $672\arcsec$ are held fixed, since interior to this we have removed the nucleus and exterior to this the surface brightness and sky become comparable.

For the $\chi^2$ tests, we step in semi-major axis units of  $94\arcsec$ ($\sim 0.38$ kpc) and in angular units of $9^{\circ}$.  We measure out to a maximum semi-major axis distance of $31.2\arcmin$ 
($\sim 7.5$ kpc) and use the center of each areal bin to determine the corresponding surface brightness.  Since our semi-major axis steps are not coincident with those of \citet{cho02} isophotes, we interpolate to obtain an intermediate ellipticity and position angle that is complementary to our step.  
The unweighted radial surface brightness profile of the simulated distribution is then found by summing the number of tracer particles within each predetermined isophotal annulus, while the unweighted angular surface brightness profile is measured by totaling the number of tracer particles contained within each angular beam.  
The unweighted $\chi^2$ statistic for both the radial and angular surface brightness profiles can then be computed using Eqn \ref{eq:chi}, setting each weight $w_{i}$ equal to 1, normalizing our observed distribution $O_{i}$ to total the number of tracer particles used in the simulation, and setting $\sigma_{i}^2$ equal to $O_{i}$, the number of expected particles in the bin.  
For the weighted $\chi^2$ tests, the same formula is followed, except that the weighting function given by Eqn \ref{eq:weight} is used to determine $w_{i}$.   

Each of the 4 surface brightness profile tests discussed above are evaluated independently and combined to update the GA fitness.  Similar to the velocity profile test in \S\,\ref{sssec_wtvelo}, tests with probabilities $\ge 10^{-10}$ are considered reasonable fits and are assigned a value of $0.1+0.08\times$probability.  Profiles resulting in probabilities $< 10^{-10}$ are designated bad fits and assigned a value of $-1.0/$log$[Q_{j}(\chi^{2}|\nu)]$.  The updated fitness then becomes,
\begin{displaymath}
\mbox{Fitness} =  f_{3} + \sum_{j=4}^{7} f_{j}
\end{displaymath}
\begin{equation}
f_{j} = \left\{ \begin{array}{cl}
0.10 + 0.08\times Q_{j}(\chi^{2}|\nu), & \mbox{if $Q_{j} \ge 10^{-10}$} \\[.08in]
\displaystyle\frac{-1.0}{{\rm log}[Q_{j}(\chi^{2}|\nu) ]}, & \mbox{if $Q_{j}< 10^{-10}$} \\[.08in]
\end{array}\right.
\end{equation}
where the $j$ index specifies which of the 4 surface brightness profile tests, and $f_{3}$ refers to the weighted velocity profile fitness from Eqn \ref{eq:fit_velo}.
Hence, each $\chi^2$ test can contribute a maximum of $18\%$ to the total fitness, resulting in a maximum fitness value of 1.
Once the fitness of a orbit has been determined, the GA can use the information to create new, improved orbits.

   \section{RESULTS}\label{sec_results}
In this section, we present the genetic algorithm's results for the three NGC~205 models outlined in \S\,\ref{ssec_particles}: the rotating cold disk (\S\,\ref{ssec_colddisk}), the rotating warm disk (\S\,\ref{ssec_warmdisk}) and the non-rotating hot spheroid (\S\,\ref{ssec_bulge}).  
We perform 1000 GA runs on each model (with different initial random number seeds for each run) in order to thoroughly explore the parameter space and expose possible degeneracies.  Each GA run consists of 1000 generations of 5 individual orbits and produces a best orbit (highest fitness) from these 5000 orbits.  Hence, for each model, $5\times10^6$ orbits are explored and $10^3$ best orbits are returned.   
Below, the resulting direction of NGC~205's approach on the sky plane, parameter values and single best orbit are given for these three models.

\subsection{Rotating Cold Disk}\label{ssec_colddisk}
We initialize NGC~205 as a rotating, cold exponential disk of mass-less test particles, serving as tracers of surface brightness.  The resulting direction of approach (\S\,\ref{sssec_cold_approach}), preferred parameter values (\S\,\ref{sssec_cold_params}) and best orbit (\S\,\ref{sssec_cold_best}) for the cold disk simulations are outlined below.

\subsubsection{Direction of NGC~205's Approach (Cold Disk)}\label{sssec_cold_approach}
\downscale
\begin{figure*}\includegraphics[trim=0.5in 3in 0in 3in,clip]{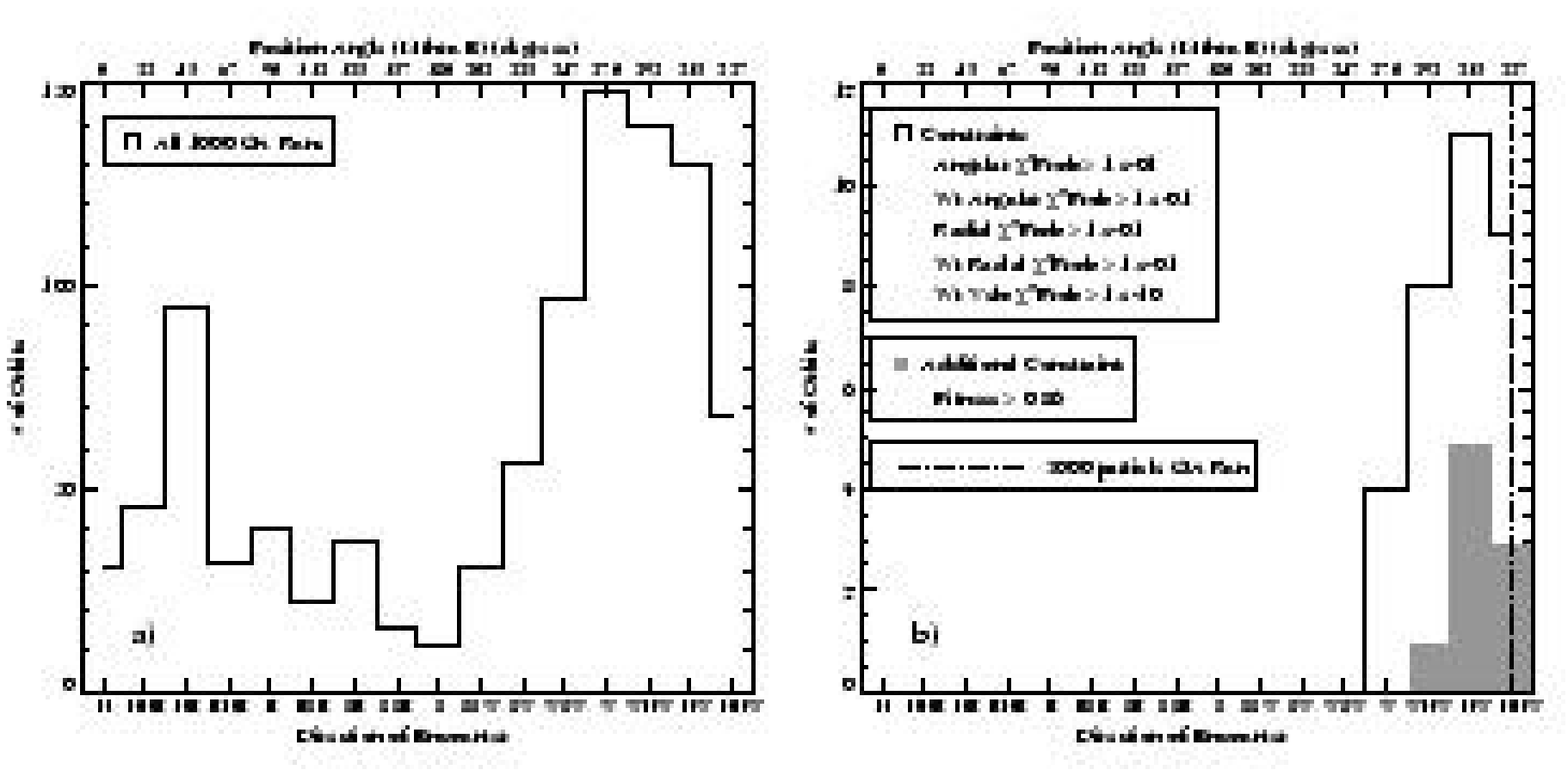}
\includegraphics[trim=0.5in 3in 0in 3in,clip]{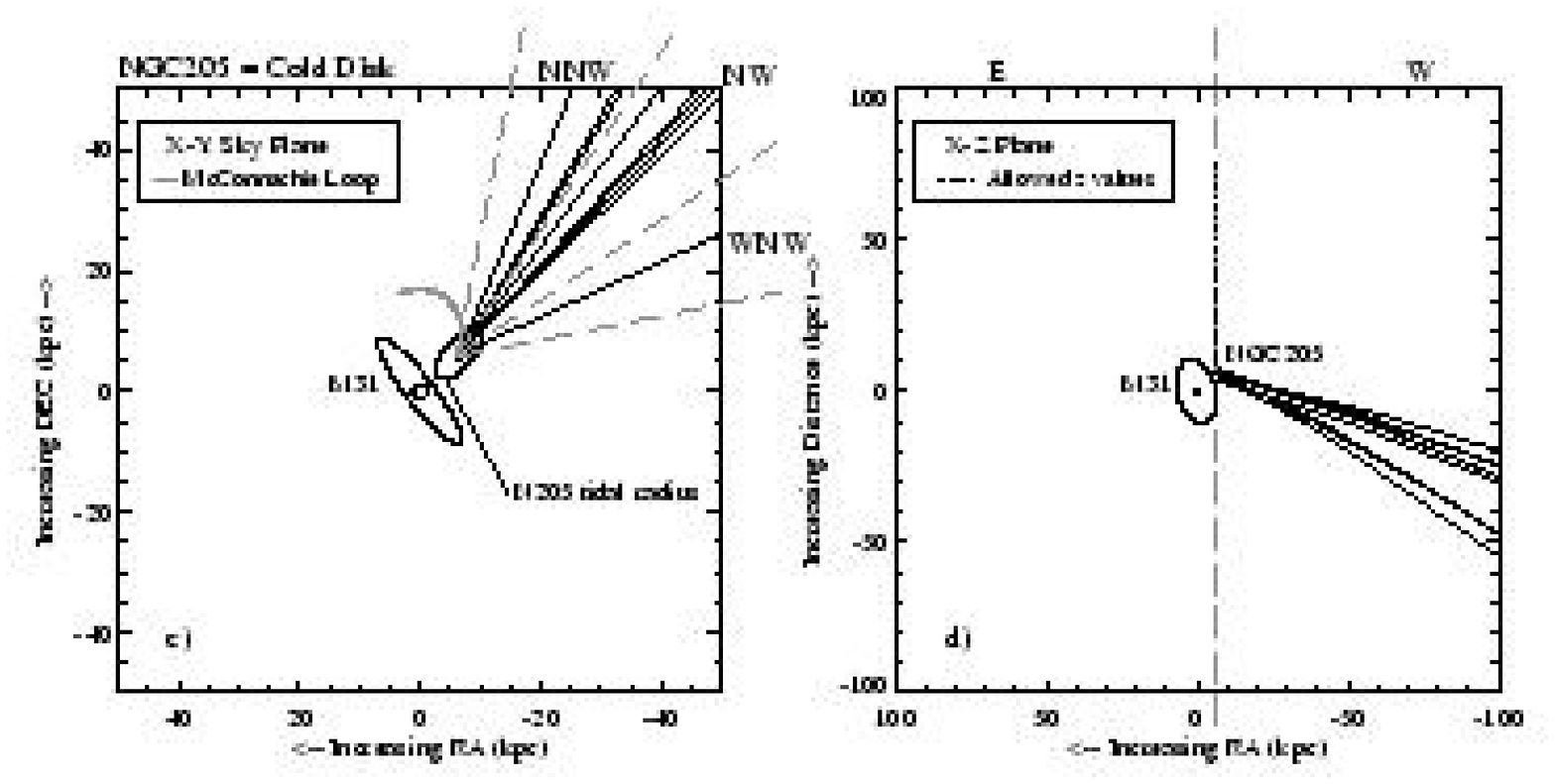}
\caption{Cold Disk Model. Direction of approach for NGC~205 modeled as an initially rotating, cold disk. (a) The best orbit from each of the 1000 GA runs. (b) Orbits with additional photometric, kinematic and fitness constraints.  Included is the resulting best orbit from a single GA run with 3000 particles (dashed line).  (c) The sky projected ($x-y$) orbits from the shaded region of the histogram.  Also plotted are M31's bulge and disk at two scale radii (subtending $5\arcmin$ and $45\arcmin$, respectively), an approximation of \citet{mcc04} stellar arc and two surface brightness isophotes, one at $270\arcsec$ (the onset of NGC~205's tidal distortion) and the other at $19.\arcmin77$ (NGC~205's tidal radius).  Note, none of these orbits trace the stellar arc observed in M31's northwest quadrant.  (d) The line-of-sight ($x-z$) orbits from the shaded region of the histogram.  Also included are M31's bulge and disk at two scale radii and the investigated $z$ values ($2-76$ kpc).}
\label{fig:cold_direct}
\end{figure*}
\returnscale
Figure \ref{fig:cold_direct} shows histograms of NGC~205's varying directions of approach on the plane of the sky and selected spatial projections for the cold, rotating disk model.  The histogram in Figure \ref{fig:cold_direct}a depicts the best orbit from each of the 1000 GA runs.  Even without any further constraints, a clear directional preference emerges from the 1000 orbits.  The histogram peaks at 148 orbits with NGC~205 advancing from the west (W), followed closely by 140 orbits from the west-northwest (WNW) and 130 orbits from the northwest (NW).  Of these 1000 orbits, 752 are bound to M31, 500 are prograde, 280 are retrograde and 220 are radial (with radial orbits defined by $\arccos|\hat{r}\cdot\hat{v}|\le 5^\circ$) .

Figure \ref{fig:cold_direct}b shows orbits with additional photometric, kinematic, and fitness constraints.  The outlined histogram imposes that the weighted velocity $\chi^2$ probability $\ge 10^{-10}$ and that each of the 4 surface brightness $\chi^2$ probabilities return values $\ge 0.1$ (\S\,\ref{sssec_wtvelo}--\S\,\ref{sssec_photo}).   This constraint significantly reduces the initial 1000 orbits to 32 orbits and returns a direction of approach that lies somewhere between the north-northwest (NNW) and west (W), with the peak lying in the NW.  Furthermore, none of these orbits are retrograde (in fact, none of the retrograde orbits have a fitness $> 0.25$).  The enclosed shaded region further imposes the condition that fitness $\ge 0.86$, a value selected to be just below that of the top 10 orbits (or $1\%$).  This results in a reduction from 32 orbits to 9, meaning that one of the top 10 orbits fails to satisfy all the imposed photometric and kinematic constraints.   The remaining directions of approach continue to peak in the NW, with only W approaches now ruled out.  Hence, the directional preference initially suggested by the histogram of 1000 orbits is reinforced by the addition of photometric, kinematic and fitness requirements.  

More importantly, these constraints rule out other possible orbits for the cold disk model.  This includes \citet{mcc04} favored $1^{\circ}$ long stellar arc-like feature (observed in the northwest quadrant of M31) as a tidal stellar stream originating from NGC~205.  Modeling this trail as tidal debris from the satellite suggests encroachment from somewhere in between the north-northeast (NNE) and east (E) region ($22.5^{\circ}$ to $90^{\circ}$) on the plane of the sky.   While $21\%$ of the orbits from the 1000 GA runs fall within this 4-bin region, we find that these orbits all have poor angular surface brightness distributions (i.e they cannot simultaneously match both a weighted and unweighted angular $\chi^{2}$ probability constraint of $0.1$).  This result indicates that orbits approaching from the NNE through E are ruled out as solutions by the observed isophotal twisting in NGC~205 (\S\,\ref{ssec_photo}).

To further test this conclusion, we perform a single cold disk GA run with 3000 particles.  While numerous GA runs with $>1000$ particles are too expensive for our purposes, a single GA run allows us to check our results against a more complex system.  The 3000 particle GA run returns a direction of approach from the NNW, an orbit in qualitative agreement with our 1000 particle runs.  This result is denoted in Figure \ref{fig:cold_direct}b with a dashed line. 
\begin{figure*}\includegraphics[trim=0in 1.2in 0in 1.2in,clip]{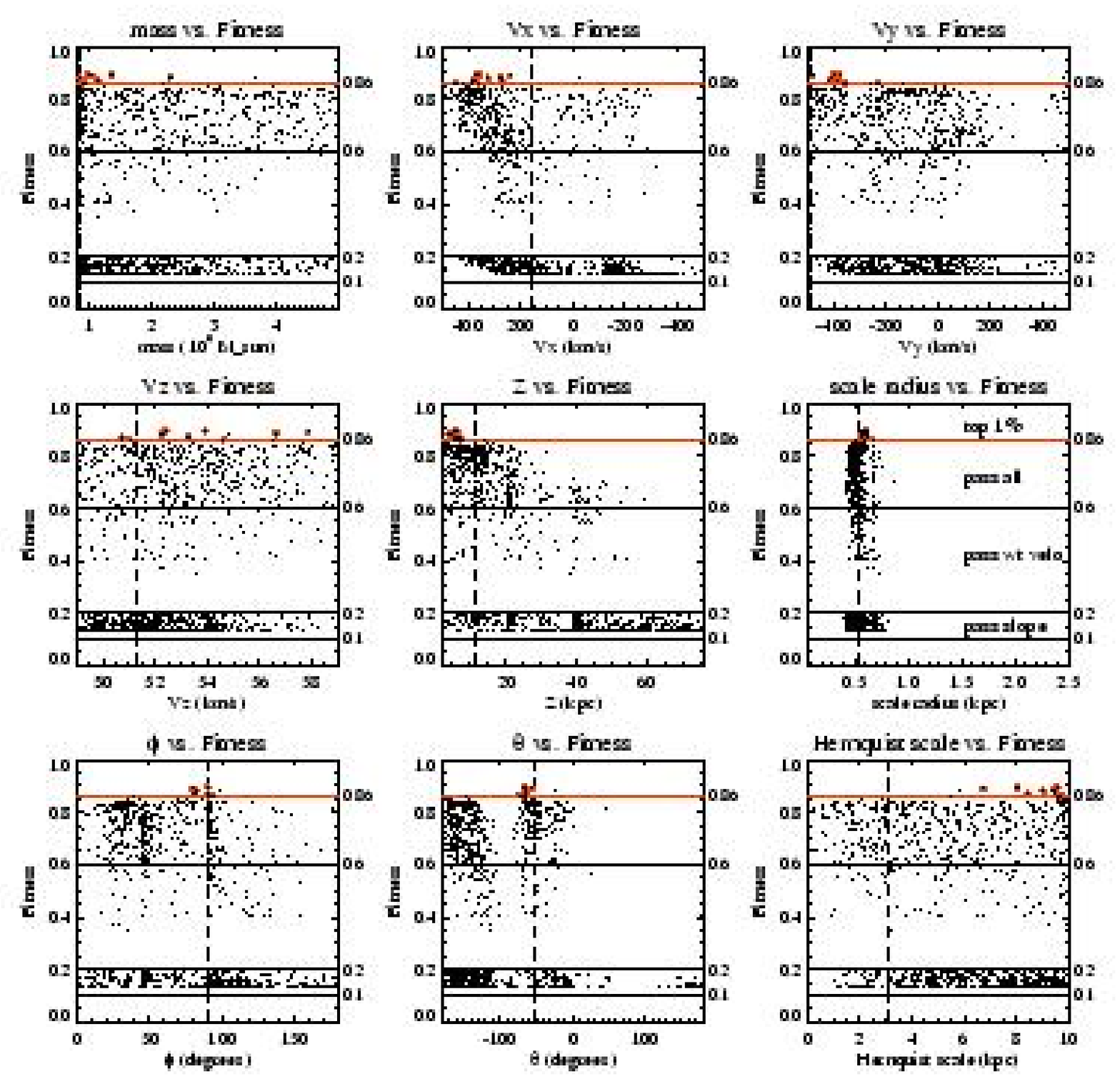}
\caption{Cold Disk Model. Plotted are the parameter values vs. fitness from the 1000 GA runs when NGC~205 is modeled as a rotating, cold disk.  Each point corresponds to the best orbit from an individual GA run and a particular initial configuration (structure, dynamics, orientation) of the progenitor.  The included lines denote passing criteria for the various fitness tests:  slope test (Fitness $\ge 0.1$), weighted velocity $\chi^{2}$ test (Fitness $\ge 0.2$), all five $\chi^{2}$ tests (Fitness $\ge 0.6$), top 10 ($1\%$) orbits (red line, Fitness $\ge 0.86$), and nine constrained orbits (highlighted in red).  Orbits that fail are poor fits to photometry and/or kinematics and reside below the Fitness $=0.6$ line.  Also included are the parameter values of the resulting best orbit from a single GA run with 3000 particles (dashed line).}
\label{fig:cold_params}
\end{figure*}

The plots at the bottom of Figure \ref{fig:cold_direct} show the trajectories of the constrained nine orbits corresponding to the shaded region in Figure \ref{fig:cold_direct}b.  All nine orbits have very large tangential motions and are not bound to M31.  Included in the $x-y$ sky projection plot (Figure \ref{fig:cold_direct}c) are tracers of the nine orbits, an approximation of \citet{mcc04} stellar arc, M31's bulge and disk at two scale radii (subtending $5\arcmin$ and $45\arcmin$, respectively), and two surface brightness isophotes, one at $270\arcsec$ corresponding to the onset of NGC~205's tidal distortion and the other at $19.\arcmin77$ indicating NGC~205's tidal radius.  As discussed, none of the constrained orbits trace the stellar arc observed in the northwest quadrant of M31.  Included in the $x-z$ line-of-sight plot (Figure \ref{fig:cold_direct}d) are tracers of the nine orbits, M31's bulge and disk at two scale radii, and the range of allowed $z$ values (2--76 kpc).  All nine orbits converge at distances very close to M31 and follow roughly similar trajectories.  Furthermore, the nine orbits begin on nearly radial courses with $0^{\circ} < \mu < 1.5^{\circ}$, where $\mu$ is the angle between $\vec{r}$ and $\vec{v}$, the position vector from the center of NGC~205 to M31 and the velocity vector of NGC~205, respectively.  Ergo, the radial paths of these nine simulated satellites precludes prograde or retrograde encounters.

\subsubsection{Preferred Parameter Values (Cold Disk)}\label{sssec_cold_params}
\downscale
\begin{figure*}\includegraphics[trim=0in 3.5in 0in 3.5in,clip]{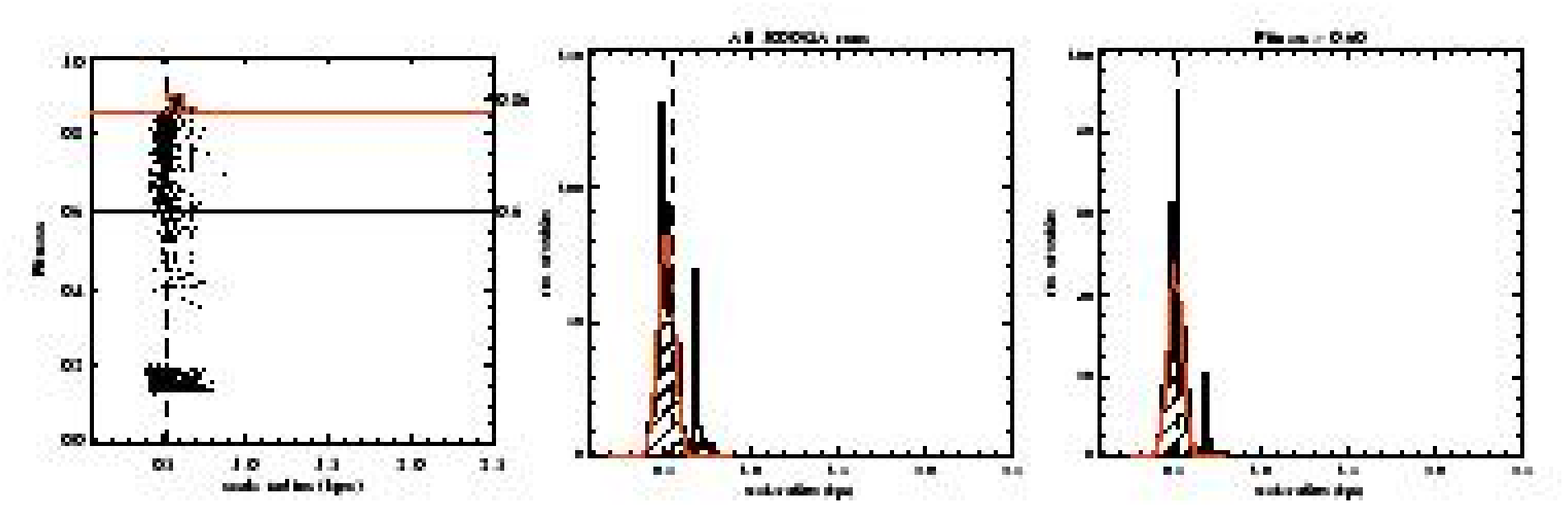}
\includegraphics[trim=0in 3.7in 0in 3.7in,clip]{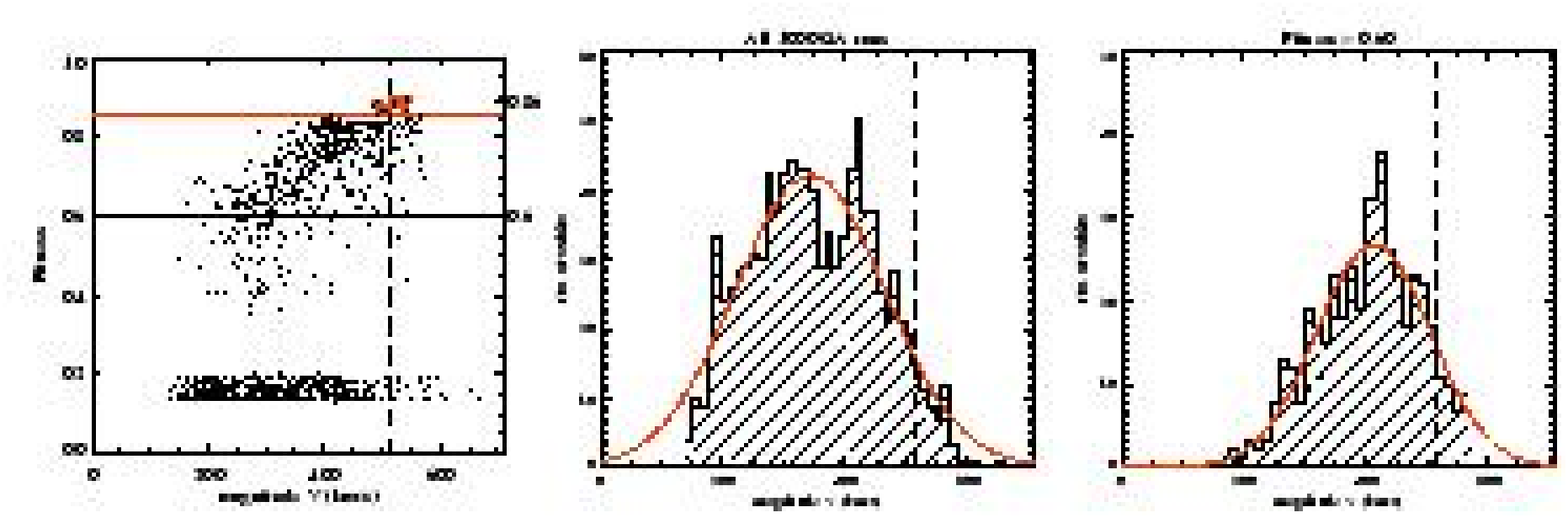}
\includegraphics[trim=0in 3.5in 0in 3.5in,clip]{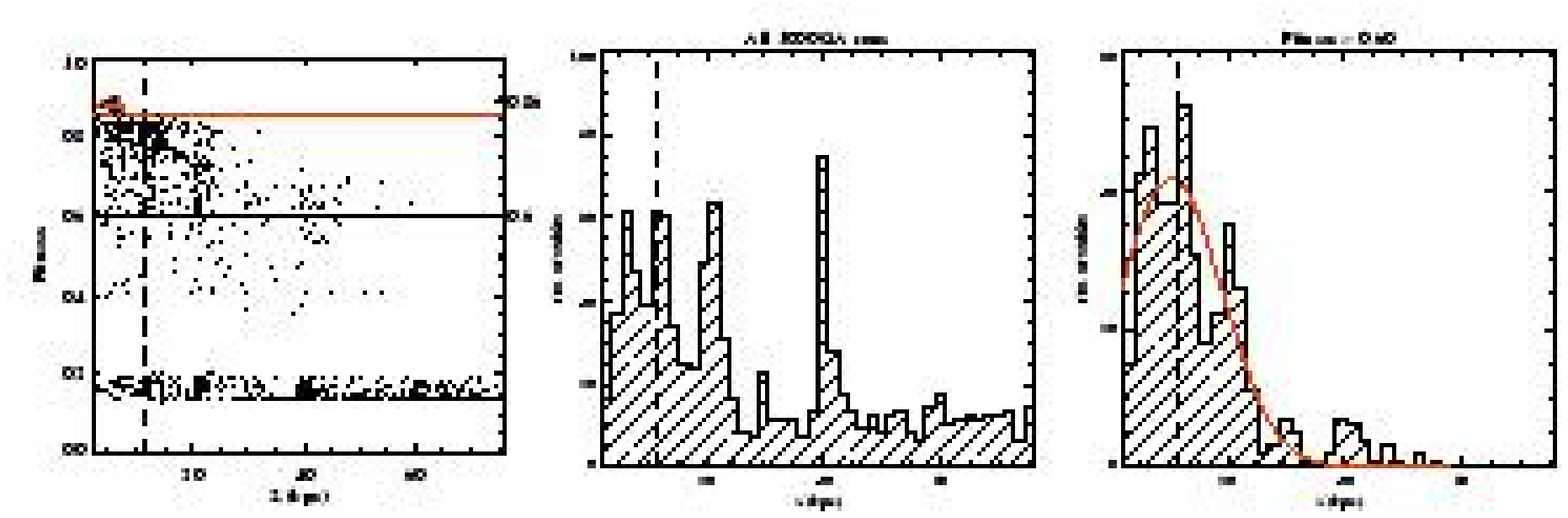}
\caption{Cold Disk Model. Shown here are three parameters, for the rotating, cold disk model, that are well constrained by the GA.  From top to bottom: initial scale length $r_{0,i}$, velocity magnitude $|v|=(v_{x}^{2}+v_{y}^{2}+v_{z}^{2})^{1/2}$, and line-of-sight distance $z$.  From left to right: parameter values vs. Fitness for the 1000 GA runs, histograms and fitted Gaussians (except $z$) for all 1000 GA runs, and histograms and fitted Gaussians for orbits satisfying Fitness $\ge 0.6$.  Orbits with Fitness $\ge 0.6$ give $r_{0,i}=128\pm10\arcsec$, $|v|=414\pm94$ \kms, $z=11\pm9$ kpc.  Also included are the parameter values of the resulting best orbit from a single GA run with 3000 particles (dashed line).}
\label{fig:cold_z}
\end{figure*}
\returnscale
Figure \ref{fig:cold_params} shows the evolved parameter values resulting from the 1000 cold disk GA runs.  Each best orbit's parameter and corresponding fitness are plotted.  Since the fitness is determined by combining numerous tests, the line illustrating passing criteria for each fitness step is also included, with the sole exception of the 2DKS test for which all 1000 orbits quantitatively pass.  The nine constrained orbits from Figure \ref{fig:cold_direct}b and the cutoff fitness value line for the top 10 orbits are highlighted in red.  Additionally, the resulting best orbit's parameter values from the 3000 particle GA run are included (dashed line).

The GA's preference for certain parameter values emerges in Figure \ref{fig:cold_params}.
The most notable concentration is in the initial scale radius $r_{0,i}$ of NGC~205, shown in detail in Figure \ref{fig:cold_z} (top row).  The 1000 simulations converge at $r_{0,i} = 129 \pm 12\arcsec$ ($0.51 \pm 0.05$ kpc), with orbits passing all five $\chi^2$ tests (Fitness $\ge 0.6$) at $r_{0,i} = 128 \pm 10\arcsec$ ($0.51 \pm 0.04$ kpc) and the constrained nine orbits (highlighted in red) focused at $148 \pm 7\arcsec$ ($0.59 \pm 0.03$ kpc), a value equivalent to the satellite's present scale radius.  This implies that a large portion of NGC~205 is unaffected by tides and the internal regions of the satellite have experienced little, if any, distortion.  
The remaining parameters experience a greater amount of scatter. However, definite trends still exist in the data.  For example, the simulations favor orbits with very large velocities of $414 \pm 94$ \kms\  (middle row of Figure \ref{fig:cold_z}, Fitness $\ge 0.6$), preferentially moving NGC~205 towards the south ($v_y=-346\pm385$\kms) and to the east ($v_x=333\pm89$\kms). 
In addition, the GA prefers $z$ distances very close to M31 with $z = 11 \pm 9$ kpc (bottom row of Figure \ref{fig:cold_z}, Fitness $\ge 0.6$).  An orbit's fitness beyond this distance declines rapidly.  Also included in Figure \ref{fig:cold_z} are the resulting best orbit's $r_{0,i}$, $|v|$ and $z$ values from the 3000 particle GA run (dashed line).  Note that these values are in qualitative agreement with the 1000 particle GA runs, with a slightly higher $|v|$ value that is comparable to the velocities of the nine constrained orbits.
 
Conversely, some of the parameters cannot be further reduced by the simulations.  The huge scatter in the radial velocity, $v_{z}$, indicates that that the GA is unable to constrain this value beyond current observations (Figure \ref{fig:cold_params}, middle row, first column).  This result is not surprising since the observed error on $v_z$ is quite small.  
Additionally, the GA is unable to reach convergence for the satellite's mass, $M_{205}$, and Hernquist scale length, $a_{205}$.  One explanation for this is that significant degeneracies might exist for $M_{205}$ and $a_{205}$ in the parameter space.  A second possibility is that perhaps the photometric and kinematic fitness tests are not as sensitive to $M_{205}$ and $a_{205}$ as they are to the other parameters.  Hence, the GA both produces inconclusive results for $M_{205}$ and $a_{205}$ and does not further reduce the observed error on $v_{z}$.  However, it is able to tightly constrain NGC~205's initial scale radius and place more general bounds on the remaining 5 parameters.  

\subsubsection{Best Orbit (Cold Disk)}\label{sssec_cold_best}
\downscale
\begin{figure*}\includegraphics[trim=0in 2in 0in 2.5in,clip]{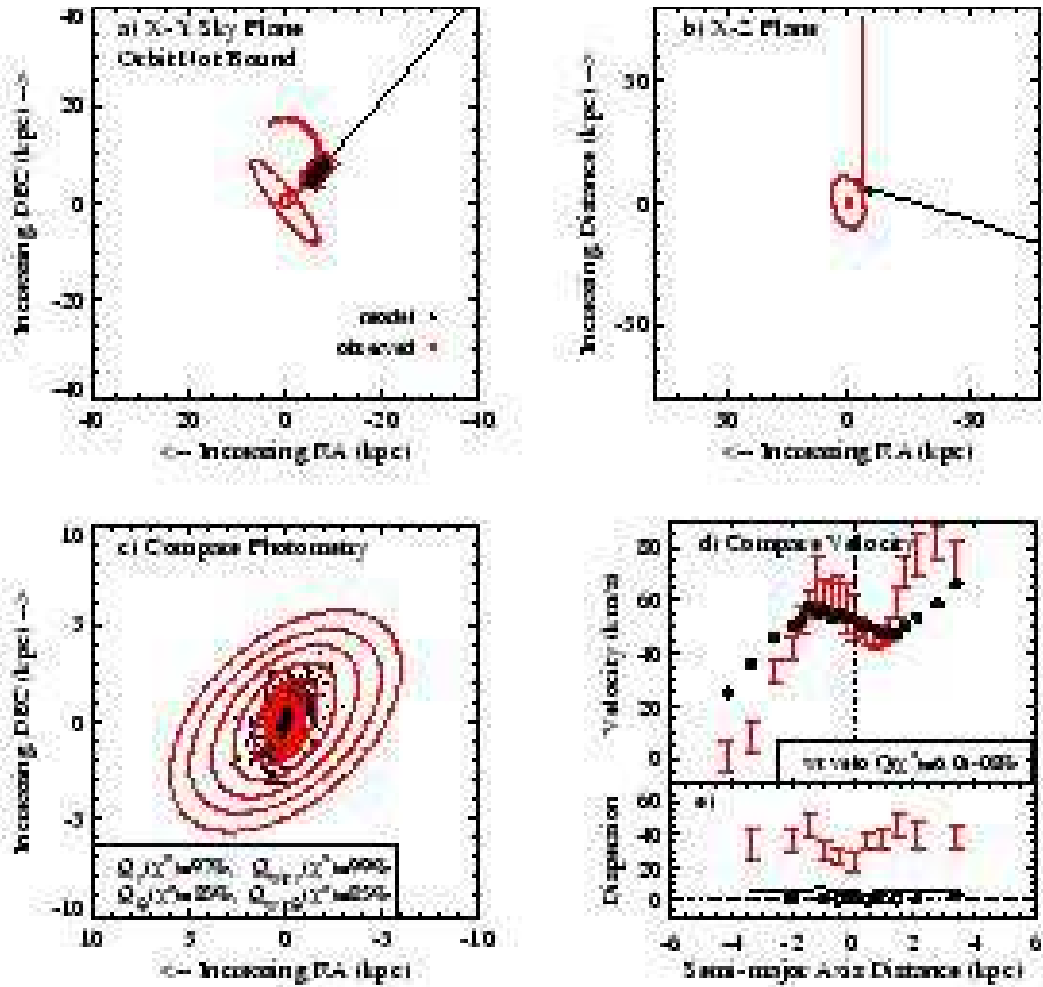}
\caption{Cold Disk Model.  The resulting best orbit, with a fitness of 0.90, from the 1000 GA runs when NGC~205 is modeled as a rotating, cold disk.   (a) Illustration of the orbit on the sky plane (black) + observations of M31, NGC~205 and the stellar loop (red).  (b) Orbit in the $x-z$ plane (black) + allowed $z$ values (vertical red).  (c) Integrated mass-less test particles (black) + \citet{cho02} isophotes (red).  (d) Simulated radial velocity profile (black) + \citet{geh06} corresponding radial velocity profile (red).  (e) Simulated velocity dispersion + \citet{geh06} corresponding dispersion profile (red).  Note, the open circles in the simulated velocity dispersion profile denote radii bins with velocity distributions that cannot be fit with a Gaussian.  Instead, the standard deviation of the distribution is given.  The solid black line denotes the results of a simulation using identical parameter values and 5500 test particles.}
\label{fig:cold_best}
\end{figure*}
\returnscale

\upscale
\begin{figure*}\includegraphics[width=7in,trim=0in 3.7in 0in 3.7in,clip]{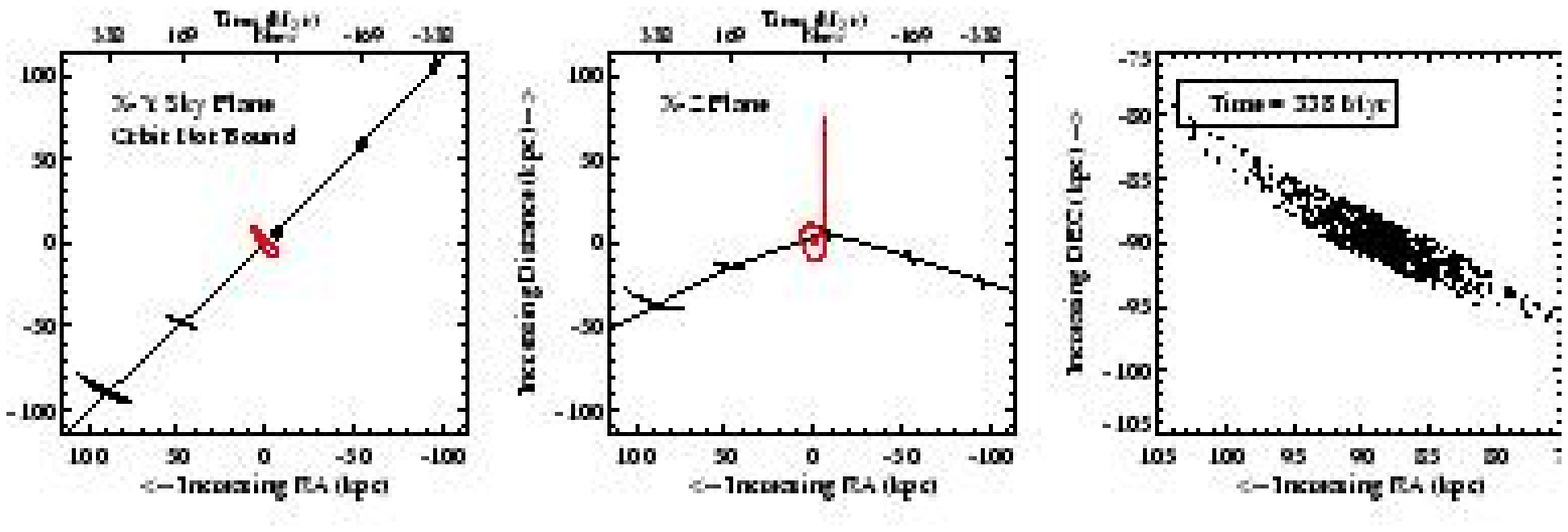}
\caption{Cold Disk Model.  Past, present and future projections of NGC~205 when the satellite is modeled as a initially rotating, cold disk.  The left panel shows the predictions in the plane of the sky.  The middle panel views the same orbit in the $x-z$ plane.  Note that the apparent kink in the orbit is merely a resolution effect and not indicative of an abrupt change in velocity.  The positions are given at times $-$338 Myr, $-$169 Myr, present, 169 Myr and 338 Myr.  The panel on the right is a zoomed in view of NGC~205 at 338 Myr on the plane of the sky.}
\label{fig:cold_future}
\end{figure*}
\returnscale
Figure \ref{fig:cold_best} shows the resulting best orbit from the 1000 GA runs using the cold disk model.  This best orbit approaches from the NW, is unbound to M31 (with $|v|= 546$\kms\  and $v_{\rm esc}=480$\kms), and has a fitness value of 0.90.  Figure \ref{fig:cold_best}a displays the simulated galaxy and its trajectory on the plane of the sky (in black).  Also included (in red) are M31's bulge and disk, at $2r_b$ and $2R_d$, and the predicted particle distribution for NGC~205, derived from \citet{cho02} isophotes and our surface brightness profile (see \S\,\ref{sec_observ}, Figure \ref{fig:intensity}).  Figure \ref{fig:cold_best}b shows this same orbit in the $x-z$ plane, with the addition of a vertical red line denoting the allowed $z$ parameter values ($2-76$ kpc).  Shown in the bottom plots are the simulated galaxy's velocity dispersion (circles) and performance on the five $\chi^2$ tests, with \citet{cho02} isophotes and \citet{geh06} velocity profile + errors depicted in red.  The simulated surface brightness $\chi^2$ tests result in probabilities of 0.979, 0.856, 0.999 and 0.852 for the radial, angular, weighted radial and weighted angular tests, respectively, and in a $\chi^2$ probability of $6.0\times10^{-10}$ for the weighted velocity profile.  Note that while the weighted velocity profile probability seems quite low, it satisfies the goal of the simulation by matching the turnover and reversal in the semi-major axis velocity profile.  Furthermore, although this satellite began as a purely rotating disk with no dispersion, after the simulated interaction with M31 a Gaussian fit to the combined velocity distribution for all particles along the semi-major axis is 3 \kms.  Even though this simulated average dispersion is much lower than the observed average dispersion of 42\kms\  \citep{geh06}, its presence indicates that some of the observed dispersion could have been tidally induced by M31.  In order to reduce the Poisson noise in the dispersion profile, we reran the orbit with 5500 particles.  The resulting 5500 particle Gaussian fitted velocity dispersion profile is given in Figure \ref{fig:cold_best}e (solid black line).  The parameter values of the 1000 particle orbit are $M_{205}=1.38\times10^{9}$\Msun, $v_{x}=369$\kms, $v_{y}=-399$\kms, $v_{z}=52$\kms, $z=6.3$ kpc, $r_{0,i}=0.6$ kpc, $\phi=90^{\circ}$, $\theta=-66^{\circ}$, and $a_{205}=9.6$ kpc.  This orbit is projected to pass within $\approx 9$ kpc of M31's center in the plane of M31's disk.  However, the radial path of the orbit precludes any sense of a prograde or retrograde encounter.

Using these orbital parameters, we roughly predict the position of the satellite and its particles at given points in time.  Figure \ref{fig:cold_future} illustrates the past, present and future predictions for NGC~205 when it is modeled as a cold, rotating disk.  The distribution of particles and positions in space are given for times $-$338 Myr, $-$169 Myr, present, 169 Myr and 338 Myr.  The panel to the far right in Figure \ref{fig:cold_future} shows NGC~205 in the plane of the sky at 338 Myr.  The simulated satellite experiences a significant amount tidal distortion after passing through the disk of M31.

\downscale
\begin{figure*}\includegraphics[trim=0.5in 3in 0in 3in,clip]{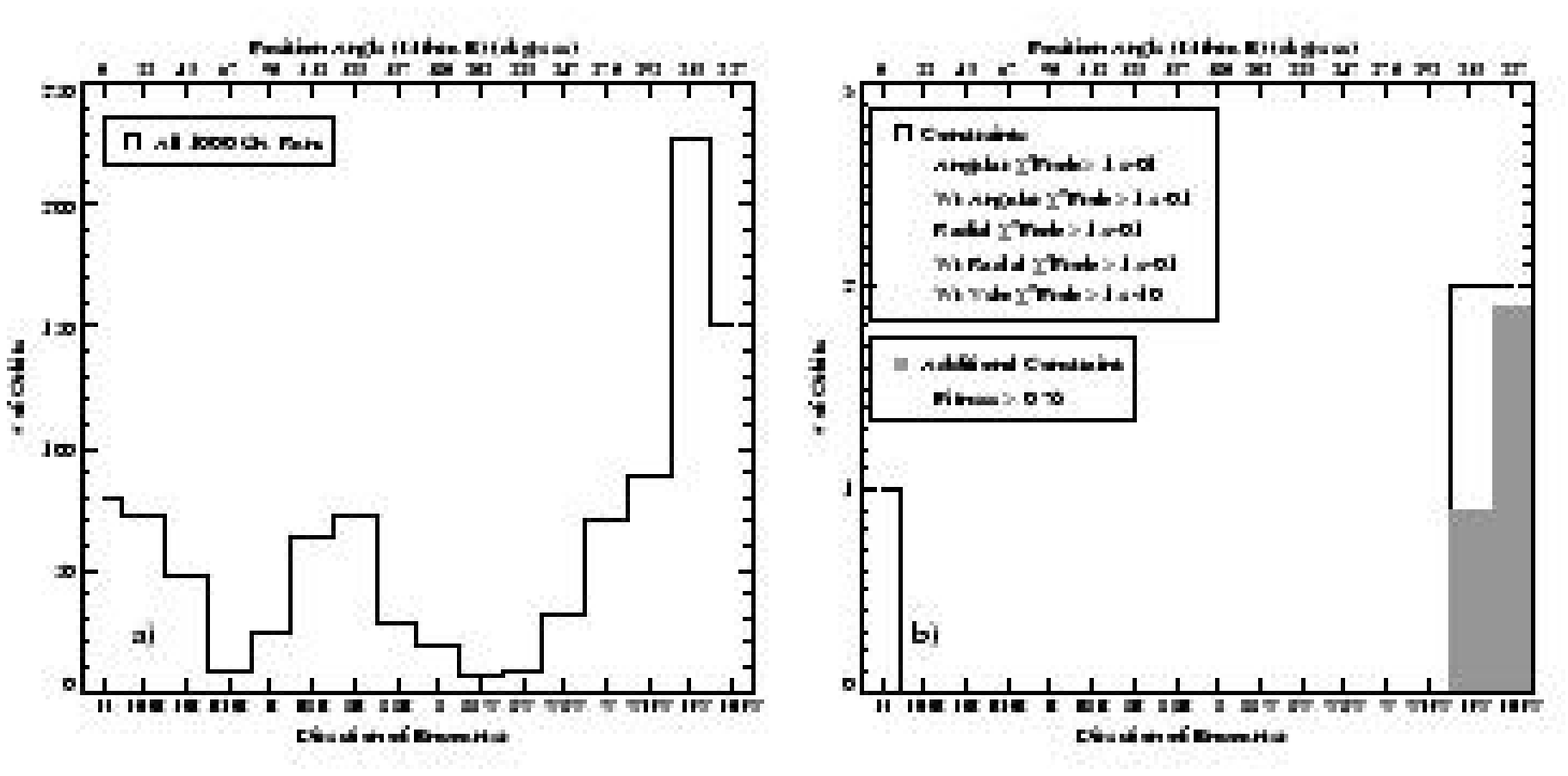}
\includegraphics[trim=0.5in 3in 0in 3in,clip]{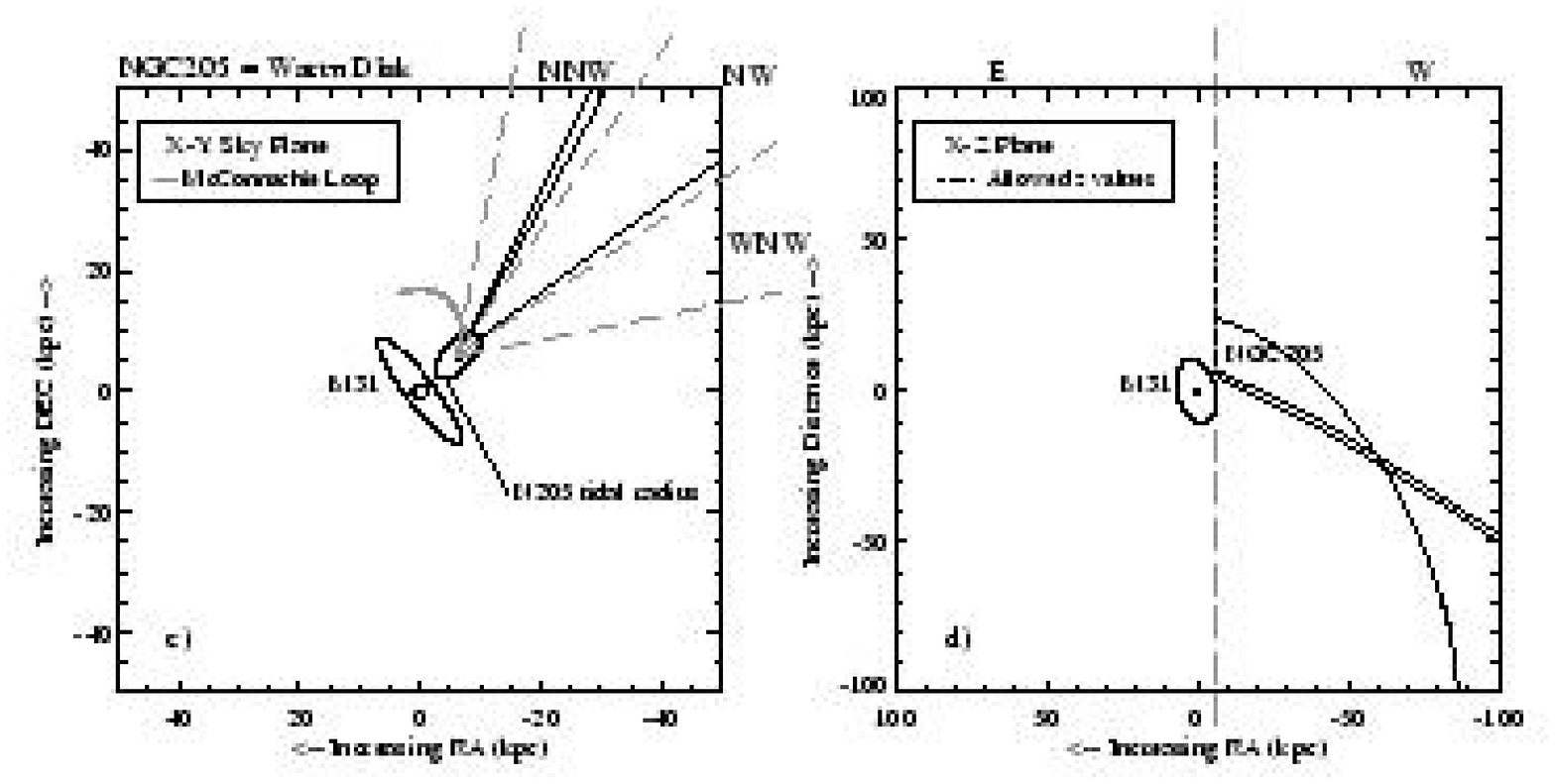}
\caption{Warm Disk Model. Same as Figure \ref{fig:cold_direct} but for NGC~205 modeled as a warm disk with both rotational and anisotropic velocities.  Note that the directions of approach on the plane of the sky are similar to those from the cold disk models.}
\label{fig:warm_direct}
\end{figure*}
\returnscale
\subsection{Rotating Warm Disk}\label{ssec_warmdisk}
This section is identical to \S\,\ref{ssec_colddisk} but for NGC~205 modeled as a rotating, warm exponential disk of mass-less test particles supported by a combination or rotational and anisotropic velocities.

\subsubsection{Direction of NGC~205's Approach (Warm Disk)}\label{sssec_warm_approach}

In Figure \ref{fig:warm_direct}a 
a distinct directional preference emerges.  The histogram peaks at 228 orbits with NGC~205 advancing from the northwest (NW), followed by 151 orbits from the north-northwest (NNW) and 89 orbits from the west-northwest (WNW). Almost half the orbits are contained within these three directional bins.  Furthermore, these directions of approach reinforce those from the 1000 cold disk model runs, which peaked in the west (W).  Of these 1000 orbits, 808 are bound to M31, 233 are prograde, 593 are retrograde and 174 are radial.  

Figure \ref{fig:warm_direct}b shows 
an effective reduction of 
1000 orbits to 5 and returns a direction of approach lying somewhere between the north (N) and northwest (NW), with the peaks in both the NW and NNW.  Of these 5 orbits, 2 are prograde, 1 is retrograde, and 2 are radial.  The shaded region enclosed within the histogram further imposes that the Fitness $\ge 0.76$, a value selected to be just below that of the top 10 orbits (or $1\%$).  This constraint eliminates the retrograde orbit and reduces the 5 orbits to 3, indicating that 7 of the top 10 orbits fail to satisfy all the imposed photometric and kinematic constraints given above.   The remaining directions of approach continue to peak in the NNW, with only N approaches now ruled out, a result similar to that of the cold disk model's.  Hence, the initial directional preference from the 1000 orbits histogram is reinforced with the addition of photometric, kinematic and fitness constraints.  

These restrictions rule out other possible orbits, including those tracing the stellar arc-like feature seen to the north of M31.  The warm disk model simulations place $15\%$ of the 1000 GA orbits in bins NNE through E.  However, only 2 of these 153 orbits meet both a weighted and unweighted angular $\chi^{2}$ constraint of $0.1$ and fail to meet a weighted and unweighted radial $\chi^{2}$ constraint of $0.1$.  As with the cold disk model, orbits tracing the observed stellar arc are ruled out as solutions by NGC~205's photometric profile.

\subsubsection{Preferred Parameter Values (Warm Disk)}\label{sssec_warm_params}
\begin{figure*}\includegraphics[trim=0in 1.2in 0in 1.2in,clip]{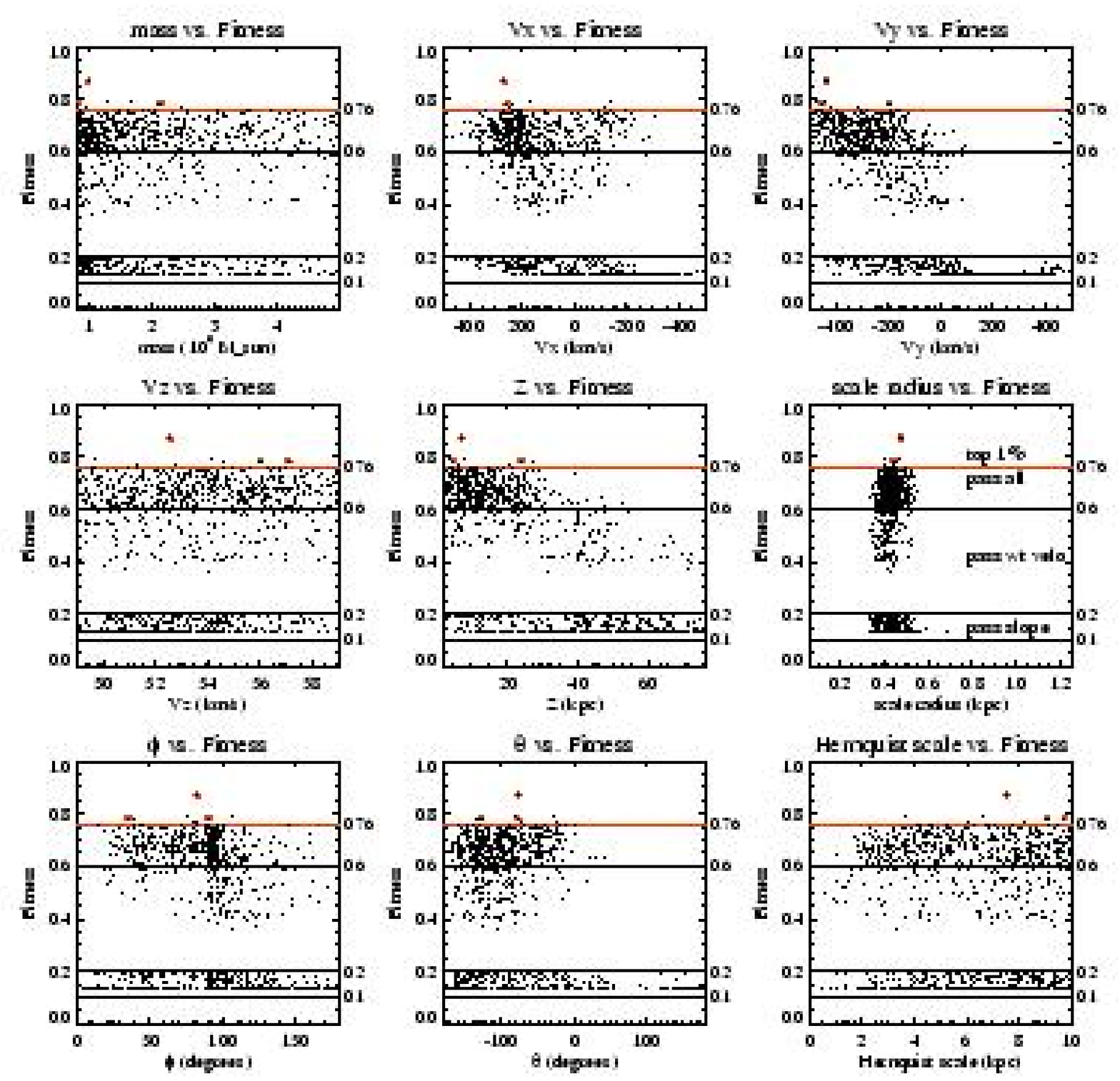}
\caption{Warm Disk Model. Same as Figure \ref{fig:cold_params} but for the warm disk model with only three constrained orbits, highlighted in red.}
\label{fig:warm_params}
\end{figure*}
\returnscale
\downscale
\begin{figure*}\includegraphics[trim=0in 3.5in 0in 3.5in,clip]{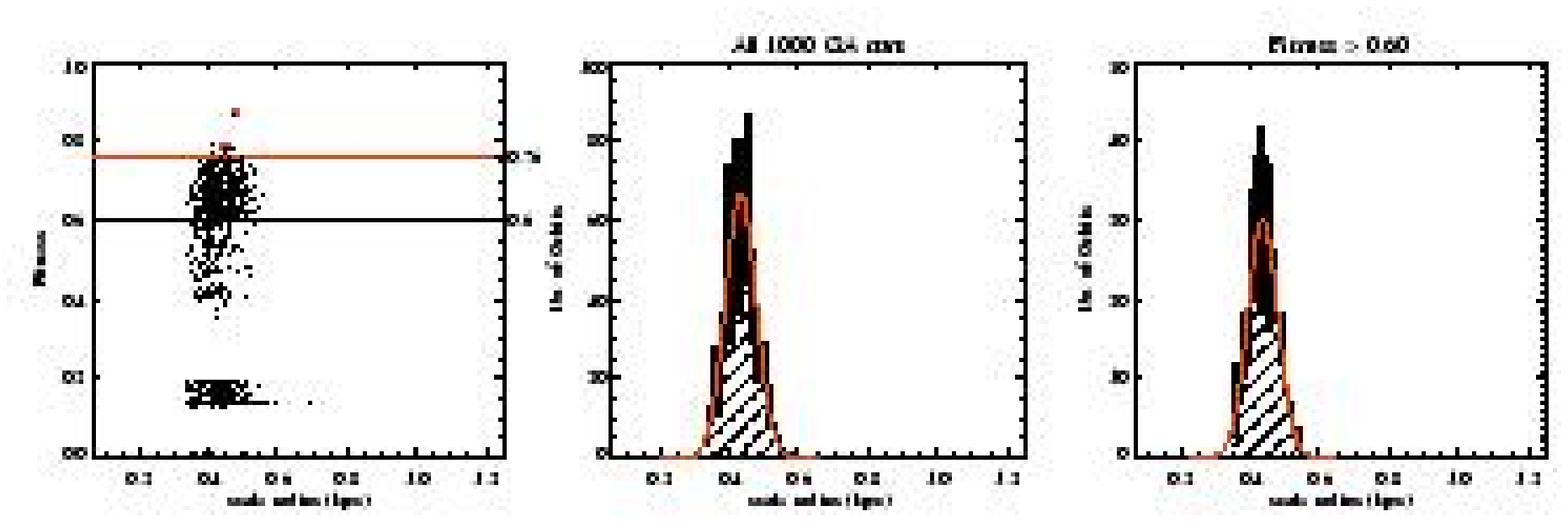}
\includegraphics[trim=0in 3.7in 0in 3.7in,clip]{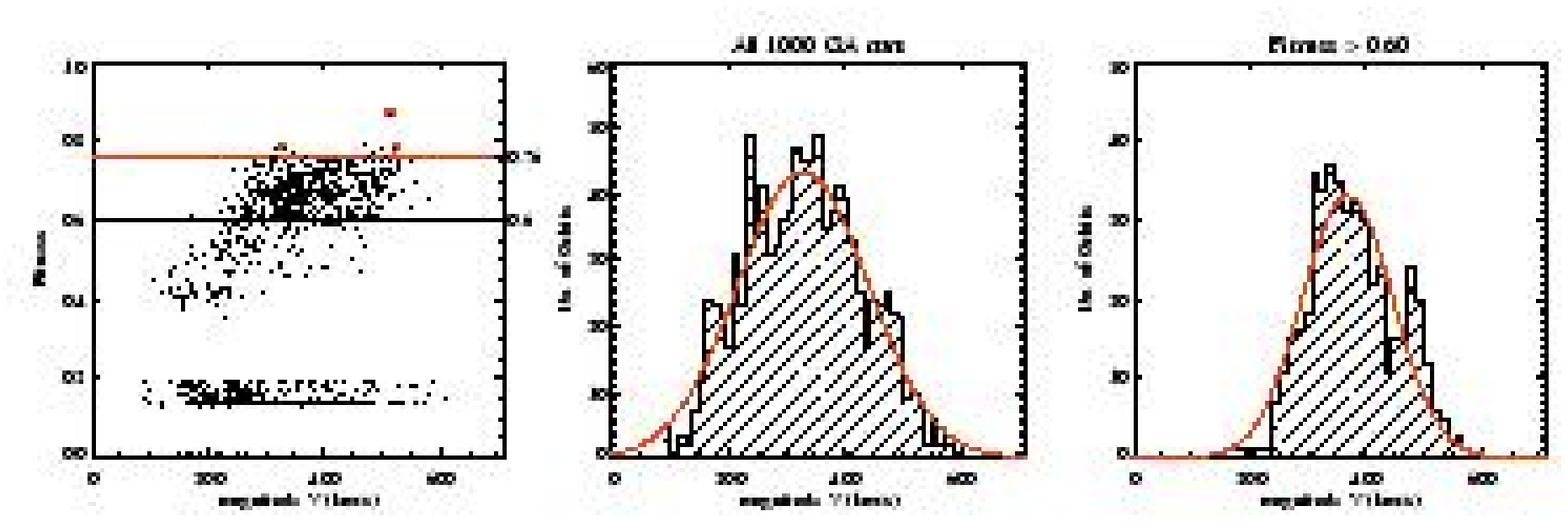}
\includegraphics[trim=0in 3.5in 0in 3.5in,clip]{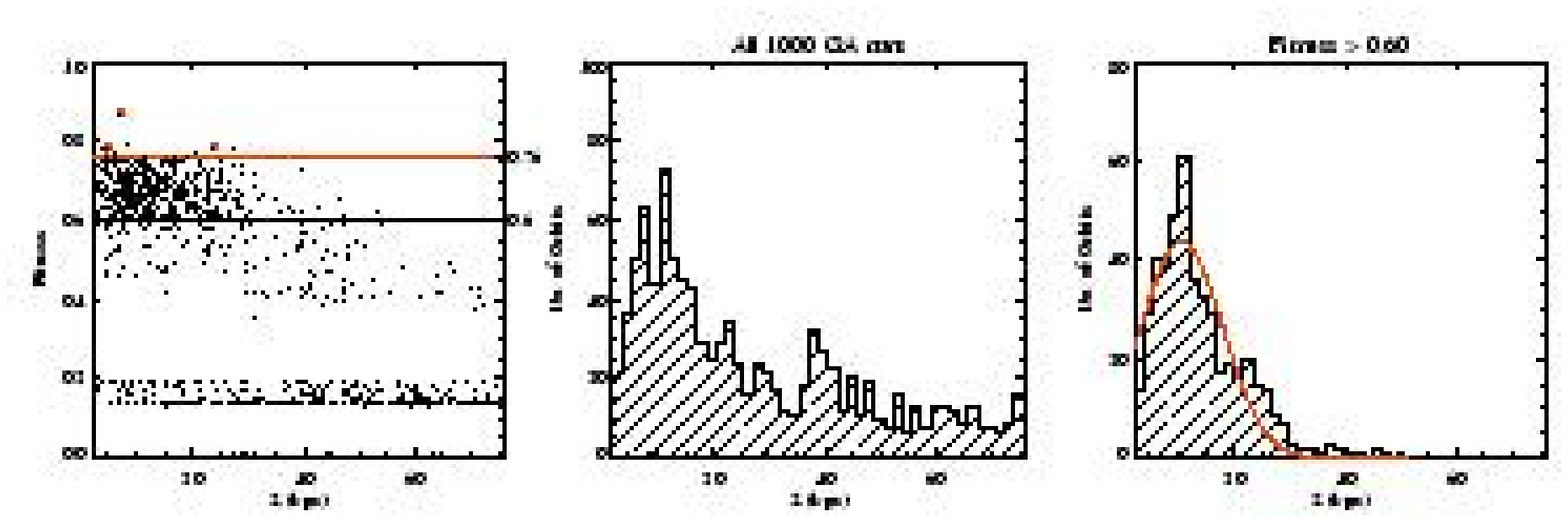}
\caption{Warm Disk Model. Same as Figure \ref{fig:cold_z} but for the warm disk model.  Orbits with Fitness $\ge 0.6$ give $r_{0,i}=109\pm11\arcsec$, $|v|=368\pm79$ \kms, $z=11\pm8$ kpc, values nearly identical to those from the cold disk models.}
\label{fig:warm_z}
\end{figure*}
Preferences by the GA for various parameter values can be seen in  Figure \ref{fig:warm_params}.
As with the cold disk model, the most notable focus is in the initial scale radius $r_{0,i}$ of NGC~205, shown in detail in Figure \ref{fig:warm_z} (top row).  Both the 1000 simulations and the orbits passing all five $\chi^2$ tests (fitness $\ge 0.6$) converge at $r_{0,i} =  108\pm 11\arcsec$ ($0.43\pm 0.04$ kpc), with the constrained three orbits (highlighted in red) focused at $113 \pm 13\arcsec$ ($0.45 \pm 0.05$ kpc).  These GA selected scale radii are close to the satellite's present scale radius of $148\arcsec$ (0.59 kpc), which implies that the internal regions of the satellite have experienced very little distortion. 
There is a greater amount of scatter in the remaining parameters, however, distinct trends still exist in the data.  Shown in the middle row of Figure \ref{fig:warm_z}, orbits with very large velocities of $368 \pm 78$ \kms\  (Fitness $\ge 0.6$) are favored, moving NGC~205 primarily towards the southeast ($v_x=229\pm54$ \kms\ and $v_y=-312\pm132$ \kms).  
In addition, the GA prefers $z$ distances very close to M31 with $z =11 \pm 8$ kpc (bottom row of Figure \ref{fig:warm_z}, Fitness $\ge 0.6$).  Beyond this distance an orbit's fitness declines rapidly. 
\downscale
\begin{figure*}\includegraphics[trim=0in 2in 0in 2.5in,clip]{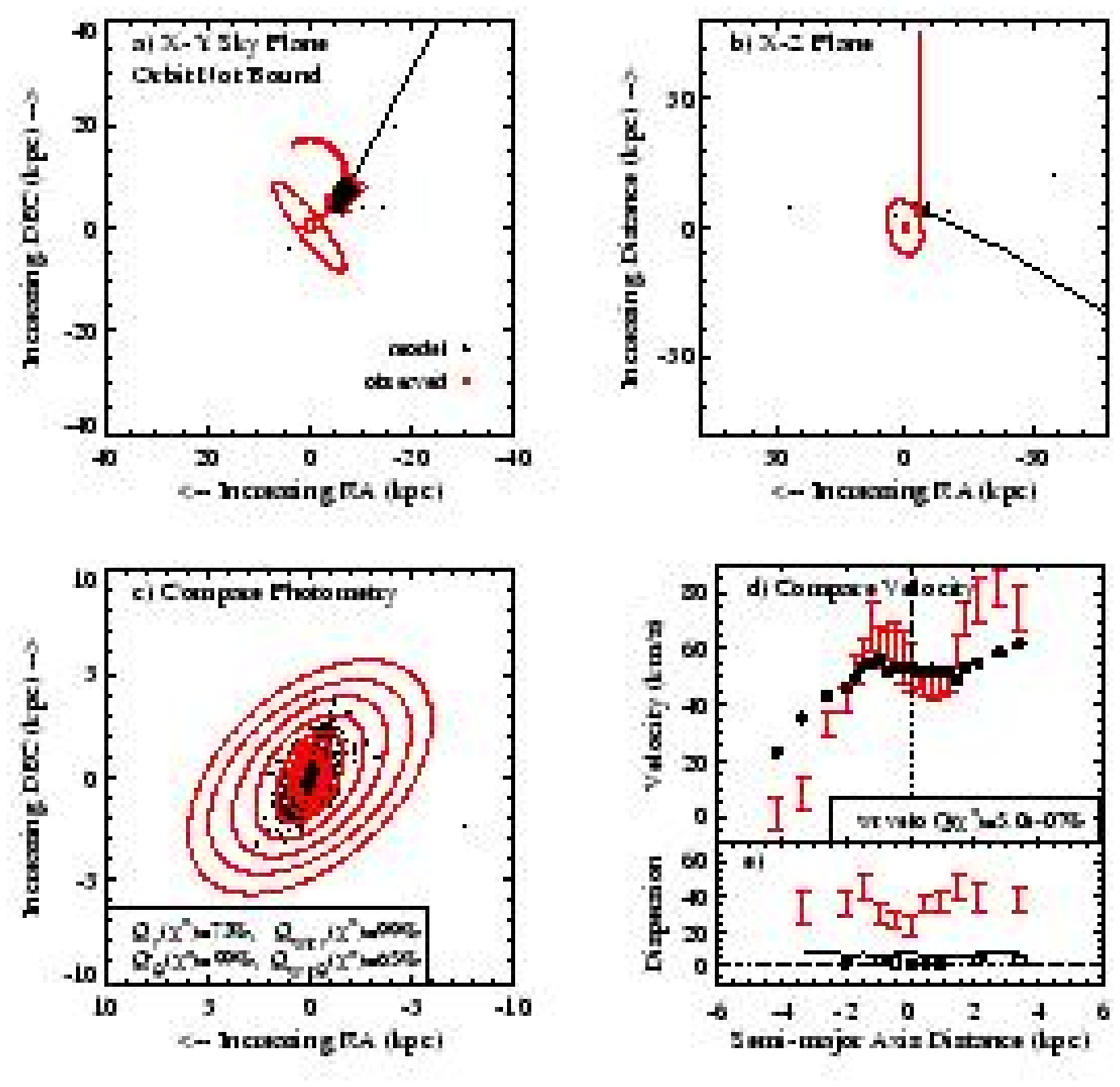}
\caption{Warm Disk Model. Same as Figure \ref{fig:cold_best} but for the warm disk model whose resulting best orbit has a fitness of 0.87.}
\label{fig:warm_best}
\end{figure*}
\returnscale
However, some parameters are not well constrained by the simulations.  As with the cold disk model, the radial velocity $v_{z}$ contains a significant amount of scatter amongst the allowed parameter values.     
In addition, the satellite's mass, $M_{205}$, and Hernquist scale length, $a_{205}$, are also unresolved parameters.  Hence, as with the cold disk results, the GA both produces inconclusive results for $M_{205}$ and $a_{205}$ and does not further reduce the observed error on $v_{z}$.  However, it is able to tightly constrain NGC~205's initial scale radius and place more general bounds on the remaining five parameters, which also happen to closely match the values from the cold disk simulations.  

\subsubsection{Best Orbit (Warm Disk)}\label{sssec_warm_best}
\upscale
\begin{figure*}\includegraphics[width=7in,trim=0in 3.7in 0in 3.7in,clip]{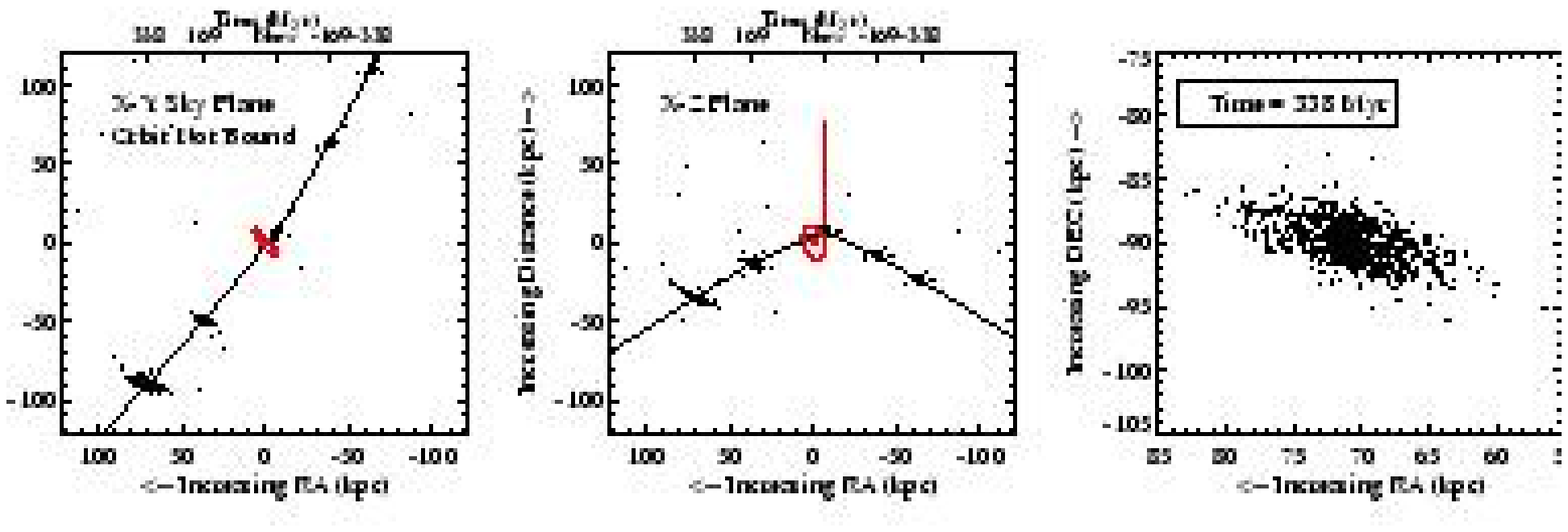}
\caption{Warm Disk Model. Same as Figure \ref{fig:cold_future} but for the warm disk model.}
\label{fig:warm_future}
\end{figure*}
\returnscale
Figure \ref{fig:warm_best} displays the resulting best orbit from the 1000 GA runs for the warm disk model.  This orbit approaches M31 from the NNW, is unbound (with $|v|= 514$\kms\  and $v_{\rm esc}=475$ \kms), and has a fitness of 0.87.  
The simulated surface brightness $\chi^2$ tests result in probabilities of 0.730, 0.994, 0.999 and 0.657 for the radial, angular, weighted radial and weighted angular tests, respectively, and in a $\chi^2$ probability of $5.0\times10^{-10}$ for the weighted velocity profile.   
This satellite is initialized with a radial dispersion at one scale length of $\sigma_{R=r_{0,i}}=3.7$\kms\  and a vertical dispersion of $\sigma_{z=r_{0,i}}=5.5$ \kms. After the simulated interaction with M31, the semi-major axis velocity profile contains an average observable dispersion of 6 \kms\  (which is slightly greater than its initial dispersion), with a maximum dispersion of 4 \kms\  at a radius of 2.1 kpc.  The parameter values of this orbit are $M_{205}=1.0\times10^{9}$\Msun, $v_{x}=269$\kms, $v_{y}=-435$\kms, $v_{z}=52$\kms, $z=7.2$ kpc, $r_{0,i}=0.5$ kpc, $\phi=83^{\circ}$, $\theta=-77^{\circ}$, and $a_{205}=7.5$ kpc.  As with the cold disk model, this orbit is projected to pass within $\approx 9$ kpc of M31's center in the plane of M31's disk.  Also, the radial path of this orbit precludes any sense of a prograde or retrograde encounter.

Figure \ref{fig:warm_future} illustrates the past, present and future predictions for NGC~205 when it is modeled as a warm disk supported by rotation and anisotropic velocities.  The distribution of particles and positions in space are given for times $-$338 Myr, $-$169 Myr, present, 169 Myr and 338 Myr.  The panel to the far right in Figure \ref{fig:warm_future} shows NGC~205 in the plane of the sky at 338 Myr.  The simulated satellite experiences a significant amount of tidal distortion after passing through the disk of M31 .

\subsection{Non-Rotating Hot Spheroid}\label{ssec_bulge}
This section is similar to \S\,\ref{ssec_colddisk} \& \S\,\ref{ssec_warmdisk} except that here NGC~205 is modeled as a non-rotating, hot spheroid of mass-less test particles supported by isotropic velocities.  Since this configuration of particles has increased velocity dispersion (compared to the disk models), we include and additional discussion about our ability to match this observed component in \S\,\ref{sssec_dispersion}.
\downscale
\begin{figure*}\includegraphics[trim=0.5in 3in 0in 3in,clip]{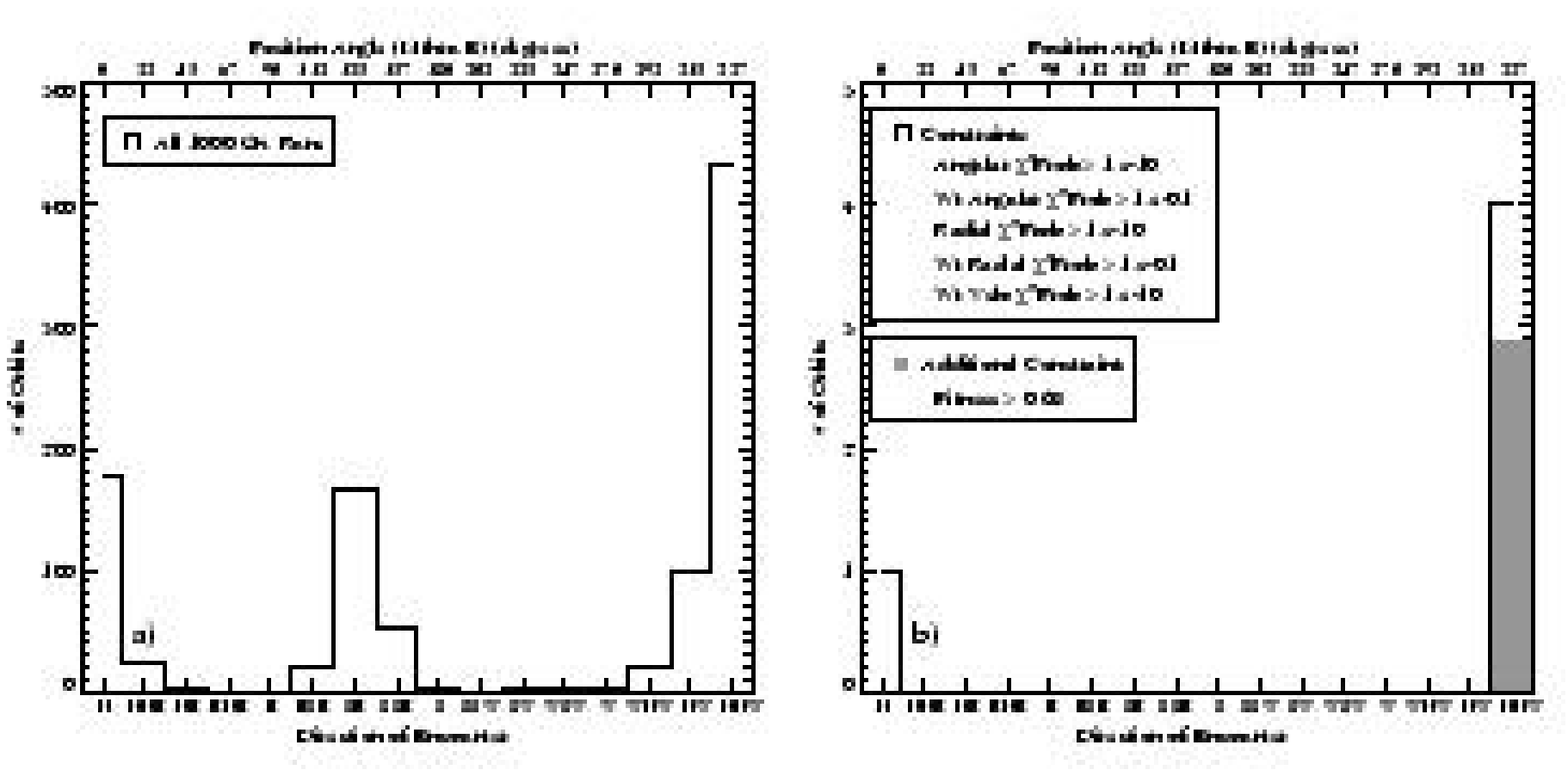}
\includegraphics[trim=0.5in 3in 0in 3in,clip]{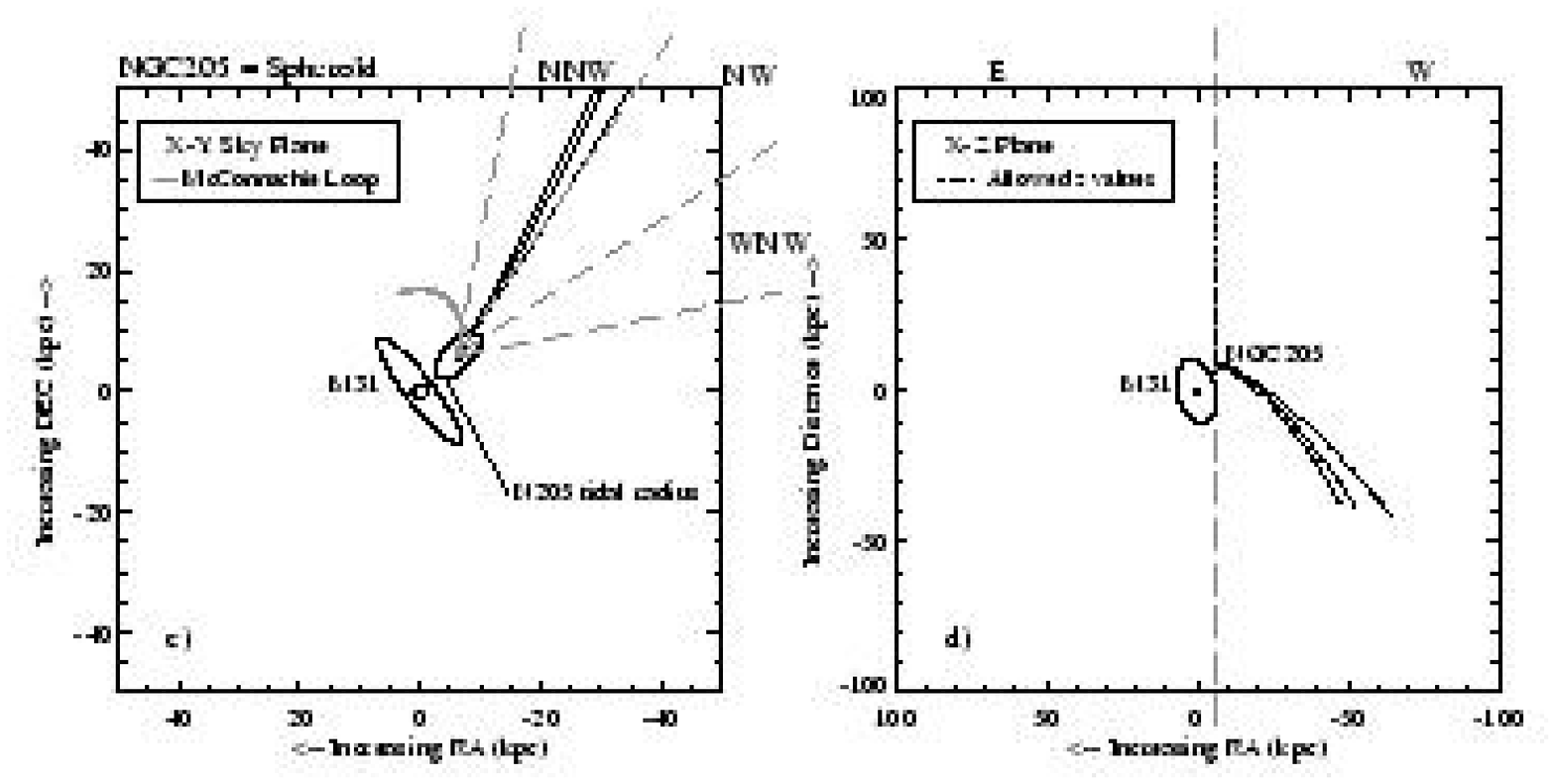}
\caption{Hot Spheroid Model. Similar to Figures \ref{fig:cold_direct} and \ref{fig:warm_direct} but for NGC~205 modeled as a non-rotating hot spheroid.  For the spheroid model the unweighted photometric constraints are set to $10^{-10}$, which are more relaxed than the disk models' constraint of 0.1.  Note that the directions of approach on the plane of the sky are similar to those found for the disk models.}
\label{fig:hot_direct}
\end{figure*}
\returnscale
\subsubsection{Direction of NGC~205's Approach (Hot Spheroid)}\label{sssec_bulge_approach}
In Figure \ref{fig:hot_direct} a clear directional preference is seen in the initial histogram of 1000 orbits.  The histogram peaks at 432 orbits with NGC~205 advancing from the north-northwest (NNW), followed by 178 orbits from the north (N). More than half the orbits are contained within these two directional bins.  Furthermore, these directions of approach reinforce those from the disk model runs, which peaked in the W (cold) and NW (warm).  Of these 1000 orbits, 653 are bound to M31 and 308 are on radial trajectories.

The top-right panel of Figure \ref{fig:hot_direct} shows a histogram of orbits with added photometric, kinematic, and fitness constraints.  The outlined histogram imposes that weighted surface brightness $\chi^2$ probabilities return values $\ge 0.1$, and that the unweighted surface brightnes and weighted velocity $\chi^2$ probabilities $\ge 10^{-10}$.  Notice that these constraints differ from those applied to the exponential disk models.  Since the hot spheroidal model follows a Hernquist profile characterized by an $r^{1/4}$ law (and not an exponential profile), we relax the surface brightness $\chi^2$ conditions that are more sensitivite to the satellite's inner regions, a portion that has experienced little, if any, tidal distortion.  These imposed constraints reduce 1000 orbits to 5 and return a direction of approach lying somewhere between the north (N) and north-northwest (NNW), with a peak of 4 orbits in the NNW.  The shaded region enclosed within the histogram further imposes that the Fitness $\ge 0.685$, a value selected to be just below that of the top 10 orbits (or $1\%$).  This constraint reduces the 5 orbits to 3, indicating that 7 of the top 10 orbits fail to satisfy all the imposed photometric and kinematic constraints given above.   The remaining directions of approach all fall in the NNW, ruling out N approaches, a result similar to the disk models' findings.  Hence, the initial directional preference suggested by the 1000 orbits histogram is reinforced with the addition of photometric, kinematic and fitness constraints. 

These restrictions rule out other possible orbits, including those tracing the stellar arc-like feature seen to the north of M31.  The hot spheroid simulations place a mere $2.5\%$ of the 1000 GA orbits in the 4 bins NNE through E.  None of these orbits meet a weighted angular $\chi^{2}$ constraint of $0.1$.  Hence, orbits tracing the observed stellar arc are ruled out as solutions by the twisting of elliptical isophotes in the tidally distorted regions of NGC~205.

\subsubsection{Preferred Parameter Values (Hot Spheroid)}\label{sssec_bulge_params}
\begin{figure*}\includegraphics[trim=0in 2.2in 0in 2.2in,clip]{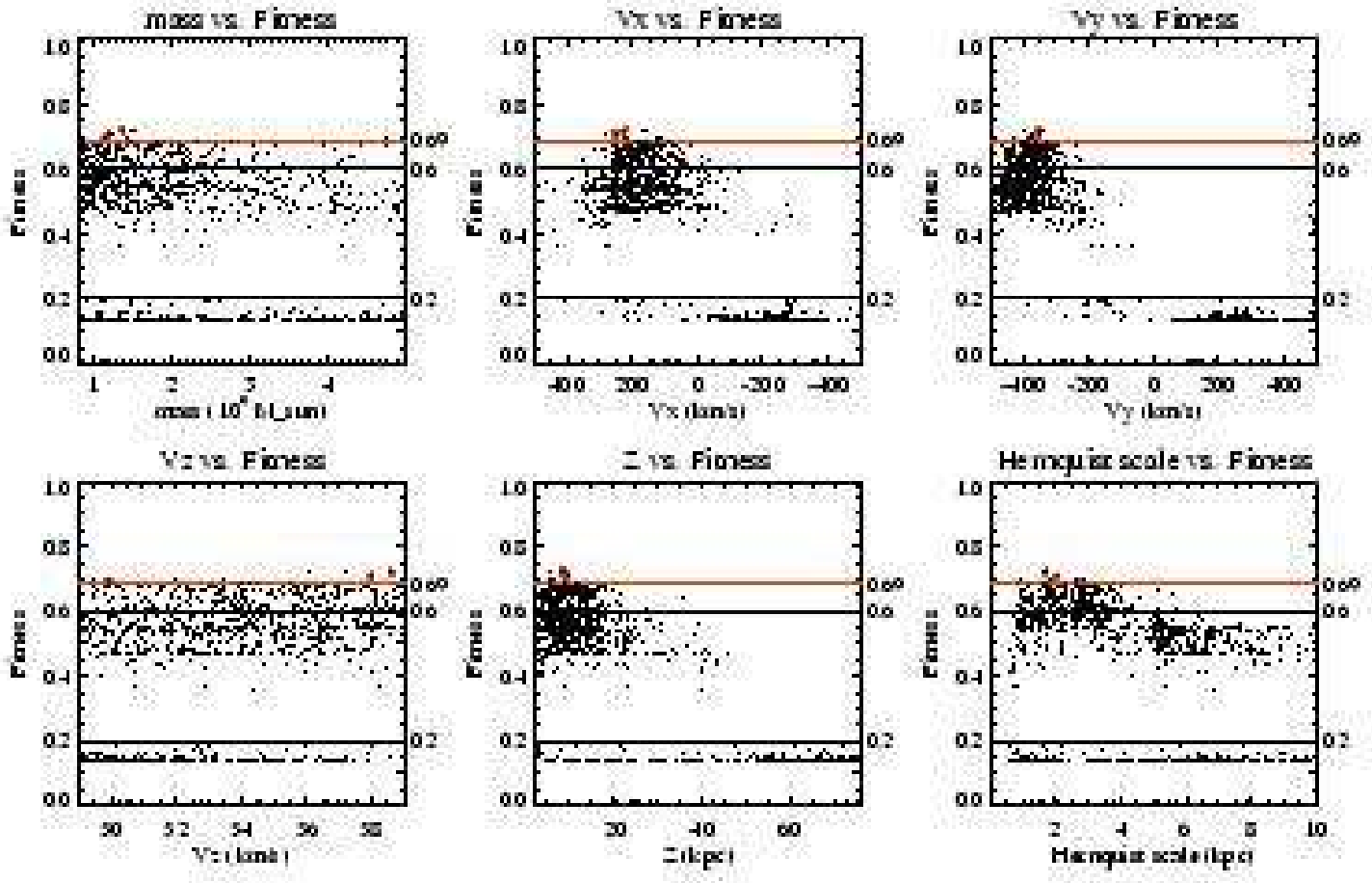}
\caption{Hot spheroid. Same as Figures \ref{fig:cold_params} and \ref{fig:warm_params} but for the hot spheroid model with only three constrained orbits, highlighted in red. Notice that the spherical nature and density profile of this model reduces the parameters from 9 to 6.}
\label{fig:hot_params}
\end{figure*}
\downscale
\begin{figure*}\includegraphics[trim=0in 3.5in 0in 3.5in,clip]{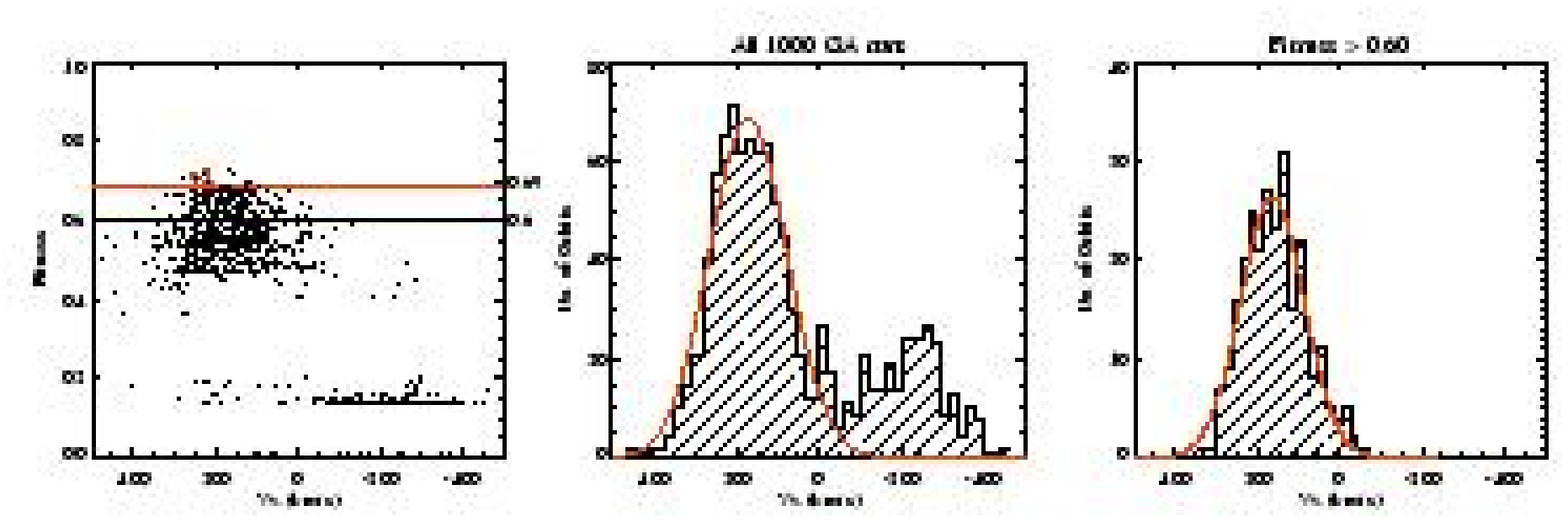}
\includegraphics[trim=0in 3.7in 0in 3.7in,clip]{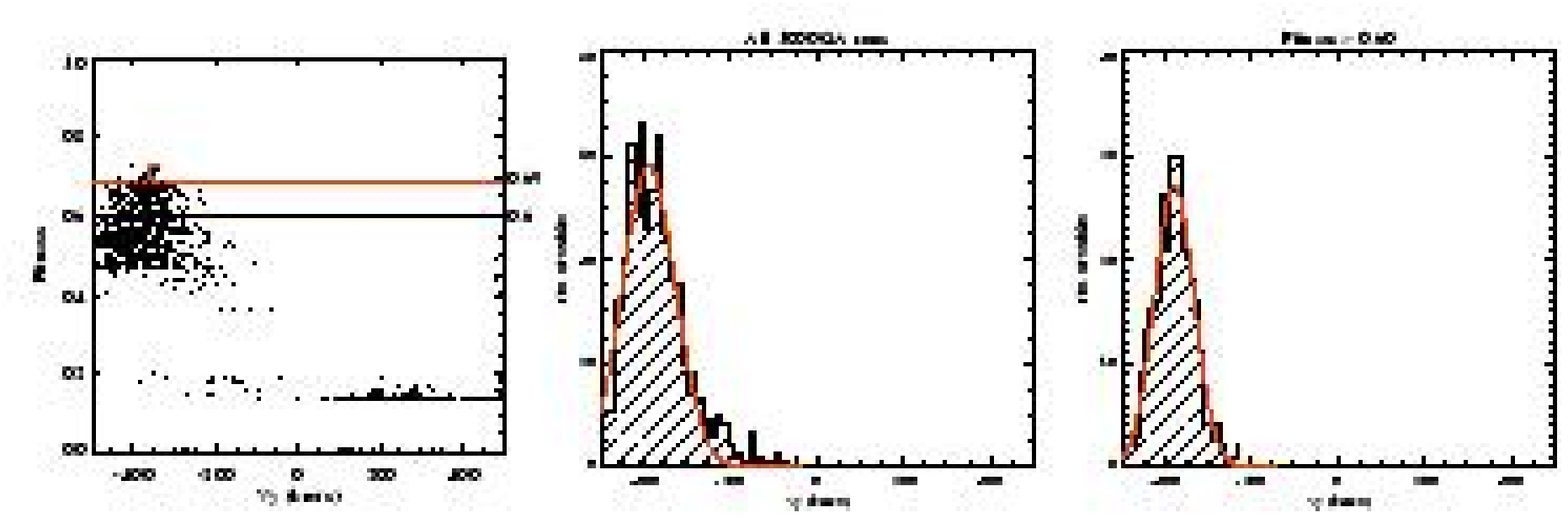}
\includegraphics[trim=0in 3.5in 0in 3.5in,clip]{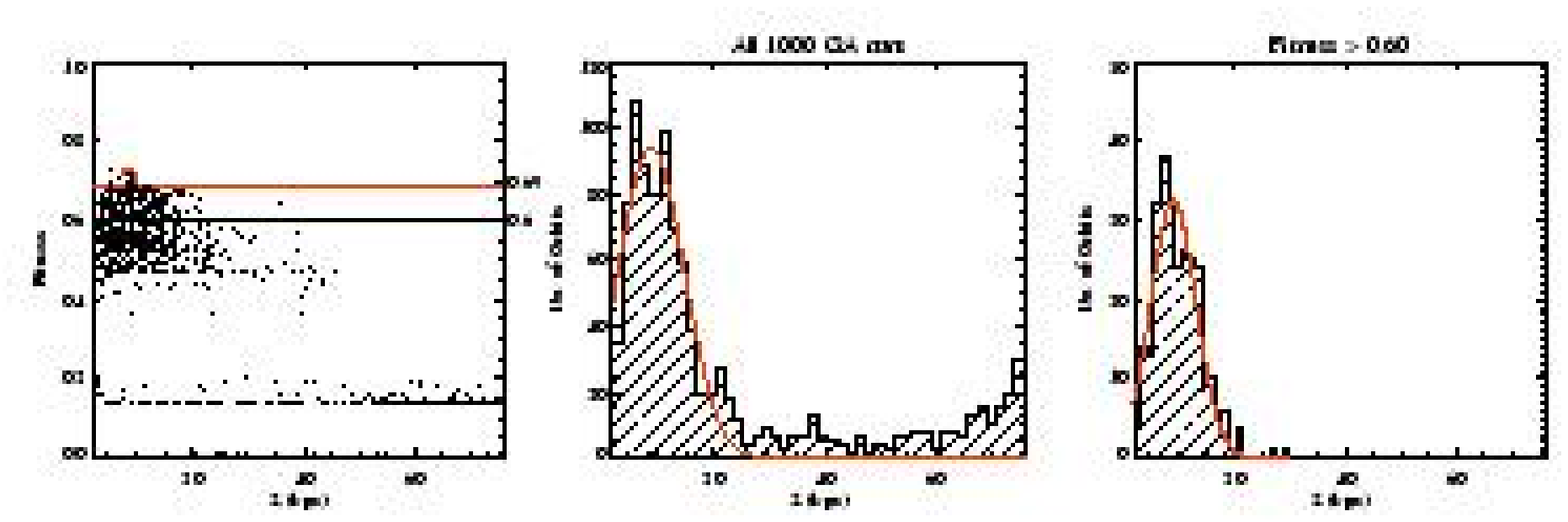}
\caption{Hot Spheroid Model. Similar to Figures \ref{fig:cold_z} and \ref{fig:warm_z} but instead showing $v_x$, $v_y$ and $z$ for the hot spheroid model.  Orbits with Fitness $\ge 0.6$ give $v_{x}=163\pm77$ \kms, $v_{y}=-376\pm46$ \kms, $|v|=417\pm33$ \kms\ (not shown), and $z=9\pm5$ kpc, values nearly identical to those from the disk models.}
\label{fig:hot_z}
\end{figure*}
\returnscale
The resultant parameter values from the 1000 hot spheroid GA runs are shown in Figure \ref{fig:hot_params}.  The spherical nature and density profile of this model effectively reduces the number of parameters from 9 to 6, eliminating $r_{0,i}$, $\phi$, and $\theta$.    

The GA's preference for certain parameter values is quantified in Figure \ref{fig:hot_z}.
The top two rows (Fitness $\ge 0.6$) show the well resolved tangential components of the velocity vector with $v_x=163\pm77$ \kms\ and $v_y=-376\pm46$ \kms, effectively moving NGC~205 towards the southeast.  This results in a net velocity of $|v|=417\pm33$ \kms\ (Fitness $\ge 0.6$).  Contrastly, only the magnitude of the velocity could be well resolved in the case of the cold disk model.
The GA also prefers $z$ distances very close to M31 with $z =9 \pm 5$ kpc (bottom row of Figure \ref{fig:hot_z}, Fitness $\ge 0.6$).  As with the disk models, an orbit's fitness beyond this distance declines rapidly. 
In the bottom-right plot of Figure \ref{fig:hot_params}, a correlation of higher fitness to lower $a_{205}$ values can be seen.  This is a result of the initial particle distribution's dependence on $a_{205}$ in Eqn \ref{eq:hern_rho}, a dependence that did not exist for the disk models.  Hence, the GA's attempt to reproduce the exponential surface brightness profile results in this correlation. 
\downscale
\begin{figure*}\includegraphics[trim=0in 2in 0in 2.5in,clip]{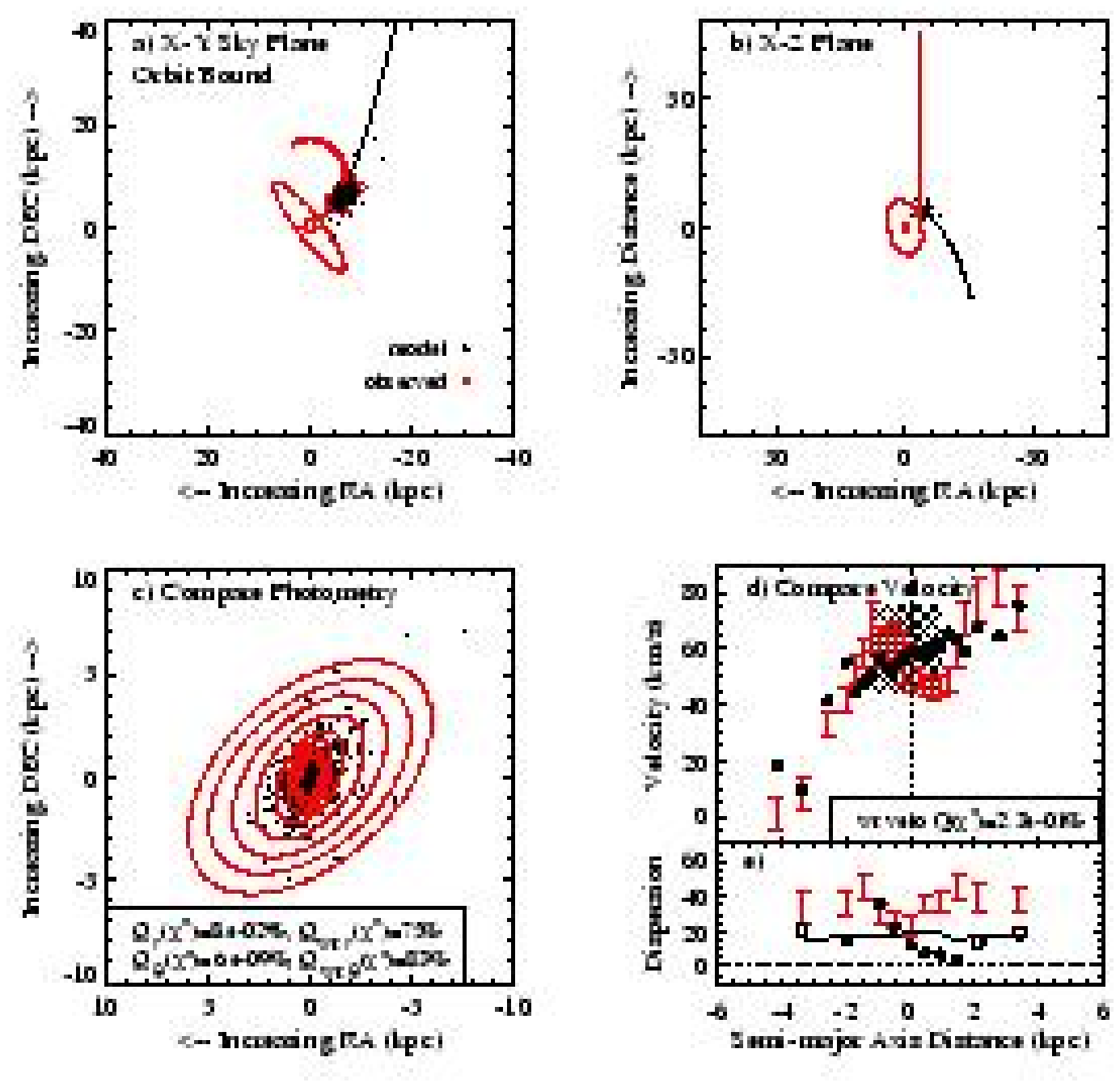}
\caption{Hot Spheroid Model. Same as Figures \ref{fig:cold_best} and \ref{fig:warm_best} but for the hot spheroid model whose resulting best orbit has a fitness of 0.73.  (d) Note that the inner velocity profile is hatched out indicating that it was not used to determine the orbit's fitness.}
\label{fig:hot_best}
\end{figure*}
\returnscale
However, Figure \ref{fig:hot_params} also demonstrates that some of the parameters are not well constrained by the simulations.  As with the disk models, the radial velocity $v_{z}$ contains a significant amount of scatter and is likely not constrainable beyond observations.   
In addition, the satellite's mass, $M_{205}$ is unresolved. 
 Hence, as with the disk results, the GA produces inconclusive results for $M_{205}$ and does not further reduce the observed error on $v_{z}$.  However, it is able to reasonable constrain NGC~205's tangential velocity and line-of-sight distance, values which also happen to closely match those from the disk simulations.  

\subsubsection{Best Orbit (Hot Spheroid)}\label{sssec_bulge_best}
\upscale
\begin{figure*}\includegraphics[width=7in,trim=0in 3.7in 0in 3.7in,clip]{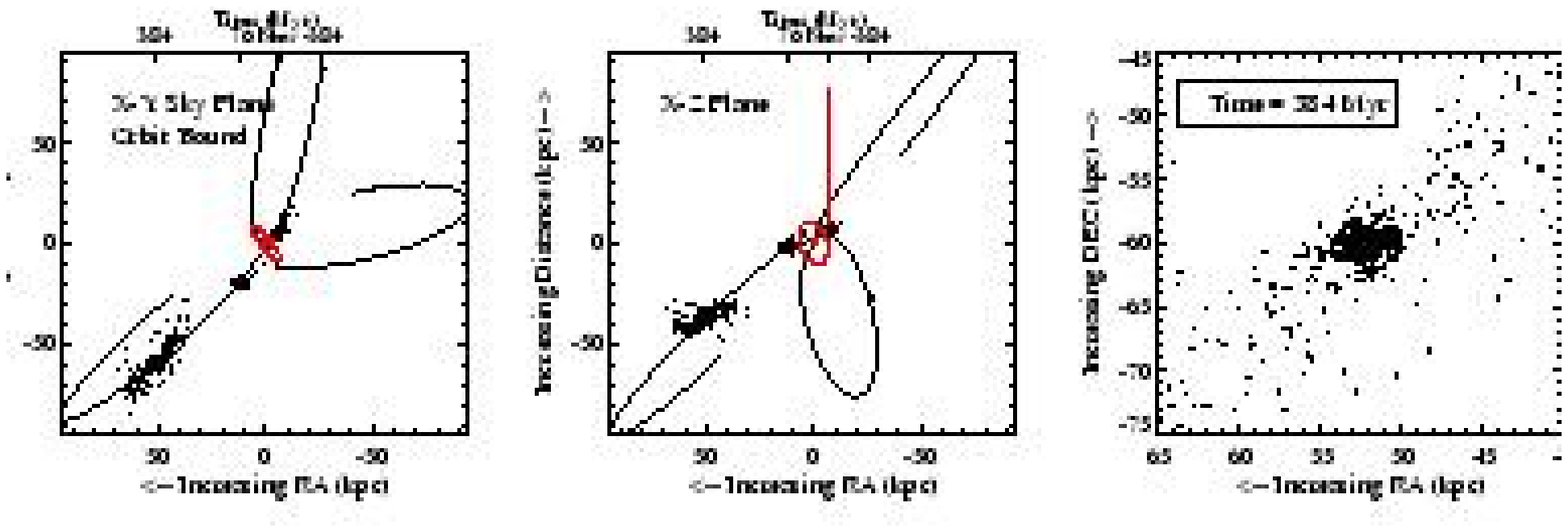}
\caption{Hot Spheroid Model. Same as Figures \ref{fig:cold_future} and \ref{fig:warm_future} 
but for the hot spheroid model. The positions are given at times $-$384 Myr, present, 76 Myr and 384 Myr.  In this model, NGC~205 traces out a loop orbit.  The panel on the right is a zoomed in view of NGC~205 at 384 Myr on the plane of the sky.}
\label{fig:hot_future}
\end{figure*}
Figure \ref{fig:hot_best} displays the hot spheroid's resulting best orbit from the 1000 GA runs.  This orbit approaches M31 from the NNW, is just barely bound (with $|v|= 434$\kms\  and $v_{\rm esc}=483$ \kms), and has a Fitness of 0.73.  Notice that on the bottom-right plot the inner velocity profile is hatched out indicating that the slope test was not used for this model.  The simulated surface brightness $\chi^2$ tests result in probabilities of $8\times10^{-10}$, $6\times10^{-10}$, $0.757$ and $0.83$ for the radial, angular, weighted radial and weighted angular tests, respectively, and in a $\chi^2$ probability of 0.002 for the weighted velocity profile.  The orbit's poor performance on the unweighted surface brightness $\chi^2$ tests is a direct result of comparing the simulated spheroid, constructed using a Hernquist density profile, to the observed NGC~205, which follows an exponential profile.  Despite this fact, the orbit performs quite well on the weighted $\chi^2$ tests that model the tidally distorted regions of NGC~205.  After the simulated interaction with M31, the semi-major axis velocity profile contains an average observable dispersion of 17 \kms\  (which is lower than the observed dispersion of 42 \kms), with a maximum dispersion of 36 \kms\  at a radius of 1.0 kpc.  The parameter values of this orbit are $M_{205}=1.5\times10^{9}$\Msun, $v_{x}=168$\kms, $v_{y}=-397$\kms, $v_{z}=57$\kms, $z=5.8$ kpc, and $a_{205}=1.7$ kpc.
As with both disk models, this orbit is projected to pass within $\approx 9$ kpc of M31's center in the plane of M31's disk. Since this satellite is not rotating, prograde or retrograde definitions do not apply.

Figure \ref{fig:hot_future} illustrates the past, present and future predictions for NGC~205 when it is modeled as a hot spheroid supported by isotropic velocities.  The distribution of particles are given for times $-$384 Myr, present, 76 Myr and 384 Myr.  The gravitationally bound satellite traces out a loop orbit.  The panel to the far right in Figure \ref{fig:hot_future} shows NGC~205 in the plane of the sky at 384 Myr.  The simulated satellite experiences a significant amount of tidal distortion after passing through the disk of M31 . 

\subsubsection{Semi-Major Axis Velocity Dispersion}\label{sssec_dispersion}
Figures \ref{fig:cold_best}, \ref{fig:warm_best} and \ref{fig:hot_best} show the best fit orbit for each of the three models.  However, none these orbits produce profiles that match the observed velocity dispersion profile.
This is expected since our fitness tests do not contain a velocity dispersion profile test, meaning our orbits are not encouraged to match this observable.  The main reason this feature is not used is because the low number of particles ($1,000$) used for each orbit makes it nearly impossible to fit a histogram and recover an accurate measurement for the velocity dispersion at all points along the semi-major axis of NGC 205. 

\begin{figure*}\includegraphics[trim=0in 3.5in 0in 3.5in,clip]{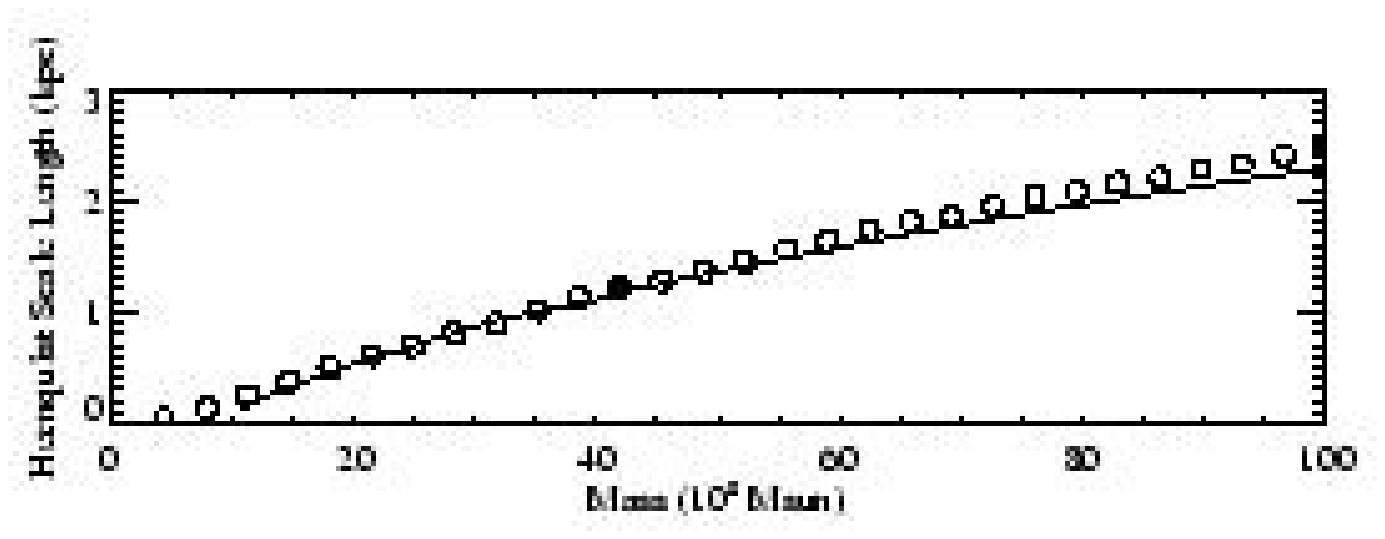}
\includegraphics[trim=0in 2in 0in 2.3in,clip]{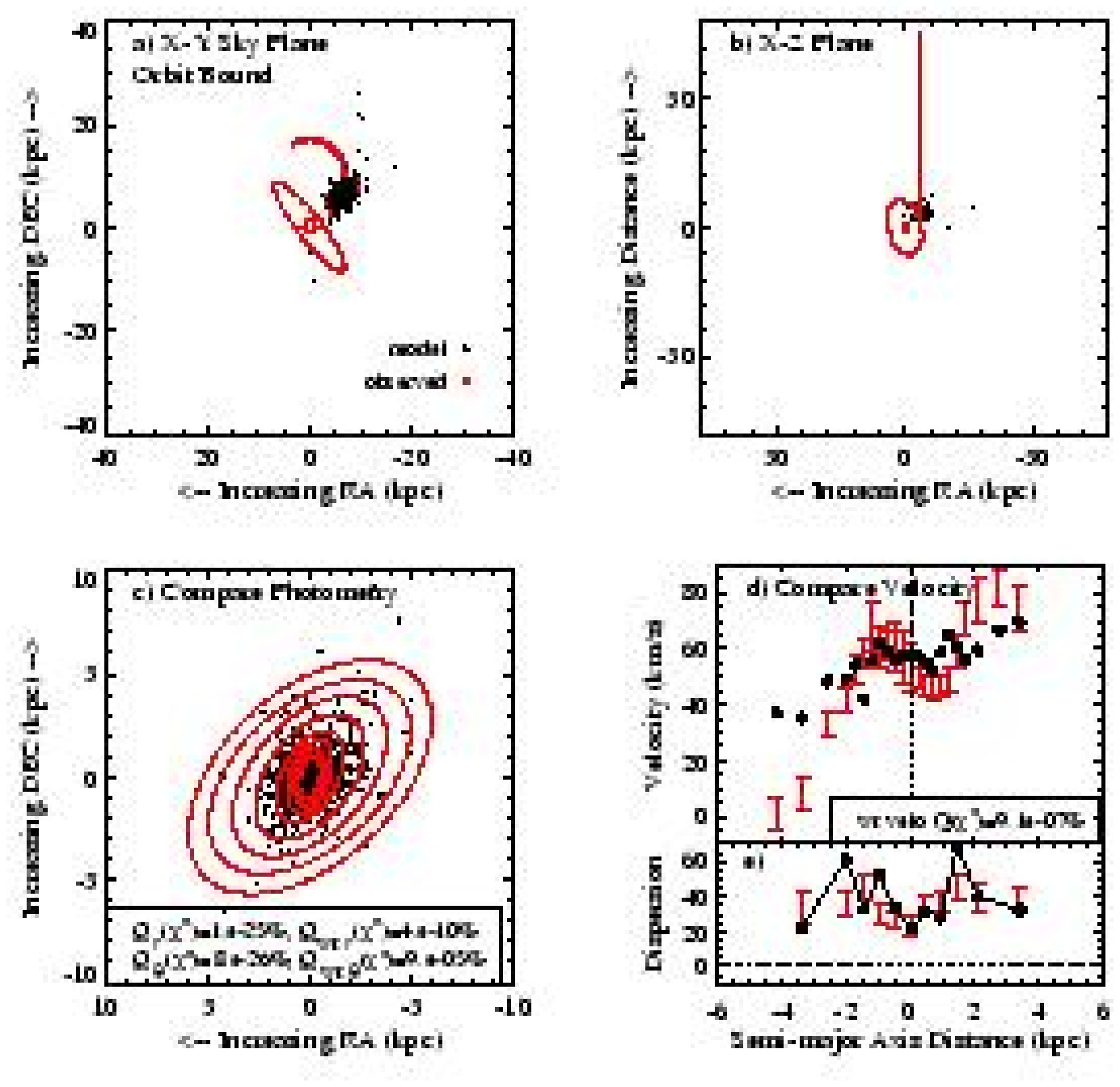}
\caption{Hot Spheroid Model.  The circles in the top panel show, for the tidally distorted best-fit hot spheroid orbit, the  combinations of the mass $M_{205}$ and Hernquist scale length $a_{h}$ parameters that reproduce the observed mean velocity dispersion of 35\kms.  The solid line gives analytic predictions for these parameter in a tidally undistorted satellite with a projected velocity dispersion of 35 km s$^{-1}$ at a radius of 1 kpc. The bottom four panels correspond to the orbit denoted by the black point in the top panel, and are similar to Figure 18, but with 3000 particles, a mass of $4.2 \times 10^{9} \Msun$, and Hernquist scale length of $1.2$ kpc. Note that this model progenitor reproduces the observed velocity dispersion profile.}
\label{fig:m_vs_ah}
\end{figure*}

We expect that of the three models the hot spheroid model, supported completely by isotropic velocities, is most capable of reproducing the observed dispersion.  
To test this, we use $3,000$ mass-less test particles plus the best fit hot spheroid model's velocity and line of sight distance parameters given in \S\,\ref{sssec_bulge_best}, and we conduct a coarse, iterative search in the mass ($M_{205}$) and Hernquist scale length ($a_{205}$) dimensions.  We search these two parameters because they are unconstrained in all three models and heavily influence the simulated velocity dispersion profile.  After producing discrete relations between $a_{h}$ and the mean velocity dispersion, we fit a power law line and interpolate to find the values of $a_h$, at each value of $M_{205}$, that result in a mean dispersion of 35 \kms. This result is displayed as circles in the top panel of Figure \ref{fig:m_vs_ah}.  The solid line in the top panel of Figure \ref{fig:m_vs_ah} indicates the analytic solution for these parameter values in a tidally undistorted satellite with a projected velocity dispersion of 35 km s$^{-1}$ at a radius of 1 kpc (\citet{her90}, Eqn [42]).   Although this line measures the velocity dispersion at a given radius, rather than the mean velocity dispersion, the result suggests that tidal interactions between at hot spheroid progenitor and parent galaxy have not significantly impacted the velocity dispersion within a 1 kpc radius.

In addition to matching the observed mean dispersion, we find that all these combinations of $M_{205}$ and $a_h$ sufficiently reproduce the velocity dispersion profile, including the observed dip at the center.  We randomly select one of these points ($M_{205} = 4.2 \times 10^{9} \Msun$, $a_{h}=1.2$ kpc), denote it with a black circle and display the orbit in the bottom 4 panels of Figure \ref{fig:m_vs_ah}.  The poor fitness values reported by this orbit are not concerning since this orbit was not a true result of the optimization scheme.  Hence, while the mass and Hernquist scale length parameters are not constrained in the optimized parameter space search, their relative values are suggested by the observed velocity dispersion.

\downscale
\returnscale

\section{DISCUSSION}\label{sec_discussion}
This paper presents the methods and results of an optimized, restricted $N$-body search for NGC~205's orbit and system parameters.  We account for uncertainties in the parameter space by placing very liberal upper and lower limits on our parameters, using three dynamically distinct models for the initial configuration of NGC~205, and running our optimization algorithm (GA) 1000 times on each model.  We find that out of the $10^{22}$ possible orbits, not only are certain orbits and portions of parameter space carved out, but this convergence is also model independent.  These findings are outlined below:

\begin{itemize}

\item The simulations indicate that NGC~205 is approaching from the NW region on the plane of the sky, moving towards the SE (i.e. increasing in RA, decreasing in DEC), with the cold disk model favoring an approach from the NW and the warm disk \& hot spheroid models favoring the NNW.  

\item These orbits do not trace out nor do they intersect the $1^{\circ}$ long stellar arc observed to north of M31, which was hypothesized as a tidal stream emanating from NGC~205 by \citet{mcc04}.  Orbits that trace out this region on the plane of the sky (i.e. approaches from the NNE through E region) poorly match the photometry of NGC~205, specifically, the isophotal twisting. Instead, approaches from the NW-NNW region are preferred, thus ruling out the stellar arc as a trail of debris from NGC~205.

\item Large velocities in the range 300--500 \kms\  are highly favored for NGC~205.   
While the majority of the orbits found by the GA are bound, the best fitting orbits are either very close to escape velocity or unbound.  However, such velocities are not completely uncommon for local group satellites. Proper motion measurements reveal that the Large Magellanic Cloud is very near escape velocity and likely on its first passage about the Milky Way \citep{kal06, bes07}. The dwarf spheroidal galaxies And XIV and And XII are near, or exceed, their local escape speeds and are presumably falling into the Local Group for the first time \citep{maj07, cha07}.  The large tangential component of NGC~205's simulated velocities translates into a proper motion of $\sim 0.1$ mas yr$^{-1}$.

\item The GA is able to better fit NGC~205's observed kinematic profile when the satellite's line-of-sight distance is very close to M31. The resulting fits report relative distances between NGC~205 and M31 in the range 2--20 kpc.  Note that 2 kpc is our lower bound on the line sight distance parameter.  
These distances correspond to a change of $0.077^{+0.018}_{-0.015}$ mag in \citet{mcc05} measurements of NGC~205's red giant branch tip (TRGB), analogous to either an increase in their adopted TRGB absolute magnitude ($M_{I}^{\rm{TRGB}}$) and/or reddening correction ($A_{I}$), or in a decrease in the TRGB observed magnitude ($I_{\rm{TRGB}}$).  Furthermore, this magnitude difference is in agreement with the $\pm 0.07$ mag errors quoted by \citet{mcc05} on the individual distance measurements to NGC~205 and M31. 
As an aside, a magnitude difference could also be accounted for by NGC~205's composite stellar population \citep{sal05} resulting from a recent ($\sim 0.5$ Gyr) burst of star formation \citep{mar06}.  Given these possible corrections and the close relative distances found by the simulation, the question is raised: \textit{Could NGC~205 be in front of M31 rather than behind it?}

\item The disk models return tightly constrained initial scale lengths in the range 0.39--0.55 kpc, values approximately equal to NGC~205's current scale length of 0.59 kpc.  This result is consistent with the photometric and kinematic observations which suggest the central region of NGC~205 is largely unaffected by tides.  Hence, both the inner kinematics and photometry are determined by the initial parameters of the system.

\item  
The cold disk model demonstrates its plausibility as a configuration for NGC~205 by the induction of velocity dispersions into its initially cold profile via tidal interactions.  Hence, all three models contain some dispersion along their final semi-major axis velocity profile. 

\item For the best fit hot spheroid model, we can reproduce the observed mean velocity dispersion of 35\kms and the semi-major axis velocity dispersion profile by using specific combinations of two otherwise unconstrained parameters, satellite mass ($M_{205}$) and Hernquist scale length ($a_{205}$).

\item For all three models, NGC~205's large tangential velocities and close approaches to M31 indicate that it is difficult to distort a satellite to the extent that NGC~205 is observed.  In order to reproduce the abrupt turnover and reversal in the velocity profile, NGC~205 must come very near to M31. While the photometric profile can be reproduced at large distances, the resulting velocity turnover is less abrupt and the reversal is not as steep. However, at very close approaches ($< 20$ kpc), it is very easy to disrupt NGC~205.  In order to avoid complete disruption and/or long tidal streamers, the satellite must approach M31 with a very high velocity.  Hence, the combined photometric and kinematic observations have managed to remove degeneracies in some of the parameter space.

\item The constrained orbits are primarily radial and projected to pass within 10 kpc of M31's center. For the cold disk model, orbits that are not radial and have a fitness $> 0.25$ are prograde.  In contrast, a larger fraction of the warm disk model's orbits with fitness $> 0.25$ are retrograde.  This result is not surprising since the warm disk model contains some initial velocity dispersion, allowing its satellites to be more easily distorted.  Thus, some of the warm disk satellites are able to reproduced milder photometric distortions by approaching on retrograde orbits \citep{rea06}. 

\item Given that many of the best fitting orbits are both radial and unbound, it is not surprising that NGC~205 has only experienced a small amount of distortion.   That is, an unbound orbit indicates that NGC~205 is on its first passage while satellites on radial orbits are more difficult to disrupt than those on prograde orbits \citep{rea06}.  Although the future projections show NGC~205 whizzing through M31, inclusion of dynamical friction would change this result by slowing the satellite after hitting M31 for the first time \citep{seg94}.  This braking mechanism is so efficient that it is likely NGC~205 will eventually merge with M31. 

\end{itemize}

Although the GA's search was quite successful, it was unable to constrain the mass and dark matter content.  This latter result could be due to either degeneracies in the parameter space or an insensitivity to these parameters by the fitness tests. Furthermore, it appears that to some degree the algorithm did get stuck in local optima, a result evidenced by the scatter in the solutions and the fact that some orbits had very low fitness.  Although we attempt to circumvent this undesirable result by running the GA numerous times, we cannot be 100\% certain that we converge on the ``best orbit'' in the parameter space.  However, we do find that other GA and fitness test configurations result in declining levels of a best orbit fitness' and are thus confident that our method is effective. While other optimization tools are available (i.e. simulated annealing), they too have their problems and limitations, and cannot offer guaranteed improvements.

\section{FUTURE DIRECTIONS}\label{sec_future}
We have presented the results of optimized restricted $N$-body simulations of NGC~205's tidal interaction with M31 using initially self-consistent satellite models and parameter spaces tailored specifically to the system.  Although this simple model produces satisfactory results, improvements to the simulations will generate a more detailed and accurate account of the interaction, possibly produce tighter constraints on the parameters, and provide greater insight into the past, present and future of NGC~205.  
In this final section we discuss two observational advances that can be made and a handful of numerical improvements that can be applied to the NGC~205-M31 system in order to better and more accurately model their interaction.

The simulations can be significantly improved by reducing the observational errors.
As discussed, the current value for the line-of-sight distance between NGC~205 and M31 is estimated to be $39\pm37$ kpc \citep{mcc05}.  
An improvement on the relative distance between NGC~205 and M31 would significantly constrain the parameter space.
In addition, if NGC~205 indeed has a large tangential motion of $300-500$ \kms, the satellite's proper motion could be resolved at long wavelengths over the course of a decade or two, thus providing another significant constraint on the system.  These two measurements, in conjunction with current sky positions, radial velocities, and models of M31's mass distribution, would provide a complete description of NGC~205's orbit.

Our simple model of NGC~205's interaction with M31 yields results that are consistent and relatively independent of model type.  An obvious improvement to the current simulation is to use velocity dispersion as a constraint on the system.  Another improvement, applying to only the spheroid model, would be to weight the brightness of the mass-less test particles such that the initial surface brightness profile matches the observed exponential profile of NGC~205.  Furthermore, this system can be better modeled with the inclusion of dynamical friction and mass loss, effects which act to slow the system down and increase the level of distortion in the satellite.  The addition of these processes effectively transforms our time-independent model into a time-dependent problem.  Another modification is to construct NGC~205 as a self gravitating model with both dark and luminous particles, thus switching from a restricted $N$-body to a full $N$-body simulation.  However, this more realistic model changes the problem from $\mathcal{O}(N)$ to a $\mathcal{O}(N^{2})$, leading to a notable increase in computation time and making it extremely costly to run the simulation along with a genetic algorithm. Despite this complication, a self-gravitating model has a significant advantage in that it allows for evolution in NGC~205.  This construction provides the necessary platform to test various formation scenarios for the satellite since it can be initialized as a spiral or dIrr and then transformed into a dE through tidal interactions.  If this improvement were to be made, one might also consider adding an initial gas mass to the progenitor galaxy and performing hydrodynamic simulations for completeness.  Last but not least, NGC~205's orbit can be improved by adding the perturbations from M31's other satellites, such as M32, as well as the potential of the Milky Way.   

   \acknowledgments 

We thank P. Choi for generously providing his photometric data table and acknowledge Sara Ellison and Henk Hoekstra for allowing us access to the University of Victoria's computing cluster during the early phases of this project.  K.M.H. thanks D. Dearborn, C. Rockosi, N. Konidaris, G. Novak and I. Dobbs-Dixon for the many productive conversations, as well as LLNL, KITP, and a Eugene Cota Robles Fellowship from the University of California Santa Cruz for support during this project.  K.M.H. and P.G. gratefully acknowledge partial funding support for this work from NSF grants AST-0307966 and AST-0607852 and NASA/STScI grant GO-10794.02.  G.L. and K.V.J. were supported by NSF CAREER awards AST-0449986 and AST-0133617, respectively.

\end{document}